\documentclass{ws-ijmpa}
\usepackage{cite}
\usepackage{graphicx}
\usepackage{amsmath}
\usepackage{amssymb}
\usepackage{slashed}
\usepackage{bbm}
\begin{document}
\markboth{M.~Moinester and S.~Scherer}{Electromagnetic
Polarizabilies of Pions}

\catchline{}{}{}{}{}
%
%%%%%%%%%%%%%%%%%%%%%%%%%%%%%%%%%%%%%%%%%%%%%%%%%%%%%%%%%%%%%%%%%%%%

\title{COMPTON SCATTERING OFF PIONS AND ELECTROMAGNETIC POLARIZABILITIES}

\author{Murray Moinester}

\address{School of Physics and Astronomy, R.~\& B.~Sackler Faculty of Exact Sciences, Tel Aviv University\\
69978 Tel Aviv, Israel\\
murray.moinester@gmail.com}

\author{Stefan Scherer}

\address{Institute for Nuclear Physics, Johannes Gutenberg University Mainz, J.~J.~Becher-Weg 45\\
D 55128 Mainz, Germany\\
stefan.scherer@uni-mainz.de}

\maketitle

\begin{history}
\received{14 May 2019}
%\revised{}
\end{history}

\begin{abstract}
      The electric ($\alpha_\pi$) and magnetic ($\beta_\pi$) Compton polarizabilities of both the charged and the
neutral pion are of fundamental interest in the low-energy sector of quantum chromodynamics (QCD).
   Pion polarizabilities affect the shape of the $\gamma\pi\to\gamma\pi$ Compton scattering angular distribution at back scattering angles and
$\gamma\gamma\to\pi\pi$ absolute cross sections.
   Theory derivations are given of the $\gamma\pi\to\gamma\pi$  Compton scattering differential cross section, dispersion relations, and sum rules
in terms of the polarizabilities.
   We review experimental charged and neutral polarizability studies and theoretical predictions.
   The $\pi^0$ polarizabilities were deduced from DESY Crystal Ball $\gamma\gamma\to\pi^0\pi^0$ data, but with large uncertainties.
   The charged pion polarizabilities were deduced most recently from (1) radiative pion Primakoff scattering $\pi^- Z \to \pi^-Z\gamma$ at CERN COMPASS,
(2) two-photon pion pair production $\gamma\gamma\to\pi^+\pi^-$
at SLAC Mark II, and (3) radiative pion photoproduction $\gamma p\to \gamma \pi ^+ n$ from the proton at MAMI in Mainz.
   A stringent test of chiral perturbation theory (ChPT) is possible based on comparisons of precision experimental charged pion polarizabilities with
ChPT predictions.
   Only the CERN COMPASS charged pion polarizability measurement has acceptably small uncertainties.
   Its value $\alpha_{\pi^\pm}-\beta_{\pi^\pm} = (4.0\pm 1.8)\times 10^{-4}\,\text{fm}^3$ agrees well with the two-loop ChPT prediction
$\alpha_{\pi^\pm}-\beta_{\pi^\pm}=(5.7\pm 1.0)\times 10^{-4}\,\text{fm}^3$, strengthening the
identification of the pion with the Goldstone boson of chiral symmetry breaking in QCD.

\end{abstract}

\ccode{PACS numbers:
11.55.Fv Dispersion relations,
12.38.Qk Experimental tests,
12.39.Fe  Chiral Lagrangians,
13.40.-f  Electromagnetic processes and properties,
13.60.Fz  Elastic and Compton scattering,
13.60.Le  Meson production,
13.60.-r  Photon and charged-lepton interactions with hadrons,
13.75.Lb  Meson-meson interactions
14.40.Aq  pi, K, and eta mesons,
14.70.Bh  Photons
}

%\tableofcontents

\section{Introduction}
   The pion is the lightest strongly interacting composite particle \cite{Tanabashi:2018oca}.
   It belongs to the lowest-lying pseudoscalar meson octet and is believed to be one of the Goldstone bosons associated with spontaneously broken chiral symmetry.
   The electric ($\alpha_\pi$) and magnetic ($\beta_\pi$) Compton polarizabilities of both the charged and the neutral pion are of fundamental interest
in the low-energy sector of quantum chromodynamics (QCD).
   Pion polarizabilities affect the shape of the $\gamma\pi$ Compton scattering angular distribution.
   The polarizability dipole moments are induced during the Compton scattering process via the interaction of the $\gamma$'s electromagnetic field
with the quark substructure of the pion.
   In particular, $\alpha_\pi$ is the proportionality constant between the electric field and the induced electric dipole moment, while
$\beta_\pi$ is similarly related to the magnetic field and the induced magnetic dipole moment
\cite{Klein:1955zz,Holstein:1990qy,Lvov:1993fp,Moinester:1997um,Scherer:1999yw,Holstein:2013kia}.
   A systematic method of incorporating the conditions imposed by chiral symmetry and its spontaneous breakdown in QCD is provided by
chiral perturbation theory (ChPT), which, therefore, is expected to successfully describe the electromagnetic interactions of pions in the low-energy regime.
   A stringent test of ChPT is possible by comparing the experimental charged-pion polarizabilities with the ChPT two-loop predictions
$\alpha_{\pi^\pm}-\beta_{\pi^\pm} = (5.7\pm 1.0)\times 10^{-4}\,\text{fm}^3$ and $\alpha_{\pi^\pm}+\beta_{\pi^\pm} = 0.16\times 10^{-4}\,\text{fm}^3$ \cite{Gasser:2006qa}.

  We review experimental and theoretical polarizability studies.
  After discussing Thomson scattering and the concept of electromagnetic polarizabilities at the classical level,
we provide a nonrelativistic quantum-mechanical description of Compton scattering for both point particles
and composite systems.
   The latter will allow us to establish the impact of electromagnetic polarizabilities on the differential
Compton scattering cross section.
   We then turn to the discussion of the (virtual) Compton tensor in relativistic quantum field theory,
with particular emphasis on the role of gauge invariance, crossing symmetry, and the discrete symmetries.
   After deriving the low-energy behavior of the Compton tensor, we consider dispersion relations and sum rules
for the Compton scattering process.
   Subsequently, we describe how ChPT uses a systematic QCD-based effective Lagrangian to establish relationships between different processes in terms of
a common set of physical (renormalized) parameters.
   Thereby, in this framework, via a perturbative derivative and quark-mass expansion of the Lagrangian limited to ${\cal O}(p^4)$,
radiative pion beta-decay data is used to make a firm one-loop prediction of the charged-pion polarizabilities.
   Subsequent two-loop corrections at ${\cal O}(p^6)$ are also described.
   We review the determination of neutral pion polarizabilities from $\gamma\gamma\to\pi^0\pi^0$ data.
   The combination $\alpha_{\pi^\pm}-\beta_{\pi^\pm}$ for charged pions was most recently measured by: (1) CERN COMPASS via radiative pion Primakoff scattering
(bremsstrahlung of 190 GeV/$c$ negative pions) in the nuclear Coulomb field, $\pi^-Z\to\pi^- Z\gamma$ \cite{Adolph:2014kgj},
(2) SLAC PEP Mark II via two-photon production of
pion pairs, $\gamma\gamma\to\pi^+\pi^-$, via the $e^+e^-\to e^+e^-\pi^+\pi^-$ reaction, focusing on events below $M_{\pi^+\pi^-} = 0.5$~GeV/$c^2$ \cite{Boyer:1990vu}, and (3)
Mainz Microtron via radiative pion photoproduction from the proton,
$\gamma p\to\gamma\pi^+n$, via events in which the incident $\gamma$ ray is scattered off an off-shell pion \cite{Ahrens:2004mg}.
   We describe a planned JLab polarizability experiment via Primakoff scattering of high-energy $\gamma$'s in the nuclear Coulomb field leading to
two-photon fusion production of pion pairs, $\gamma\gamma\to\pi\pi$ \cite{Aleksejevs:2013}.
   The COMPASS data are equivalent to $\gamma\pi\to\gamma\pi$ Compton scattering for laboratory $\gamma$'s having momenta of order 1 GeV/$c$
incident on a target pion at rest.
   In the reference frame of this target pion, the cross section is sensitive to $\alpha_{\pi^\pm}-\beta_{\pi^\pm}$ at backward angles of the scattered
$\gamma$'s.
   To date, only the COMPASS polarizability measurement has acceptably small uncertainties. Its value $\alpha_{\pi^\pm}-\beta_{\pi^\pm} =
(4.0\pm 1.8)\times 10^{-4}\,\text{fm}^3$ agrees well
with the two-loop ChPT prediction $\alpha_{\pi^\pm}-\beta_{\pi^\pm} = (5.7\pm 1.0)\times 10^{-4}\,\text{fm}^3$,
strengthening the identification of the pion with the Goldstone boson of QCD.
   Finally, in the appendix, we also give an outlook to the concept of generalized polarizabilities.

\section{Classical Description}
   Compton scattering at low energies provides a tool for extracting
the electromagnetic polarizabilities of particles.
   These quantities provide information on the structure of composite particles
beyond their static properties, such as mass, charge, and magnetic moment
\cite{Klein:1955zz,Holstein:1990qy,Lvov:1993fp,Moinester:1997um,Scherer:1999yw,Holstein:2013kia}.
   Before addressing the quantum-mechanical description of the scattering
process, we will have a look at the scattering off free charged particles
in the framework of classical electrodynamics \cite{Landau}.
   Any consistent description of Compton scattering in the quantum
regime should result in the Thomson formula in the low-energy limit
\cite{Thirring:1950cy,Low:1954kd,GellMann:1954kc}.

\subsection{The classical Thomson formula}
    Let us consider an incident monochromatic electromagnetic
plane wave with wave vector $\vec q=q\hat e_z$, $q=|\vec q\,|$, and polarization vector $\vec \epsilon$,
\begin{align}
\vec{E}(t,\vec r)&=E_0 \cos(qz-\omega t)\vec{\epsilon} ,
\quad \omega=cq,\quad
\hat{e}_z\cdot\vec{\epsilon}=0,\nonumber\\
\vec{B}&=\hat{e}_z\times \vec{E}.
\label{EBincident}
\end{align}
   The force acting on a point particle with mass $M$ and charge $q$ at the position $\vec{r}$
is given by the Lorentz force equation,\footnote{In this section, we make use of Gaussian units, where
the fine-structure constant is given by $\alpha=e^2/(\hbar c)\approx 1/137$.}
\begin{equation}
M\frac{d^2 \vec{r}}{dt^2}=\vec{F}=q\left(\vec{E}+\frac{\vec{v}}{c}\times \vec{B}\right).
\end{equation}
   The electromagnetic fields induce an acceleration of the charged particle,
which is then responsible for producing an outgoing scattering wave.
   We assume that the velocity $v=|\vec{v}\,|$ of the charged particle, induced
by the incident electromagnetic wave, is negligible in comparison with the speed of
light, i.e., $v\ll c$.
   Furthermore, the displacement of the charge from the origin is always
assumed to be negligible in comparison with
the wave length, i.e., $|\vec{q}\,||\vec{r}\,|\ll 1$.
   With these assumptions we obtain a differential equation for the
electric dipole moment $\vec{d}=q \vec{r}$ with respect to the origin,
\begin{equation}
\ddot{\vec{d}}=\frac{q^2}{M}E_0\cos(\omega t)\vec\epsilon.
\label{ddotd}
\end{equation}
   At $\vec R=R\hat n$ far away from the oscillating charge, the dipole radiation resulting from $\ddot{\vec{d}}$ is given
by \cite{Landau}
\begin{align*}
\vec B'(t,\vec R)=\frac{1}{c^2 R}\ddot{\vec d}\times\hat n,\quad \vec E'(t,\vec R)
=\vec B'\times\hat n.
\end{align*}
   For the scattering cross section we need the Poynting vectors of the incident
and the scattered waves, respectively,
\begin{align}
\vec S&=\frac{c}{4\pi}\vec E\times\vec B=\frac{c}{4\pi}E_0^2\cos^2(qz-\omega t)\hat e_z,\\
\vec S'&=\frac{c}{4\pi}\vec E'\times\vec B'=\frac{1}{4\pi c^3 R^2}\left(\ddot{\vec d}\times\hat n\right)^2\hat n.
\end{align}
   Using Eqs.~(\ref{EBincident}) and (\ref{ddotd}), the corresponding time averages,
with $T=2\pi/\omega$, are given by
\begin{align*}
\langle \vec S\rangle&=\frac{1}{T}\int_0^T dt\,\vec{S}(t)=\frac{c}{8\pi}E_0^2\hat e_z,\\
\langle \vec S'\rangle&=\frac{c}{8\pi}E_0^2 \frac{1}{R^2}\left(\frac{q^2}{M c^2}\right)^2\left(\vec\epsilon\times\hat n\right)^2\hat n.
\end{align*}
   The cross section differential is defined as
\begin{displaymath}
d\sigma=\frac{\mbox{Energy radiated in the solid angle
$d\Omega$/unit time}}{\mbox{Incident energy/unit area/unit time}}
=\frac{\langle\vec{S}\,'\rangle\cdot d\vec{a}}{|\langle\vec{S}\,\rangle|}.
\end{displaymath}
   Making use of $d\vec{a}=R^2 \hat{n}d\Omega$, we obtain
\begin{displaymath}
\frac{d\sigma}{d\Omega}
=\left(\frac{q^2}{M c^2}\right)^2 (\vec{\epsilon}\times \hat{n})^2.
\end{displaymath}
   The differential cross section does not depend on the frequency of the electromagnetic
wave.

   Let $\vartheta$ denote the angle between $\hat{n}$ and $\vec{\epsilon}$ such
that $(\vec{\epsilon}\times \hat{n})^2=\sin^2(\vartheta)$.
   In order to determine the total cross section, we choose a Cartesian coordinate frame with
$\hat e_3=\vec\epsilon$ and make use of polar coordinates,
\begin{displaymath}
\int_0^{2\pi} d\varphi\int_0^\pi d \vartheta \sin^3(\vartheta)=\frac{8\pi}{3}.
\end{displaymath}
   The total cross section, obtained by integrating over the entire
solid angle, results in the classical Thomson cross section
denoted by $\sigma_T$,
\begin{equation}
\label{sigmat}
\sigma_T=\frac{8\pi}{3}\frac{q^4}{M^2 c^4}.
\end{equation}
   Numerical values of the Thomson cross section for the electron, charged
pion, and the proton are shown in Table \ref{tcs}.

\begin{table}[t]
\tbl{Thomson cross section $\sigma_T$ for
the electron, charged pion, and the proton.}
{\begin{tabular}{@{}cc@{}}
\toprule
Particle&$\sigma_T$\\
\colrule
Electron & 0.665 barn\\
Pion & 8.84 $\mu$barn\\
Proton & 197 nbarn\\
\botrule
\end{tabular}
\label{tcs}}
\end{table}

\subsection{Electromagnetic polarizabilities}
\label{subsection_electromagnetic_polarizabilities}
   The phenomenon of electric polarization is well-known from the electrostatics of
macroscopic media.
   Given a collection of atoms or molecules, the application of
an external electric field may lead to three different types
of polarization \cite{bd}.
   In the first case, the so-called electronic polarization, the negative cloud
of electrons is shifted in the direction of the field and the positive
atomic nucleus is shifted opposite to the direction of the field.
   In the case of orientation polarization, preexisting but initially randomly
oriented permanent dipole moments are lined up by the applied field.
   Finally, in the case of ionic polarization, the ions of an ionic crystal are shifted
by the electric field.
   In anisotropic materials, even though the polarization still depends linearly
on the field, its direction is not necessarily parallel to the applied field but
is determined in terms of a polarizability tensor of second rank.

   In the first of the above cases, the electric polarizability $\alpha_E$ of a
system is simply the constant of proportionality between the applied static and
uniform field $\vec E$ and the induced electric dipole moment $\vec d$,
\begin{equation}
\vec d=\alpha_E\vec E.
\label{ddef}
\end{equation}
   A simple and pedagogical example for illustrating the concept of an electric
polarizability is a system of two harmonically bound point particles of
masses $m_1$ and $m_2$ with charges $q_1$ and $q_2$.
   Introducing center-of-mass coordinates and relative coordinates as
\begin{align}
\label{Rr}
\vec R&=\frac{m_1}{m_1+m_2}\vec r_1+\frac{m_2}{m_1+m_2}\vec r_2´,\quad \vec r=\vec r_1-\vec r_2,\\
\label{Pp}
\vec P&=M\dot{\vec R},\quad \vec p=\mu\dot{\vec r}=\frac{m_2\vec{p}_1-m_1\vec{p}_2}{M},\\
\label{Mmu}
M&=m_1+m_2,\quad\mu=\frac{m_1 m_2}{m_1+m_2},
\end{align}
   the Hamiltonian can be written as
\begin{equation}
H=\frac{{\vec{p}_1}\,\!\!^2}{2m_1}+\frac{\vec{p}_2\,\!\!^2}{2m_2}+\frac{\mu\omega^2_0}{2}(\vec r_1-\vec r_2)^2
=\frac{\vec P^2}{2M}+\frac{\vec p\,^2}{2\mu}+\frac{\mu\omega^2_0}{2}
\vec{r}\,^2,
\label{Hho}
\end{equation}
   where we neglect the Coulomb interaction between the charges.
   If a static and uniform external electric field
$$\vec{E}=E\hat{e}_z$$
   is applied to this system, the equilibrium position in the
center-of-mass frame, $\vec R=\vec 0$,
is determined by
\begin{displaymath}
\mu\ddot{z}=-\mu\omega_0^2z+\tilde{q}E \stackrel{!}{=}0,\quad
\tilde{q}=\frac{m_2}{M}q_1-\frac{m_1}{M}q_2,
\end{displaymath}
leading to the solution
   $$z_0=\frac{\tilde{q}E}{\mu\omega_0^2}.$$
      Using
\begin{displaymath}
\vec d=q_1\vec r_1+q_2\vec r_2=q_1\frac{m_2 z_0}{M}\hat e_z-q_2\frac{m_1 z_0}{M}\hat e_z=\tilde q z_0\hat e_z
\end{displaymath}
in combination with Eq.~(\ref{ddef}) for the induced intrinsic dipole moment,
\begin{displaymath}
\vec{d}=\alpha_E E \hat e_z,
\end{displaymath}
   the electric polarizability $\alpha_E$ is
proportional to the inverse of the spring constant $C=\mu\omega_0^2$,
\begin{equation}
\label{alphaho}
\alpha_E=\frac{\tilde{q}^2}{\mu\omega^2_0},
\end{equation}
i.e., it is a measure of the stiffness or
rigidity of the system \cite{Holstein:1990qy}.
   In order to obtain an order-of-magnitude estimate for $\alpha_E^{\pi^+}$,
let us insert $M=M_{\pi^+}=140\,\,\text{MeV}$, $m_1=m_2=\frac{M}{2}$, $q_1=\frac{2}{3}e$ for the up quark,
$q_2=\frac{1}{3}e$ for the down antiquark, and $\hbar\omega_0=500$ MeV as a typical hadronic excitation
energy into Eq.~(\ref{alphaho}),
\begin{equation}
\alpha_E^{\pi^+}=\frac{e^2}{36}\frac{2}{M}\frac{1}{\omega_0^2}=
\frac{1}{18}\frac{e^2}{\hbar c}\frac{(\hbar c)^3}{M_{\pi^+}c^2}\frac{1}{(\hbar\omega_0)^2}
=0.89\times 10^{-4}\,\text{fm}^3.
\end{equation}
   This result is by a factor of three smaller than the experimental result.\footnote{In fact,
using $\hbar\omega_0=300$ MeV yields $\alpha_E^{\pi^+}=2.47\times 10^{-4}\,\text{fm}^3$,
surprisingly close to the experimental result.}

   The potential energy associated with the induced electric dipole moment
reads
\begin{equation}
\label{valpha}
V_E=-\frac{1}{2}\alpha_E\vec{E}^2=-\frac{1}{2}\vec{d}\cdot\vec{E},
\end{equation}
   where the factor $\frac{1}{2}$ results from the interaction
of an induced rather than a permanent electric dipole moment with
the external field.

   Similarly, the potential of an induced magnetic dipole,
$\vec{m}=\beta_M \vec{B}$, is given by
\begin{equation}
\label{vbeta}
V_M=-\frac{1}{2}\beta_M \vec{B}^2=-\frac{1}{2}\vec m\cdot\vec B.
\end{equation}
   The magnetic polarizability $\beta_M$ receives contributions from two
mechanisms.
   If the fundamental constituents of the system themselves
possess intrinsic magnetic dipole moments, they tend to align in the
direction of the applied field, producing a positive (paramagnetic) effect.
   On the other hand, by Lenz's law, an applied field induces currents which
produce an induced magnetic moment opposite to this field, yielding a
negative (diamagnetic) effect.
   Classically, the second effect can be estimated in terms of the system
of Eq.~(\ref{Hho}) via turning on a uniform magnetizing field
$\vec B=B (t)\,\hat{e}_z$ \cite{fnm}.
   The changing magnetic field leads via Faraday's law,
\begin{displaymath}
\vec\nabla\times\vec E=-\frac{1}{c}\frac{\partial\vec B}{\partial t},
\end{displaymath}
to a circulating electric
field $\vec E=E(\rho,t)\,\hat{e}_\phi$, which we determine by applying Stokes's theorem,
\begin{displaymath}
\int_S (\vec\nabla\times\vec E)\cdot d\vec a=\oint_C\vec E\cdot d\vec l.
\end{displaymath}
   By considering a closed circle of radius $\rho$ in the $(x,y)$ plane concentric with the origin,
we obtain
\begin{displaymath}
2\pi\rho E(\rho,t)=-\frac{1}{c}\frac{d}{dt}\left[\pi \rho^2B(t)\right],
\end{displaymath}
resulting in
\begin{equation}
E=-\frac{\rho}{2c}\frac{dB}{dt}.
\end{equation}
   We assume that the center-of-mass of the two-particle system is at the origin and that the
relative motion takes place in the $(x,y)$ plane, i.e., $\vec r=\rho\hat e_\rho$.
   For each particle, the electric field results in a torque, generating a change of the relative angular momentum,
\begin{align}
\frac{d\vec l}{dt}
&=\vec r_1\times\vec F(t,\vec r_1)+\vec r_2\times\vec F(t,\vec r_2)
=q_1\frac{m_2}{M}\vec r\times \vec E\left(t,\frac{m_2}{M}\vec r\right)-q_2\frac{m_1}{M}\vec r\times \vec E\left(t,-\frac{m_1}{M}\vec r\right)\nonumber\\
&=-\left(q_1\frac{m_2^2}{M^2}+q_2\frac{m_1^2}{M^2}\right)\frac{\rho^2}{2c}\frac{dB}{dt}\hat e_z.
\end{align}
   Integrating with respect to time from zero field, results in an extra angular momentum
\begin{equation}
\Delta\vec l=\Delta\vec l_1+\Delta\vec l_2=-\left(q_1\frac{m_2^2}{M^2}+q_2\frac{m_1^2}{M^2}\right)\frac{\rho^2}{2c}B\hat e_z,
\end{equation}
producing an additional orbital magnetic moment
\begin{equation}
\label{betaMdia}
\Delta\vec m=\frac{q_1}{2c m_1}\Delta\vec l_1+\frac{q_2}{2 c m_2}\Delta\vec l_2
=-\frac{\rho^2}{4c^2\mu M^3}\left(q_1^2m_2^3+q_2^2m_1^3\right)B\hat e_z
=\beta_M^{\rm dia}B\hat e_z.
\end{equation}
   According to Lenz's law, the added moment is indeed opposite to the
magnetic field.
   For an order-of-magnitude estimate of the diamagnetic contribution to the magnetic polarizability
of the charged pion, we make use of the same parameters as in the estimate for the electric polarizability.
   Furthermore, we replace $\rho^2$ by $\frac{2}{3} r_E^2$, where $r_E^2$ is the mean-square
charge radius.
   Inserting the empirical value $r_E^2=0.452\,\,\text{fm}^2$ \cite{Tanabashi:2018oca}, one obtains
\begin{equation}
\label{betadiapiest}
\beta_M^{\pi^+}=-\frac{2}{3}r^2_E\frac{1}{4c^2}\frac{1}{M}\frac{5}{9}e^2
=-\frac{5}{54}\frac{e^2}{\hbar c}\frac{\hbar c r^2_E}{M_{\pi^+}c^2}
=-4.30\times 10^{-4}\,\,\text{fm}^3.
\end{equation}
   This result is roughly a factor 2 times the empirical result.

    Polarizabilities are better known as associated with the Rayleigh scattering cross section of sunlight photons on atomic electrons in atmospheric N$_2$ and O$_2$.
    The oscillating electric field of sunlight photons forces the atomic electrons to vibrate.
    Applying a harmonic electric field in the $x$ direction, $\vec E(t)=E_0 \cos(\omega t) \hat{e}_x$, the simple model of the
forced oscillator with damping results in
\begin{equation}
\label{fod}
\ddot x+\gamma\dot x+\omega_0^2 x=-\frac{e}{m}E_0\cos(\omega t).
\end{equation}
   For simplicity, we consider just one oscillator frequency and also neglect the damping \cite{fnm2}.
   For the dipole moment $\vec d(t)=-ex(t)\hat{e}_x=d(t)\hat{e}_x$, we obtain from Eq.~(\ref{fod})
\begin{equation}
d(t)=\frac{e^2}{m}\frac{1}{\omega_0^2-\omega^2}E_0\cos(\omega t)
\end{equation}
and, thus, a frequency-dependent electric polarizability
\begin{equation}
\alpha_E(\omega)=\frac{e^2}{m}\frac{1}{\omega_0^2-\omega^2}.
\end{equation}
   For sufficiently small $\omega$, we obtain $\alpha_E\approx e^2/(m\omega_0^2)$, with the polarizability
independent of the exciting frequency [static polarizability, see Eq.~(\ref{alphaho})].
   The scattering cross section may be obtained from Eqs.~(\ref{ddotd}) and (\ref{sigmat})
by replacing $q^2/M$ with $-\omega^2\alpha_E(\omega)$, resulting in the Rayleigh cross section
\begin{equation}
\sigma_R=\frac{8\pi}{3}\frac{\omega^4\alpha_E^2(\omega)}{c^4}=\frac{8\pi}{3}\frac{1}{\lambda^4}\alpha_E^2(\omega).
\end{equation}
   For scattering of optical wavelengths, the incident photon energies $\hbar \omega$ are
in the range 1.6 to 3.2 eV.
   The low-frequency condition discussed above is then satisfied, this energy being sufficiently small
in comparison with typical $\hbar\omega_0$ electronic binding energies of the order of tens
of eV.
   Since the cross section for the scattering of light depends on $\lambda^{-4}$,
the intensity of scattered and transmitted sunlight is dominated by blue and red,
respectively.
   Such scattering is denoted by Rayleigh scattering, following Rayleigh's explanation
of blue skies and red sunrises and sunsets.

\section{Compton Scattering in Nonrelativistic Quantum Mechanics}
\subsection{Compton scattering off a charged point particle}
\subsubsection{Kinematics and notation}
\begin{figure}[t]
\centerline{\includegraphics[width=3.8cm]{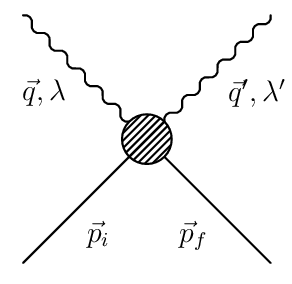}}
\caption{Kinematical variables for Compton scattering. $\lambda$ and
$\lambda'$ denote the polarizations of the initial photon  and the
final photon, resepectively. \label{figurekin}}
\end{figure}

   The kinematical variables for real Compton scattering (RCS),
$\gamma \pi^+\to\gamma \pi^+$, are defined in Fig.\ \ref{figurekin}.
   The invariance under translations in time results in energy conservation,
\begin{equation}
\label{energy_conservation}
E_\gamma+E_i=E'_\gamma+E_f,
\end{equation}
where $E_\gamma=\hbar\omega$ and $E_\gamma'=\hbar \omega'$.
   Depending on whether one uses a nonrelativistic or a relativistic framework,
the energy-momentum relation of the particle is given by
\begin{align*}
E(\vec p)&=\frac{\vec p\,^2}{2M}\quad (\text{nonrelativistic}),\\
E(\vec p)&=\sqrt{M^2 c^4+\vec{p}\,^2 c^2}\quad(\text{relativistic}).
\end{align*}
   For real photons, the dispersion relation reads
\begin{displaymath}
\omega(\vec k) =c|\vec k|,
\end{displaymath}
where $\vec k$ denotes the wave vector of the photon.
   Because of translational invariance in space, the total three
momentum is conserved,
\begin{equation}
\hbar\vec q+\vec p_i=\hbar\vec q\,'+\vec p_f.
\end{equation}
   For the description of the RCS amplitude one requires two kinematical
variables, e.g., the energy of the initial photon,
$E_\gamma=\hbar\omega$, and the scattering angle between the initial photon and the
scattered photon, $\cos(\theta)=\hat{q}\cdot\hat{q}\,'$.
   Using relativistic kinematics, the energy of the scattered photon in the lab frame ($\vec p_i=\vec 0$)
is given by
\begin{equation}
\label{omegap}
E_\gamma'=\frac{E_\gamma}{1+\frac{E_\gamma}{M c^2}[1-\cos(\theta)]}\leq E_\gamma.
\end{equation}
   From Eq.\ (\ref{omegap}) one obtains the well-known result for
the wavelength shift of the Compton effect,
$$\Delta \lambda=\lambda'-\lambda=\frac{4\pi\hbar}{M c}\sin^2\left(\frac{\theta}{2}\right),$$
which is independent of the frequency of the photon.

   On the other hand, using the nonrelativistic energy-momentum relation results
in
\begin{align}
\label{egammapnonrel}
E_\gamma'&=-M c^2+E_\gamma\cos(\theta)+M c^2\sqrt{1+2\frac{E_\gamma}{Mc^2}[1-\cos(\theta)]
-\frac{E_\gamma^2}{M^2c^4}\sin^2(\theta)}.
\end{align}
   Expanding in powers of $E_\gamma/(Mc^2)$, the nonrelativistic expression and the relativistic
expression agree up to and including terms of second order,
\begin{equation}
\label{Egammapexp}
E_\gamma'=E_\gamma\left\{1-\frac{E_\gamma}{Mc^2}[1-\cos(\theta)]+\frac{E_\gamma^2}{M^2c^4}
[1-\cos(\theta)]^2+{\cal O}\left[\left(\frac{E_\gamma}{Mc^2}\right)^3\right]\right\}.
\end{equation}

\subsubsection{Hamiltonian}
   In order to set the stage, we will first discuss, in quite some detail,
Compton scattering of real photons off a free point particle of mass $M$
and charge $e>0$ (proton charge) within the framework of nonrelativistic quantum mechanics.
   First of all, this will allow us to introduce basic concepts such as gauge
invariance, photon-crossing symmetry as well as discrete symmetries.
   Secondly, the result will define a reference point beyond which the
structure of a composite object can be studied.
   Finally, it will also allow us to discuss later on, where a relativistic
description departs from a nonrelativistic treatment.

   Consider the Hamiltonian of a single, free point particle of mass $M$
and charge $e>0$,\footnote{Except for a few cases, we will use the
same symbols for quantum-mechanical operators such as $\hat{\vec{p}}$
and the corresponding eigenvalue $\vec{p}$.}
\begin{equation}
\label{h0}
H_0=\frac{\vec{p}\,^2}{2M}.
\end{equation}
   The coupling to the electromagnetic scalar and vector potentials
$\Phi(t,\vec x)$ and $\vec A(t,\vec x)$, respectively, is generated by the well-known
minimal-substitution procedure\footnote{We still use Gaussian units.}
\begin{equation}
\label{minsub}
i\hbar\frac{\partial}{\partial t}\mapsto i\hbar\frac{\partial}{\partial t}
-e\Phi(t,\vec{x}),\quad
\vec{p}\mapsto\vec{p}-\frac{e}{c} \vec{A}(t,\vec{x}),
\end{equation}
resulting in the Schr\"{o}dinger equation
\begin{equation}
\label{schr}
i\hbar\frac{\partial \Psi(t,\vec{x})}{\partial t}=
H(\Phi,\vec{A})\Psi(t,\vec{x})=
[H_0+H_I(t)]\Psi(t,\vec{x})=
[H_0+H_1(t)+H_2(t)]\Psi(t,\vec{x}),
\end{equation}
where
\begin{equation}
H_1(t)=-e\frac{\vec{p}\cdot\vec{A}+\vec{A}\cdot{\vec{p}}}{2Mc}+e\Phi,
\quad
H_2(t)=\frac{e^2}{2Mc^2}\vec{A}\,^2.
\end{equation}
   Given a smooth real function $\chi(t,\vec x)$, gauge invariance of Eq.\ (\ref{schr}) implies that
\begin{equation}
\Psi'(t,\vec{x})=\exp\left[-i\frac{e}{\hbar c}\chi(t,\vec x)\right]\Psi(t,\vec{x})
\end{equation}
is a solution of
\begin{equation}
i\hbar\frac{\partial \Psi'(t,\vec{x})}{\partial t}=
H\left(\Phi+\frac{1}{c}\dot{\chi},\vec{A}-\vec{\nabla}\chi\right)\Psi'(t,\vec{x}),
\end{equation}
provided $\Psi(t,\vec{x})$ is a solution of Eq.\ (\ref{schr}).
   In other words, Eq.\ (\ref{schr}) remains invariant under a so-called gauge
transformation of the second kind,
\begin{equation}
\Psi\mapsto\Psi'=\exp\left(-i\frac{e}{\hbar c}\chi\right)\Psi,
\quad\Phi\mapsto\Phi'=\Phi+\frac{1}{c}\frac{\partial\chi}{\partial t},
\quad\vec A\mapsto\vec A'= \vec A-\vec\nabla\chi.
\end{equation}

\subsubsection{S matrix}
\label{subsubsectionsmatrix}
   After introducing the interaction representation,
\begin{equation}
\label{intrep}
H_I^{\rm int}(t)=e^{\frac{i}{\hbar} H_0 t} H_I(t)e^{-\frac{i}{\hbar} H_0 t},
\end{equation}
   the $S$-matrix element is obtained by evaluating the Dyson series,
\begin{align}
\label{dyson}
S&=\hat{T}\exp\left[-\frac{i}{\hbar}\int_{-\infty}^\infty dt\,H^{\rm int}_I(t)\right]\nonumber\\
&=1+ \sum_{k=1}^\infty \frac{\left(-\frac{i}{\hbar}\right)^k}{k!}
\int_{-\infty}^\infty dt_1
\int_{-\infty}^\infty dt_2\cdots
\int_{-\infty}^\infty dt_k\, \hat{T}\left[H^{\rm int}_I(t_1)\cdots H^{\rm int}_I(t_k)\right],
\end{align}
   between $|i\rangle\equiv|\vec{p}_i; \gamma(\vec q,\lambda)\rangle$ and
$\langle f|\equiv \langle \vec{p}_f; \gamma(\vec q\,', \lambda')|$, where
$\lambda$ and $\lambda'$ denote the polarizations of the initial and final photons,
respectively.
   In Eq.\ (\ref{dyson}), $\hat{T}$ refers to the time-ordering operator,
\begin{equation}
\hat{T}\left[A(t_1) B(t_2)\right]=A(t_1) B(t_2)\Theta(t_1-t_2)
+B(t_2) A(t_1)\Theta(t_2-t_1),
\end{equation}
   where $\Theta$ is the Heaviside step function,
\begin{displaymath}
\Theta(t)=\begin{cases} 0\,\,\text{for}\,\, t<0,\\
1\,\,\text{for}\,\,t\geq 0.
\end{cases}
\end{displaymath}
   The generalization to an arbitrary number of operators is straightforward.

   For the quantization of the radiation field in the vacuum, we make use of the
Coulomb gauge
\begin{displaymath}
\vec\nabla\cdot\vec A=0
\end{displaymath}
which then also implies $\Phi=0$.
   The radiation field is expanded in a Fourier series,\footnote{The additional factor of $\sqrt{4\pi}$
in Eq.~(\ref{vecAFs}) is related to the use of Gaussian units in this section.}
\begin{equation}
\label{vecAFs}
\vec A(t,\vec x)=\sum_{\vec k}\sum_{\lambda=1}^2 \sqrt\frac{4\pi\hbar c^2}{2\omega(\vec k)V}
\left(a_\lambda(\vec k)\vec\epsilon_\lambda(\vec k)e^{-i\left(\omega(\vec k)t-\vec{k}\cdot\vec x\right)}
+a^\dagger_{\lambda}(\vec k)\vec\epsilon_\lambda\,\!\!\!^\ast(\vec k) e^{i\left(\omega(\vec k)t-\vec k\cdot\vec x\right)}\right).
\end{equation}
   For convenience, we make use of periodic boundary conditions in a box with $V=L^3$, resulting
in discrete $\vec k$ vectors \cite{Mandl:1985bg},
\begin{displaymath}
\vec k=\frac{2\pi}{L}(n_x,n_y,n_z),\quad n_x,n_y,n_z\in{\mathbbm Z}.
\end{displaymath}
   For linearly polarized photons, $\vec\epsilon_\lambda\,\!\!\!^\ast(\vec k)=\vec\epsilon_\lambda(\vec k)$, and
the triple $(\hat k,\vec\epsilon_1(\vec k),\vec\epsilon_2(\vec k))$ represents a right-handed
trihedral.
   The commutation relations of the annihilation and creation operators read
\begin{equation}
[a_\lambda(\vec k),a_{\lambda'}^\dagger(\vec k')]=\delta_{\lambda\lambda'}\delta_{\vec k,\vec k'},\quad
[a_\lambda(\vec k),a_{\lambda'}(\vec k')]=[a_\lambda^\dagger(\vec k),a_{\lambda'}^\dagger(\vec k')]=0.
\end{equation}
   Defining a single-photon state by
\begin{displaymath}
|\gamma(\vec q,\lambda)\rangle=a^\dagger_\lambda(\vec q)|0\rangle,
\end{displaymath}
we obtain the following matrix element for the radiation field evaluated between a single-photon
state and the vacuum,
\begin{equation}
\label{sqpf}
\langle 0|\vec A(t,\vec{x})|\gamma(\vec q,\lambda)\rangle=
\sqrt\frac{2\pi\hbar c^2}{\omega(\vec q)V}
\vec\epsilon_\lambda(\vec q)e^{-i\left(\omega(\vec q)t-\vec{q}\cdot\vec x\right)}.
\end{equation}
   Finally, the particle states are normalized as
\begin{equation}
\label{normstat}
\langle \vec{x}\,|\vec{p}\,\rangle=\frac{e^{\frac{i}{\hbar} \vec{p}\cdot\vec{x}}}{\sqrt{V}},
\end{equation}
where, in analogy to the photon, we have
\begin{displaymath}
\vec p=\frac{2\pi\hbar}{L}(m_x,m_y,m_z),\quad m_x,m_y,m_z\in{\mathbbm Z}.
\end{displaymath}

   The part of the scattering operator relevant to Compton scattering [${\cal O}(e^2)$] reads
\begin{equation}
\label{scomp}
S=-\frac{i}{\hbar}\int_{-\infty}^\infty dt\, H^{\rm int}_2(t)
-\frac{1}{\hbar^2}\int_{-\infty}^\infty dt_1 \int_{-\infty}^\infty dt_2\,
H_1^{\rm int}(t_1) H_1^{\rm int}(t_2)\Theta(t_1-t_2),
\end{equation}
   where the first term generates the contact-interaction contribution
or so-called sea\-gull term:\footnote{For simplicity, we omit the arguments of
the polarization vectors.}
\begin{align}
\label{scont}
S_{fi}^{\rm cont}&=
-\frac{i}{\hbar}\frac{e^2}{2Mc^2}\int_{-\infty}^\infty dt\, \langle f|
e^{\frac{i}{\hbar}H_0 t}\vec{A}\,^2(t,\hat{\vec{r}}) e^{-\frac{i}{\hbar} H_0 t}
|i\rangle\nonumber\\\
&=-\frac{i}{\hbar}(2\pi\hbar)\delta(E_\gamma'+E_f-E_\gamma-E_i)\delta_{\vec q'+\frac{\vec p_f}{\hbar},\vec{q}+\frac{\vec p_i}{\hbar}}
\frac{2\pi\hbar c^2}{V\sqrt{\omega(\vec q)\omega(\vec q\,')}}
\frac{e^2 \vec{\epsilon}\,'^\ast\cdot\vec{\epsilon}}{M c^2}.
\end{align}
   In order to obtain Eq.\ (\ref{scont}), one first contracts the photon
field operators with the photons in the initial and final states,
respectively,\footnote{Note the factor of 2 for two contractions.}
then evaluates the time integral, and, finally, makes use of
\begin{displaymath}
\langle\vec{p}\,'|f(\hat{\vec{r}})|\vec{p}\,\rangle=
\int_V d^3 r\,\langle\vec p\,'|f(\hat{\vec{r}})|\vec r\,\rangle\langle\vec r\,|\vec{p}\,\rangle
=\frac{1}{V}\int_V d^3r\, e^{\frac{i}{\hbar}(\vec{p}-\vec{p}\,')\cdot\vec{r}}f(\vec{r}),
\end{displaymath}
   with $f(\vec{r})=\exp[i(\vec{q}-\vec{q}\,')\cdot\vec{r}\,]$, to obtain
the three-momentum conservation.
   Because we are working with a box normalization, momentum conservation
is expressed in terms of the Kronecker delta.
   In the infinite-volume limit, we need to replace
\begin{displaymath}
\frac{1}{V}\delta_{\vec q'+\frac{\vec p_f}{\hbar},\vec{q}+\frac{\vec p_i}{\hbar}}
\to (2\pi)^3\delta^3\left(\vec q\,'+\frac{\vec p_f}{\hbar}-\vec{q}-\frac{\vec p_i}{\hbar}\right)
=(2\pi\hbar)^3\delta^3\left(\hbar\vec q\,'+\vec p_f-\hbar\vec{q}-\vec p_i\right).
\end{displaymath}

\begin{figure}[t]
\centerline{\includegraphics[width=\textwidth]{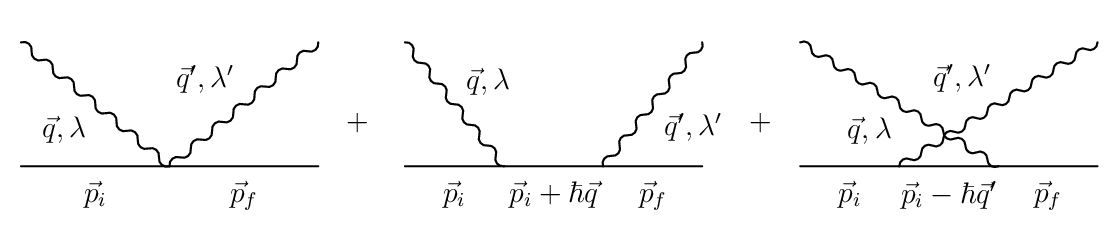}}
\caption{Seagull, direct-channel, and crossed-channel diagrams in
time-ordered perturbation theory.
\label{figure_nonrelativistic_diagrams}}
\end{figure}

   The second contribution of Eq.\ (\ref{scomp}) is evaluated by inserting a
complete set of states,
\begin{equation}
{\mathbbm 1}=\sum_{\vec p}|\vec p\,\rangle\langle\vec p\,|,
\end{equation}
between $H_1^{\rm int}(t_1)$ and $H_1^{\rm int}(t_2)$,
\begin{equation}
-\frac{1}{\hbar^2}\int_{-\infty}^\infty dt_1\int_{-\infty}^\infty dt_2\,\Theta(t_1-t_2)
\sum_{\vec p}\langle f|H^{\rm int}_1(t_1)|\vec{p}\,\rangle\langle\vec{p}\,| H^{\rm int}_1(t_2)|i\rangle.
\end{equation}
   There are two distinct possibilities to contract the photon fields,
namely, $\vec A(t_2)$ with $|\gamma(\vec q,\lambda)\rangle$ and
$\vec A(t_1)$ with $\langle\gamma(\vec q\,',\lambda')|$ and vice versa,
giving rise to the so-called direct and crossed channels, respectively.
   The dependence on time is given by
\begin{align*}
\Theta(t_1-t_2)e^{\frac{i}{\hbar}(E_f+E_\gamma'-E(\vec p))t_1}e^{\frac{i}{\hbar}(E(\vec p)-E_\gamma-E_i)t_2}&\quad
\text{(direct channel)},\\
\Theta(t_1-t_2)e^{\frac{i}{\hbar}(E_f-E_\gamma-E(\vec p))t_1}e^{\frac{i}{\hbar}(E(\vec p)+E_\gamma'-E_i)t_2}&\quad
\text{(crossed channel)}.
\end{align*}
   Making use of
\begin{displaymath}
\int_{-\infty}^\infty dt_1\int_{-\infty}^\infty dt_2\, \Theta(t_1 - t_2)
e^{iat_1} e^{ibt_2}
=\frac{2\pi i\delta(a+b)}{a+i0^+}=\frac{2\pi i\delta(a+b)}{i0^+-b},
\end{displaymath}
    one obtains for the contributions of the direct and crossed channels
\begin{align}
\label{sppdccc1}
S_{fi}^{{\rm dc}+{\rm cc}}&=-\frac{1}{\hbar^2}(2\pi\hbar)i\delta(E_\gamma'+E_f-E_\gamma-E_i)\sum_{\vec p}
\hbar\left(\frac{\langle\vec{p}_f|H^{\rm em}_1|\vec{p}\,\rangle\langle\vec{p}\,|H^{\rm abs}_1|\vec{p}_i\rangle}{
E_i+E_\gamma-E(\vec{p})+i0^+}\right.\nonumber\\
&\quad\left.
+\frac{\langle\vec{p}_f|H^{\rm abs}_1|\vec{p}\,\rangle\langle\vec{p}\,|H^{\rm em}_1|\vec{p}_i\rangle}{
E_i-E_\gamma'-E(\vec{p})+i0^+}\right).
\end{align}
   The superscripts ``abs'' and ``em'' refer to the absorption of a photon and
the emission of a photon, respectively, and the matrix elements are given by
\begin{align}
\langle\vec p_2|H_1^{\rm abs}|\vec p_1\rangle&=-e\sqrt{\frac{2\pi\hbar c^2}{\omega(\vec q)V}}
\,\delta_{\frac{\vec p_2}{\hbar},\vec q+\frac{\vec p_1}{\hbar}}\,\frac{(\vec p_2+\vec p_1)\cdot\vec \epsilon}{2M c},\\
\langle\vec p_2|H_1^{\rm em}|\vec p_1\rangle&=-e\sqrt{\frac{2\pi\hbar c^2}{\omega(\vec q)V}}
\,\delta_{\vec q\,'+\frac{\vec p_2}{\hbar},\frac{\vec p_1}{\hbar}}\,\frac{(\vec p_2+\vec p_1)\cdot\vec \epsilon\,'^\ast}{2M c}.
\end{align}
   Extracting an overall factor ${\cal N}$,
\begin{equation}
\label{factorN}
{\cal N}=2\pi\delta(E_\gamma'+E_f-E_\gamma+E_i)\delta_{\vec q'+\frac{\vec p_f}{\hbar},\vec{q}+\frac{\vec p_i}{\hbar}}
\frac{2\pi\hbar c^2}{V\sqrt{\omega(\vec q)\omega(\vec q\,')}},
\end{equation}
we may write
\begin{displaymath}
S_{fi}^{\rm cont}+S_{fi}^{{\rm dc}+{\rm cc}}=i{\cal N}t_{fi},
\end{displaymath}
where $t_{fi}$ is the transition matrix element,
\begin{align}
t_{fi}&=e^2\left\{-\frac{\vec{\epsilon}\,'^\ast\cdot\vec{\epsilon}}{M c^2}
-\left[\frac{(2\vec{p}_f+\vec{q}\,')\cdot
\vec{\epsilon}\,'^\ast}{2M c}\right]
\frac{1}{E_i+E_\gamma-E(\vec{p}_f+\vec{q}\,')+i0^+}
\left[\frac{(2\vec{p}_i+\vec{q})\cdot
\vec{\epsilon}}{2M c}\right]\right.\nonumber\\
&\left.\hspace{4em}
-\left[\frac{(2\vec{p}_f-\vec{q})\cdot\vec{\epsilon}}{2M c}\right]
\frac{1}{E_i-E_\gamma'-E(\vec{p}_f-\vec{q})+i0^+}
\left[\frac{(2\vec{p}_i-\vec{q}\,')\cdot
\vec{\epsilon}\,'^\ast}{2M c}\right]\right\}\nonumber\\
&=-\frac{e^2}{M c^2}\vec{\epsilon}\,'^\ast\cdot\vec{\epsilon}\nonumber\\
&\quad
-\frac{e^2}{M^2 c^2}\left[\frac{\vec{p}_f\cdot\vec{\epsilon}\,'^\ast\,\vec{p}_i\cdot\vec{\epsilon}}{E_i+E_\gamma-E(\vec{p}_f+\vec{q}\,')+i0^+}
+\frac{\vec{p}_f\cdot\vec{\epsilon}\,\vec{p}_i\cdot\vec{\epsilon}\,'^\ast}{E_i-E_\gamma'-E(\vec{p}_f-\vec{q})+i0^+}\right],
\end{align}
where we made use of $\vec q\cdot\vec\epsilon=\vec q\,'\cdot\vec\epsilon\,'^\ast=0$.
   A diagrammatic representation of the Compton scattering process is shown in Fig.~\ref{figure_nonrelativistic_diagrams}.

\subsubsection{Discrete symmetries}
\label{subsectiongids}

  Let us discuss a few properties of $t_{fi}$.
\begin{enumerate}
\item The transition matrix element is invariant under photon crossing \cite{GellMann:1954kc}, i.e.,
the simultaneous replacements $\vec\epsilon\leftrightarrow \vec\epsilon\,'^\ast$, $E_\gamma\leftrightarrow-E_\gamma'$,
and $\vec q\leftrightarrow-\vec q\,'$,
\begin{align*}
t_{fi}&\mapsto
-\frac{e^2}{M c^2}\vec{\epsilon}\cdot\vec{\epsilon}\,'^\ast\nonumber\\
&\quad
-\frac{e^2}{M^2 c^2}\left[\frac{\vec{p}_f\cdot\vec{\epsilon}\,\vec{p}_i\cdot\vec{\epsilon}\,'^\ast}{E_i-E_\gamma'-E(\vec{p}_f-\vec{q})+i0^+}
+\frac{\vec{p}_f\cdot\vec{\epsilon}\,'^\ast\,\vec{p}_i\cdot\vec{\epsilon}}{E_i+E_\gamma-E(\vec{p}_f+\vec{q}\,')+i0^+}\right]\\
&=t_{fi}.
\end{align*}
   In terms of the diagrams of Fig.~\ref{figure_nonrelativistic_diagrams}, we see that the seagull diagram is crossing symmetric
by itself whereas the direct channel transforms into the crossed channel and vice
versa.

\item $t_{fi}$ is invariant under $e\mapsto -e$, i.e., the Compton scattering
amplitudes for particles of charges $e$ and $-e$ are identical.
\item Under a parity transformation, all momenta and polarization vectors are multiplied by
a minus sign.
$t_{fi}$ behaves as a scalar, i.e., there
are no terms of, e.g., the type
$\epsilon_{ijk} \epsilon_i \epsilon_j'^\ast q_k$.
\item Particle crossing, $(E_i,\vec{p}_i)\leftrightarrow (-E_f,-\vec{p}_f)$,
is {\em not} a symmetry of a nonrelativistic treatment.
\end{enumerate}

\subsubsection{Cross section}
\label{subsubsection_cross_section}
   For the purpose of calculating the differential cross section,
we evaluate the $S$-matrix element in the laboratory frame, where
$\vec{p}_i=0$.
   In this case, the direct channel and the crossed channel do not
contribute and we find
\begin{displaymath}
S_{fi}=
-i2\pi\delta(E_\gamma'+E_f-E_\gamma)\delta_{\vec q'+\frac{\vec p_f}{\hbar},\vec{q}}\,
\frac{2\pi\hbar c^2}{V\sqrt{\omega(\vec q)\omega(\vec q\,')}}
\frac{e^2 \vec{\epsilon}\,'^\ast\cdot\vec{\epsilon}}{M c^2}.
\end{displaymath}
   Given the time interval $[-T/2,T/2]$, the transition rate per volume is
\begin{displaymath}
w_{fi}=\frac{|S_{fi}|^2}{TV}.
\end{displaymath}
   In the limit $T\to\infty$, we may replace
\begin{displaymath}
\left[2\pi\delta(E_\gamma'+E_f-E_\gamma)\right]^2\to\frac{T}{\hbar}2\pi\delta(E_\gamma'+E_f-E_\gamma).
\end{displaymath}
   Furthermore, making use of $\delta^2=\delta$ for the Kronecker delta, we obtain
\begin{displaymath}
w_{fi}=(2\pi)^3\hbar\delta(E_\gamma'+E_f-E_\gamma)\frac{1}{V^3}\delta_{\vec q'+\frac{\vec p_f}{\hbar},\vec{q}}
\,\frac{1}{\omega(\vec q)\omega(\vec q\,')}\frac{e^4|\vec{\epsilon}\,'^\ast\cdot\vec{\epsilon}\,|^2}{M^2}.
\end{displaymath}
   Next, we divide by the flux of projectiles, $|\vec J_{\rm in}|=\frac{c}{V}$, and by the number of
target particles per volume, $1/V$.
   Finally, we multiply by the number of final states in the volume elements $d^3 p_f$ and $d^3 q'$,
namely, $\frac{V}{(2\pi\hbar)^3}d^3p_f$ and $\frac{V}{(2\pi)^3}d^3 q'$, to obtain the following expression
for the cross section differential,
\begin{displaymath}
d\sigma =\frac{1}{(2\pi)^3}\frac{1}{\hbar^2 c}\delta(E_\gamma'+E_f-E_\gamma)\delta_{\vec q'+\frac{\vec p_f}{\hbar},\vec{q}}\,
V\frac{1}{\omega(\vec q)\omega(\vec q\,')}\frac{e^4|\vec{\epsilon}\,'^\ast\cdot\vec{\epsilon}\,|^2}{M^2}d^3p_f\,d^3q'.
\end{displaymath}
   At this stage, it is convenient to perform the infinite-volume limit, leading to the replacement
\begin{displaymath}
\delta_{\vec q'+\frac{\vec p_f}{\hbar},\vec{q}}\,
V\to(2\pi)^3\delta^3(\vec q'+\frac{\vec p_f}{\hbar}-\vec{q}).
\end{displaymath}
   The cross section differential then reads
\begin{equation}
d\sigma=\frac{\hbar}{c}\delta(E_\gamma'+E_f-E_\gamma)\delta^3(\hbar\vec q\,'+\vec p_f-\hbar\vec q)
\frac{1}{\omega(\vec q)\omega(\vec q\,')}\frac{e^4|\vec{\epsilon}\,'^\ast\cdot\vec{\epsilon}\,|^2}{M^2}d^3p_f\,d^3q'.
\end{equation}
   After integration with respect to the momentum of the particle, we make
use of $d^3q'=\frac{E_\gamma'^2dE_\gamma'}{(\hbar c)^3} d\Omega$ and obtain
\begin{displaymath}
d\sigma=\delta(E_\gamma'+E_f-E_\gamma)\frac{E_\gamma'}{E_\gamma}\frac{e^4|\vec{\epsilon}\,'^\ast\cdot\vec{\epsilon}\,|^2}{M^2 c^4}dE_\gamma'\,d\Omega.
\end{displaymath}
   Before performing the last integration with respect to $E_\gamma'$, we need the dependence
of $E_f$ on $E_\gamma'$ resulting from three-momentum conservation,
\begin{displaymath}
E_f=\frac{\vec p_f\,\!\!\!^2}{2M}=\frac{E_\gamma'^2-2E_\gamma'E_\gamma\cos(\theta)+E_\gamma^2}{2M c^2}.
\end{displaymath}
   We then make use of the expression for $\delta(f(x))$,
\begin{displaymath}
\delta(f(x))=\sum_i\frac{1}{\left|\frac{df}{dx}(x_i)\right|}\delta(x-x_i),
\end{displaymath}
where $f(x)$ is assumed to have only simple zeros, located at $x=x_i$.
  Using
\begin{align*}
f(E_\gamma')&=E_\gamma'+\frac{E_\gamma'^2-2E_\gamma'E_\gamma\cos(\theta)+E_\gamma^2}{2M c^2}-E_\gamma,\\
\frac{df}{d E_\gamma'}(E_\gamma')&=\frac{M c^2+E_\gamma'-E_\gamma\cos(\theta)}{M c^2},
\end{align*}
and Eq.~(\ref{Egammapexp}) for the relevant zero of $f$,
we obtain for the differential cross section
\begin{equation}
\label{dsdo}
\frac{d\sigma}{d\Omega}=
\left\{1-4\frac{E_\gamma}{M c^2}\sin^2\left(\frac{\theta}{2}\right)
+{\cal O}\left[\left(\frac{E_\gamma}{M c^2}\right)^2\right]\right\}
\left|\frac{e^2\vec{\epsilon}\,'^\ast\cdot\vec{\epsilon}}{M c^2}\right|^2.
\end{equation}
   Averaging and summing over initial and final photon polarizations,
respectively, is easily performed by treating
$\hat{q}=\hat{e}_z=|3\rangle$, $\vec{\epsilon}_1(\vec q)=\hat{e}_x=|1\rangle$,
and $\vec{\epsilon}_2(\vec q)=\hat{e}_y=|2\rangle$ as well as
$\hat{q}'=|3'\rangle$, $\vec{\epsilon}\,'_1(\vec q\,')=|1'\rangle$, and $\vec{\epsilon}\,'_2(\vec q\,')=|2'\rangle$
as orthonormal bases, satisfying the completeness relations
\begin{equation}
\label{completeness}
{\mathbbm 1}=\sum_{\lambda=1}^3|\lambda\rangle\langle\lambda|=\sum_{\lambda'=1}^3|\lambda'\rangle\langle\lambda'|.
\end{equation}
   We then obtain for the averaging over the initial polarizations
\begin{align*}
\lefteqn{\frac{1}{2}\sum_{\lambda=1}^2|\vec \epsilon_{\lambda'}\,\!\!\!\!'^\ast(\vec q\,')\cdot\vec\epsilon_\lambda(\vec q)|^2
=\frac{1}{2}\sum_{\lambda=1}^2|\langle\lambda'|\lambda\rangle|^2
=\frac{1}{2}\sum_{\lambda=1}^2\langle\lambda'|\lambda\rangle\langle\lambda|\lambda'\rangle}\\
&=\frac{1}{2}\sum_{\lambda=1}^3\langle\lambda'|\lambda\rangle\langle\lambda|\lambda'\rangle
-\frac{1}{2}\langle\lambda'|3\rangle\langle 3|\lambda'\rangle
=\frac{1}{2}\langle\lambda'|\lambda'\rangle-\frac{1}{2}\langle\lambda'|3\rangle\langle 3|\lambda'\rangle\\
&=\frac{1}{2}(1-\langle 3|\lambda'\rangle\langle\lambda'|3\rangle),
\end{align*}
where we made use of Eq.~(\ref{completeness}).
   The summation over the final polarizations yields
\begin{align*}
\frac{1}{2}\sum_{\lambda'=1}^2(1-\langle 3|\lambda'\rangle\langle\lambda'|3\rangle)=
\frac{1}{2}[2-(\langle 3|3\rangle-\langle 3|3'\rangle\langle 3'|3\rangle)]
=\frac{1}{2}[1+\cos^2(\theta)].
\end{align*}
Thus,
\begin{equation}
\label{sumpol}
\frac{1}{2}\sum_{\lambda,\lambda'=1}^2|\vec \epsilon_{\lambda'}\,\!\!\!\!'^\ast(\vec q\,')\cdot\vec\epsilon_\lambda(\vec q)|^2
=\frac{1}{2}[1+\cos^2(\theta)].
\end{equation}
   Let us consider the so-called Thomson limit, i.e., $E_\gamma=\hbar\omega\to 0$,
for which Eq.\ (\ref{dsdo}) in combination with Eq.\ (\ref{sumpol}) reduces
to
\begin{displaymath}
\left.\frac{d\sigma}{d\Omega}\right|_{E_\gamma=0}
=\left(\frac{e^2}{M c^2}\right)^2\frac{1+\cos^2(\theta)}{2}.
\end{displaymath}
   The total cross section, obtained by integrating over the entire
solid angle, reproduces the classical Thomson scattering cross section
denoted by $\sigma_T$ [see Eq.~(\ref{sigmat})],
\begin{equation}
\label{sigmat2}
\sigma_T=\frac{8\pi}{3}\frac{e^4}{M^2 c^4}.
\end{equation}
   Numerical values of the Thomson cross section for the electron, charged
pion, and the proton are shown in Table \ref{tcs}.

\subsection{Compton scattering off a composite system:
nonrelativistic calculation}

\subsubsection{Hamiltonian and $S$ matrix}

   Next we discuss Compton scattering off a composite system within
the framework of nonrelativistic quantum mechanics \cite{Petrunkin}.
   For the sake of simplicity, we consider a system of two particles
interacting via a central potential $V(r)$,
\begin{equation}
\label{tpp}
H_0=\frac{{\vec{p}_1}\,\!\!^2}{2m_1}+\frac{\vec{p}_2\,\!\!^2}{2m_2}
+V(|\vec{r}_1-\vec{r}_2|)
=\frac{\vec{P}\,\!^2}{2M}+\frac{\vec{p}\,^2}{2\mu}+V(r),
\end{equation}
   where the center-of-mass coordinates and the relative coordinates are given in
Eqs.~(\ref{Rr})--(\ref{Mmu}).
   As in the single-particle case, the electromagnetic interaction
is introduced via minimal coupling [see Eq.~(\ref{minsub})],
$i\hbar\partial/\partial t\to i\hbar\partial/\partial t-q_1\Phi_1-q_2\Phi_2$,
$\vec{p}_i\to\vec{p}_i-\frac{q_i}{c} \vec{A}_i$,
resulting in the interaction Hamiltonians
\begin{align*}
H_1(t)&=\sum_{i=1}^2\left[-\frac{q_i}{2m_i c}(\vec{p}_i\cdot\vec{A_i}
+\vec{A_i}\cdot\vec{p}_i)+q_i\Phi_i\right],\\
H_2(t)&=\sum_{i=1}^2\frac{q_i^2}{2m_i c^2}\vec{A_i}\,\!\!^2,
\end{align*}
where $(\Phi_i,\vec{A_i})=(\Phi(t,\vec{r}_i),\vec{A}(t,\vec{r}_i))$.

   In order to keep the expressions as simple as possible, we make
some simplifying assumptions and quote the general result in Sec.~\ref{subsubsectiondcsp}.
   First of all, we do not consider the spin of the constituents,
i.e., we omit an interaction term
\begin{displaymath}
-\sum_{i=1}^2\vec{\mu}_i\cdot\vec{B}_i,\quad
\vec{B}_i=\vec{\nabla}_i\times \vec{A}_i,
\end{displaymath}
   where $\vec{\mu}_i$ is an intrinsic magnetic dipole moment of the
$i$th constituent.
    Secondly, we take equal masses for the constituents,
$m_1=m_2=m=\frac{1}{2}M$
and assume that one has charge $q_1=e>0$ and the second one
is neutral, $q_2=0$.
   The latter assumption is only made to reduce the writing effort.
   Of course, for a constituent quark model of the $\pi^+$ one would
use $q_1=\frac{2}{3}e$ and $q_2=\frac{1}{3}e$ for the charges of
the up quark and the down antiquark, respectively.
    Finally, making use of the gauge-invariance property,
we perform the calculation within the Coulomb gauge,
$\vec\nabla_i\cdot\vec A_i=\Phi_i=0$.
   With these preliminaries, the Hamiltonian reads
\begin{displaymath}
H=H_0+H_1+H_2=H_0-\frac{e}{M c}\left(\vec{p}_1\cdot\vec{A}_1+\vec{A}_1\cdot\vec{p}_1\right)
+\frac{e^2}{M c^2}{\vec{A}_1}\,\!\!\!^2.
\end{displaymath}

   The box normalization of Eqs.~(\ref{sqpf}) and (\ref{normstat}) has served its
purpose for evaluating the differential cross section in Sec.~\ref{subsubsection_cross_section}
and we now switch to a delta-function normalization,
\begin{align*}
\langle 0|\vec A(t,\vec{x})|\gamma(\vec q,\lambda)\rangle&=N(\vec q)
\vec\epsilon_\lambda(\vec q)e^{-i\left(\omega(\vec q)t-\vec{q}\cdot\vec x\right)},
\quad N(\vec q)=\sqrt\frac{\hbar c^2}{4\pi^2\omega(\vec q)},\\
\langle \vec{R}\,|\vec{P}\,\rangle&=\frac{e^{\frac{i}{\hbar} \vec{P}\cdot\vec{R}}}{(2\pi\hbar)^\frac{3}{2}},
\end{align*}
such that
\begin{align*}
\langle \gamma(\vec q,\lambda)|\gamma(\vec q\,',\lambda')\rangle&=\delta_{\lambda\lambda'}\delta^3(\vec q-\vec q\,'),\quad
\langle\vec P|\vec P'\rangle=\delta^3(\vec P-\vec P^\prime).
\end{align*}
  The $S$-matrix element is obtained in complete analogy to the previous
section within the framework of time-dependent perturbation
theory,
\begin{equation}
\label{scp}
S_{fi}=S_{fi}^{\rm cont}+S_{fi}^{\rm dc}+S_{fi}^{\rm cc}.
\end{equation}
   In general, the seagull contribution results from the sum of the individual
contact terms and the direct-channel and crossed-channel contributions
are more complicated than in the single-particle case, since they now also
involve excitations of the composite object, which are not possible for a point particle.

   For convenience, we deviate slightly from the notation of Sec.~\ref{subsubsectionsmatrix} and denote the state kets of the initial
and final particles by $|i\rangle$ and $|f\rangle$, respectively.
   The contact contribution reads
\begin{align*}
S_{fi}^{\rm cont}&=-\frac{i}{\hbar}\int_{-\infty}^\infty dt\,
\langle f,\gamma(\vec q\,',\lambda')|H_2^{\rm int}(t)|i,\gamma(\vec q,\lambda)\rangle\\
&=-i2\pi\delta(E_\gamma'+E_f-E_\gamma-E_i)N(\vec q\,')N(\vec q)\,\frac{2e^2 \vec{\epsilon}\,'^\ast\cdot\vec{\epsilon}}{M c^2}\,
\langle f|e^{i(\vec q-\vec q\,')\cdot\hat{\vec r}_1}|i\rangle.
\end{align*}
   For the evaluation of the matrix element $\langle f|e^{i(\vec q-\vec q\,')\cdot\hat{\vec r}_1}|i\rangle$,
we make use of $\vec{r}_1=\vec{R}+\frac{1}{2}\vec{r}$ and insert complete sets of states with respect to the center-of-mass coordinate
$\vec R$ and the relative coordinate $\vec r$,
\begin{align*}
\langle f|e^{i(\vec q-\vec q\,')\cdot\hat{\vec r}_1}|i\rangle&=
\int d^3R\int d^3r\,\langle f|
e^{i(\vec q-\vec q\,')\cdot\left(\hat{\vec R}+\frac{1}{2}\hat{\vec r}\right)}|\vec R,\vec r\rangle
\langle\vec R,\vec r|i\rangle.
\end{align*}
Using
\begin{align*}
\langle\vec R,\vec r|i\rangle&=\frac{e^{\frac{i}{\hbar}\vec p_i\cdot\vec R}}{(2\pi\hbar)^\frac{3}{2}}\varphi_0(\vec r),\quad
\langle f|\vec R,\vec r\rangle=\frac{e^{-\frac{i}{\hbar}\vec p_f\cdot\vec R}}{(2\pi\hbar)^\frac{3}{2}}\varphi_0^\ast(\vec r),
\end{align*}
where $\varphi_0(\vec r)$ denotes the ground-state wave function,
we obtain
\begin{displaymath}
\langle f|e^{i(\vec q-\vec q\,')\cdot\hat{\vec r}_1}|i\rangle=
\delta^3(\hbar\vec q\,'+\vec p_f-\hbar\vec q-\vec p_i)
\int d^3 r\, |\varphi_0(\vec{r})|^2
e^{i(\vec{q}-\vec{q}\,')\cdot\frac{\vec{r}}{2}}.
\end{displaymath}
   In analogy to Eq.~(\ref{factorN}), we extract from $S_{fi}$ a factor
\begin{displaymath}
{\cal N}=2\pi\delta(E_\gamma'+E_f-E_\gamma-E_i)\delta^3(\hbar\vec q\,'+\vec p_f-\hbar\vec q-\vec p_i)N(\vec q\,')N(\vec q),
\end{displaymath}
adjusted to the delta-function normalization, and obtain
for the contact contribution ($S_{fi}^{\rm cont}=i{\cal N}t_{fi}^{\rm cont}$)
\begin{equation}
\label{tficoncoms}
t_{fi}^{\rm cont}=-\frac{2e^2 \vec{\epsilon}\,'^\ast\cdot\vec{\epsilon}}{M c^2}
\int d^3 r\, |\varphi_0(\vec{r})|^2e^{i(\vec{q}-\vec{q}\,')\cdot\frac{\vec{r}}{2}}
.
\end{equation}
   Since $q_2=0$, the integral is just the charge form factor $F$
of the ground state, evaluated for the argument $(\vec{q}-\vec{q}\,')^2$,
$$ F((\vec{q}-\vec{q}\,')^2)=1-\frac{1}{6}r_E^2(\vec{q}-\vec{q}\,')^2+\cdots.
$$
   Taking the limit $E_\gamma\to 0$ in Eq.~(\ref{tficoncoms}), which also implies $\vec q\to\vec 0$, $E_\gamma'\to 0$, and $\vec q\,'\to\vec 0$,
we note that for a composite object the contact interactions
of the constituents, in general, do not yet generate the Thomson limit.

   In analogy to Sec.~\ref{subsubsectionsmatrix}, the second contribution is evaluated by inserting a complete set of states,\footnote{Note that
$|\vec P,n\rangle=|\vec P\rangle\otimes|n\rangle$ is of the direct-product type.}
\begin{displaymath}
{\mathbbm 1}=\int d^3 P´\sum_n |\vec P, n\rangle\langle \vec P, n|
\end{displaymath}
between $H_1^{\rm int}(t_1)$ and $H_1^{\rm int}(t_2)$, resulting in
\begin{align}
\label{sdccc}
S_{fi}^{\rm dc+cc}
&=-2\pi i\delta(E_f+\omega'-E_i-\omega)\int d^3 P\sum_n\nonumber
\\
&\times\left(
\frac{\langle\vec{p}_f,0|H^{\rm em}_1|\vec{P},n\rangle\langle\vec{P},n|
H^{\rm abs}_1|\vec{p}_i,0\rangle}{E_i+E_\gamma-E_n(\vec{P})+i0^+}+
\frac{\langle\vec{p}_f,0|H^{\rm abs}_1|\vec{P},n\rangle\langle\vec{P},n|
H^{\rm em}_1|\vec{p}_i,0\rangle}{E_i-E_\gamma'-E_n(\vec{P})+i0^+}\right),\nonumber\\
\end{align}
   where, in the framework of Eq.\ (\ref{tpp}), the energy of a state $|\vec P,n\rangle$
is given by
\begin{displaymath}
E_n(\vec{P})=\frac{\vec{P}\,^2}{2M}+\epsilon_n,
\end{displaymath}
with $\epsilon_n$ denoting the eigenvalue of the Hamilton operator of the relative motion.
   In the Coulomb gauge, the corresponding Hamiltonians for the absorption and emission of photons, respectively, read
\begin{displaymath}
H_1^{\rm abs}=-\frac{2e}{M c}N(\vec q)\,\hat{\vec{p}}_1\cdot\vec{\epsilon}\,
e^{i\vec{q}\cdot\hat{\vec{r}}_1},\quad
H_1^{\rm em}=-\frac{2e}{M c}N(\vec q\,')\,\hat{\vec{p}}_1\cdot\vec{\epsilon}\,'^\ast\,
e^{-i\vec{q}\,'\cdot\hat{\vec{r}}_1}.
\end{displaymath}
   Using $\vec r_1=\vec R+\frac{1}{2}\vec r$ and $\vec p_1=\frac{1}{2}\vec P+\vec p$, it is straightforward
but tedious to evaluate the matrix elements of Eq.~(\ref{sdccc}) (see, e.g., Ref.~\cite{Pasquini:2000ue}).
   The integration over $\vec P$ generates an overall three-momentum-conserving delta function, $\delta^3(\hbar\vec q\,'+\vec p_f-\hbar\vec q-\vec p_i)$,
and the result for $t_{fi}^{\rm dc+cc}$ is given by
\begin{align}
\label{tdccccomposite}
t_{fi}^{\rm dc+cc}&=\frac{4e^2}{M^2 c^2}\sum_n\left[\frac{\langle 0|\left(\frac{1}{2}\vec q+\hat{\vec p}\right)\cdot\vec\epsilon\,'^\ast\,
e^{-i\frac{\vec q\,'\cdot\hat{\vec r}}{2}}|n\rangle\langle n|\hat{\vec p}\cdot\vec\epsilon\, e^{i\frac{\vec q\cdot\hat{\vec r}}{2}}|0\rangle}
{\epsilon_0+E_\gamma-\epsilon_n-\frac{E_\gamma^2}{2M c^2}+i0^+}
\right.\nonumber\\
&\quad\quad\quad+\left.
\frac{\langle 0|\left(-\frac{1}{2}\vec q\,'+\hat{\vec p}\right)\cdot\vec\epsilon\,
e^{i\frac{\vec q\cdot\hat{\vec r}}{2}}|n\rangle\langle n|\hat{\vec p}\cdot\vec\epsilon\,'^\ast\, e^{-i\frac{\vec q\,'\cdot\hat{\vec r}}{2}}|0\rangle}
{\epsilon_0-E_\gamma'-\epsilon_n-\frac{E_\gamma'^2}{2M c^2}+i0^+}
\right].
\end{align}
   Note that the matrix elements entering Eq.~(\ref{tdccccomposite}) depend exclusively on the wave functions of
the internal motion.
   As in the point-particle case, $t_{fi}=t_{fi}^{\rm cont}+t_{fi}^{\rm dc+cc}$ is symmetric
under photon crossing $(E_\gamma,\vec{q})\leftrightarrow (-E_\gamma',-\vec{q}\,')$
and $\vec{\epsilon}\leftrightarrow\vec{\epsilon}\,'^\ast$.

\subsubsection{Thomson limit}

   The low-energy expansion of Eq.\ (\ref{sdccc}) is obtained by expanding
the vector potentials and the denominators in $E_\gamma$, $\vec q$, $E_\gamma'$, and $\vec q\,'$.
   The explicit calculation is beyond the scope of the present treatment
and we will only quote the general result in Sec.~\ref{subsubsectiondcsp}. \cite{Petrunkin,Ericson:1973dtc}.
   However, we find it instructive to consider the limit $E_\gamma\to 0$:
\begin{align*}
\left.t_{fi}^{\rm dc+cc}\right|_{E_\gamma=0}&=
\frac{4e^2}{M^2 c^2}\sum_n\frac{ \langle 0|\hat{\vec{p}}\cdot\vec{\epsilon}\,'^\ast|n\rangle\langle n|\hat{\vec{p}}\cdot\vec{\epsilon}|0\rangle
+\langle 0|\hat{\vec{p}}\cdot\vec{\epsilon}|n\rangle\langle n|\hat{\vec{p}}\cdot\vec{\epsilon}\,'^\ast|0\rangle}
{\Delta\epsilon_n},
\end{align*}
 where $\Delta\epsilon_n=\epsilon_n-\epsilon_0$.
   Note that, because of $\langle 0|\hat{\vec p}|0\rangle=\vec 0$, the ground state does not contribute to the sum, in particular,
it does not generate a singular contribution as one might naively expect from the vanishing denominator.
  Making use of $\hat{\vec{p}}=i\mu[H_0,\hat{\vec{r}}\,]$ and applying $H_0$
appropriately to the right or left, the expression simplifies to
\begin{align}
\label{tdccccpt}
\left.t_{fi}^{\rm dc+cc}\right|_{E_\gamma=0}&=
-i\frac{4e^2\mu}{M^2 c^2}\sum_n \left(
\langle 0|\hat{\vec{r}}\cdot\vec{\epsilon}\,'^\ast|n\rangle\langle n|\hat{\vec{p}}\cdot\vec{\epsilon}|0\rangle
-\langle 0|\hat{\vec{p}}\cdot\vec{\epsilon}|n\rangle\langle n|\hat{\vec{r}}\cdot\vec{\epsilon}\,'^\ast|0\rangle\right)\nonumber\\
&=-i\frac{4e^2\mu}{M^2 c^2}\langle 0|[\hat{\vec{r}}\cdot\vec{\epsilon}\,'^\ast,
\hat{\vec{p}}\cdot\vec{\epsilon}]|0\rangle
= \frac{e^2\vec{\epsilon}\,'^\ast\cdot\vec{\epsilon}}{M c^2},
\end{align}
   where, again, we used the completeness relation,
$[\vec{a}\cdot\hat{\vec{r}},\vec{b}\cdot\hat{\vec{p}}]=i\vec{a}\cdot\vec{b}$,
and $\mu=M/4$.
   Combining this result with the contact contribution of
Eq.\ (\ref{tficoncoms}) yields the correct Thomson limit also for
a composite system,
\begin{equation}
\left.t_{fi}^{\rm cont}\right|_{E_\gamma=0}+\left.t_{fi}^{\rm dc+cc}\right|_{E_\gamma=0}
=-\frac{2e^2 \vec{\epsilon}\,'^\ast\cdot\vec{\epsilon}}{M c^2}+\frac{e^2\vec{\epsilon}\,'^\ast\cdot\vec{\epsilon}}{M c^2}
=-\frac{e^2\vec{\epsilon}\,'^\ast\cdot\vec{\epsilon}}{M c^2}.
\end{equation}
   Indeed, it was shown a long time ago in the more general framework
of quantum field theory that the scattering of photons in the limit of zero
frequency is correctly described by the classical Thomson amplitude
\cite{Thirring:1950cy,Low:1954kd,GellMann:1954kc}.
   We will come back to this point in Sec.~\ref{subsectionLET}.

\subsubsection{Differential cross section and polarizabilities}
\label{subsubsectiondcsp}
   Beyond the Thomson limit, we only quote the result for Compton scattering off a
spin-zero particle of mass $M$ and total charge $Q$ consisting of $N$ constituents with masses $m_i$, charges $q_i$, and
magnetic moments $\vec\mu_i$.
   The nonrelativistic $T$-matrix element, expanded to second order in the photon energy, reads\footnote{Note that
$|\vec q|=\omega/c$ and $|\vec q\,'|=\omega'/c$.}
\begin{equation}
\label{tfize}
t_{fi}=\vec{\epsilon}\,'^\ast\cdot\vec{\epsilon}
\left(-\frac{Q^2}{M c^2}+\bar{\alpha}_E\frac{E_\gamma E_\gamma'}{(\hbar c)^2}\right)
+\bar{\beta}_M(\vec{q}\,'\times\vec{\epsilon}\,'^\ast)
\cdot(\vec{q}\times\vec{\epsilon}),
\end{equation}
where
\begin{eqnarray}
\label{alphab}
\bar{\alpha}_E&=&\frac{Q^2 r^2_E}{3M c^2}+2\sum_{n\neq 0}
\frac{|\langle n|D_z|0\rangle|^2}{\epsilon_n-\epsilon_0},\\
\label{betab}
\bar{\beta}_M&=&-\frac{\langle \vec{D}\,^2\rangle}{2M c^2}-\frac{1}{6}
\left\langle\sum_{i=1}^N \frac{q^2_i \vec{r}_i\,^2}{m_i c^2}\right\rangle+2\sum_{n\neq 0}
\frac{|\langle n|M_z|0\rangle|^2}{\epsilon_n-\epsilon_0}
\end{eqnarray}
   denote the electric ($\bar{\alpha}_E$) and magnetic ($\bar{\beta}_M$)
polarizabilities of the system.
   In these equations
\begin{displaymath}
\vec{D}=\sum_{i=1}^N q_i(\vec{r}_i-\vec{R})
\end{displaymath}
   refers to the intrinsic electric dipole operator
and
\begin{displaymath}
\vec{M}=
\sum_{i=1}^N
\left[
\frac{q_i}{2 c m_i}(\vec{r}_i-\vec{R})\times
(\vec{p}_i-\frac{m_i}{M}\vec{P})
+\vec{\mu}_i\right]
\end{displaymath}
   to the magnetic dipole operator, where the possibility of magnetic moments
of the constituents has now been included.

   According to Eq.~(\ref{tfize}), the modification of the scattering amplitude
at order $\omega\omega'$ is encoded in the two constants $\bar{\alpha}_E$ and $\bar{\beta}_M$.
   In the nonrelativistic framework, the Compton polarizability $\bar{\alpha}_E$ receives one contribution
which is related to the extension of the particle (first term) and a second contribution related to
electric dipole transitions from the ground state to excited states of the system.
   In atomic physics, the second term of Eq.\ (\ref{alphab})
is known as the quadratic Stark effect describing the energy shift of an atom placed in an external electric field.
   In the case of the magnetic polarizability $\bar{\beta}_M$, the first term
in Eq.~(\ref{betab}) amounts to a recoil effect, the second term is exactly the
diamagnetic contribution discussed in Eqs.~(\ref{betaMdia}) and (\ref{betadiapiest}),
and the last term is the analogue of the quadratic Stark effect, but now
in terms of transitions from the ground state to excited states induced by
the magnetic dipole operator.

   Finally, let us discuss the influence of the electromagnetic
polarizabilities on the differential Compton scattering cross section.
   We restrict ourselves to the leading term due to the interference of
the Thomson amplitude with the polarizability contribution.
   The evaluation of that term requires, in addition to Eq.\ (\ref{sumpol}),
the sum
\begin{displaymath}
\sum_{\lambda,\lambda'=1}^2\text{Re}\left[\vec{\epsilon}\,'^\ast_{\lambda'}(\vec q\,')\cdot\vec{\epsilon}_\lambda(\vec q)\,
(\hat{q}\,'\times\vec{\epsilon}\,'_{\lambda'}(\vec q\,')\cdot(\hat{q}\times\vec{\epsilon}\,^\ast_\lambda(\vec q))\right]
=2\cos(\theta),
\end{displaymath}
and one obtains
\begin{align}
\label{dsdopol}
\frac{d\sigma}{d\Omega}&=
\left\{1-4\frac{E_\gamma}{M c^2}\sin^2\left(\frac{\theta}{2}\right)
+{\cal O}\left[\left(\frac{E_\gamma}{M c^2}\right)^2\right]\right\}
\left\{\frac{1}{2}[1+\cos^2(\theta)]
\left(\frac{Q^2}{M c^2}\right)^2\right.\nonumber\\
&\left. -\frac{E_\gamma E_\gamma'}{(\hbar c)^2}\frac{Q^2}{M c^2}\left\{
\bar{\alpha}_E[1+\cos^2(\theta)]+2\bar{\beta}_M\cos(\theta)\right\}
+{\cal O}\left(\frac{E_\gamma^2 E_\gamma'^2}{M^4 c^8}\right)\right\}.
\end{align}
   The differential cross sections at $\theta=0^\circ$, $90^\circ$, and
$180^\circ$ are sensitive to $\bar{\alpha}_E+\bar{\beta}_M$,
$\bar{\alpha}_E$, and $\bar{\alpha}_E-\bar{\beta}_M$, respectively.

\subsection{Polarizabilities in quantum mechanics}
\label{subsection_polarizabilies_qm}
   After having considered the amplitude for Compton scattering off a composite
particle in the framework of nonrelativistic quantum mechanics, we turn to a
discussion of the electric and magnetic polarizabilities in static and uniform external
fields.
   This will allow us to compare with both the polarizabilities in the time and space
dependent fields of the electromagnetic waves in Compton scattering as well as the classical results of
Sec.~\ref{subsection_electromagnetic_polarizabilities}.
   For that purpose, we consider the Hamiltonian for the relative motion of two particles with masses $m_1$ and $m_2$ and
charges $q_1$ and $q_2$, bound by a harmonic oscillator potential [see Eq.~(\ref{Hho})],
\begin{equation}
\label{Hhorel}
H=\frac{\vec p\,^2}{2\mu}+\frac{\mu\omega_0^2}{2}\vec r\,^2.
\end{equation}
   The discussion turns out to be particularly transparent in the Dirac operator formalism
\cite{Dirac}.
   Introducing creation and annihilation operators as
\begin{align}
a^\dagger_i&=\sqrt{\frac{\mu\omega_0}{2\hbar}}\,x_i-\frac{i}{\sqrt{2\hbar\mu\omega_0}}\,p_i,&i=1,2,3,\\
a_i&=\sqrt{\frac{\mu\omega_0}{2\hbar}}\,x_i+\frac{i}{\sqrt{2\hbar\mu\omega_0}}\,p_i,& i=1,2,3,\\
[a_i,a_j^\dagger]&=\delta_{ij},& i,j=1,2,3,
\end{align}
the Hamilton operator may be written as
\begin{equation}
H=\hbar\omega_0\left(N_1+N_2+N_3+\frac{3}{2}\right),
\end{equation}
where the number operators are given by
\begin{equation}
N_i=a_i^\dagger a_i,\quad i=1,2,3.
\end{equation}
   The eigenstates of $H$ are denoted by $|n_1,n_2,n_3\rangle$ with
\begin{equation}
H|n_1,n_2,n_3\rangle=\hbar\omega_0\left(n_1+n_2+n_3+\frac{3}{2}\right)|n_1,n_2,n_3\rangle,\quad
n_1,n_2,n_3\in{\mathbbm N}_0.
\end{equation}
   Let us apply a static and uniform electric field,
\begin{displaymath}
\vec E=E\hat e_z,
\end{displaymath}
which is included by adding to Eq.~(\ref{Hhorel}) the interaction potential
\begin{equation}
V=-\tilde q E z=-\tilde{q}E\sqrt{\frac{\hbar}{2\mu\omega_0}}\left(a_3^\dagger+a_3\right),\quad \tilde q=\frac{m_2}{M}q_1-\frac{m_1}{M}q_2.
\end{equation}
   Since we are not interested in the motion of the system as a whole, we neglect the effect
of the electric field on the center-of-mass motion.
   We calculate the energy shift using time-independent perturbation theory.
   At first order, one finds for the ground state $|0\rangle\equiv|n_1=0,n_2=0,n_3=0\rangle$,
\begin{displaymath}
\Delta \epsilon^{(1)}=\langle 0|V|0\rangle=0,
\end{displaymath}
because $\langle 0,0,0|a_3^\dagger|0,0,0\rangle=\langle 0,0,0|a_3|0,0,0\rangle=0$.
   At second order, the result reads
\begin{equation}
\label{deltae2e}
\Delta \epsilon^{(2)}=\tilde{q}^2 E^2 \frac{\hbar}{2\mu\omega_0}\sum_{(n_1,n_2,n_3)\neq(0,0,0)}
\frac{|\langle n_1,n_2,n_3|(a_3+a_3^\dagger)|0,0,0\rangle|^2}{\epsilon_0-\epsilon(n_1,n_2,n_3)}.
\end{equation}
    Here, only the state $|0,0,1\rangle=a_3^\dagger|0,0,0\rangle$ contributes to the sum,
resulting in
\begin{equation}
\Delta \epsilon^{(2)}=\tilde{q}^2 E^2 \frac{\hbar}{2\mu\omega_0}\frac{1}{(-\hbar\omega_0)}
=-\frac{1}{2}\frac{\tilde{q}^2}{\mu\omega_0}E^2=-\frac{1}{2}\alpha_E\vec {E}^2,
\end{equation}
which agrees with the classical result of Eqs.~(\ref{alphaho}) and (\ref{valpha}).
   In fact, introducing the shifted variable
\begin{displaymath}
\vec r\,'=\vec r-\frac{\tilde{q}}{\mu\omega_0^2}\vec E,
\end{displaymath}
we may rewrite
\begin{displaymath}
H=\frac{\vec p\,^2}{2\mu}+\frac{\mu\omega_0^2}{2}\vec r\,^2-\tilde{q}\vec E\cdot\vec r=
\frac{\vec p\,^2}{2\mu}+\frac{\mu\omega_0^2}{2}\vec r\,'^2+\Delta H,
\end{displaymath}
where
\begin{displaymath}
\Delta H=-\frac{\tilde{q}^2}{2\mu\omega_0^2}\vec E^2=-\frac{1}{2}\alpha_E \vec E^2.
\end{displaymath}
   In other words, all states are shifted by the same amount $\Delta\epsilon=\Delta H$.

   Let us now turn to the interaction with a static and uniform magnetic field,
\begin{displaymath}
\vec B=B\hat e_z.
\end{displaymath}
   Using $A_i=-\frac{1}{2}\epsilon_{3ij}x_j B$, the interaction potential is given by
\begin{equation}
V=-\frac{q_1}{2 c m_1}l_{1z}B-\frac{q_2}{2 c m_2}l_{2z}B
+\frac{q_1^2}{8m_1 c^2}(x_1^2+y_1^2)B^2
+\frac{q_2^2}{8m_2 c^2}(x_2^2+y_2^2)B^2,
\end{equation}
which, considering the relative internal motion only, reduces to
\begin{equation}
V_{\rm rel}=V_1+V_2=-\frac{q_1m_2^2+q_2m_1^2}{2 c\mu M^2}l_zB
+\frac{q_1^2m_2^3+q_2^2m_1^3}{8 c^2\mu M^3}\,B^2(x^2+y^2).
\end{equation}
   Note that we neglect magnetic moments of the constituents.
   Using $l_z|0\rangle=0$, we obtain for the shift of the ground-state energy
up to and including second order in perturbation theory
\begin{align}
\label{deltae2m}
\Delta\epsilon&=\Delta\epsilon^{(1)}+\Delta\epsilon^{(2)}=
\langle 0|V_1|0\rangle+\langle 0|V_2|0\rangle+
\sum_{(n_1,n_2,n_3)\neq(0,0,0)}
\frac{|\langle n_1,n_2,n_3|V_1|0,0,0\rangle|^2}{\epsilon_0-\epsilon(n_1,n_2,n_3)}
\nonumber\\
&=\frac{2}{3}r_E^2\frac{q_1^2m_2^3+q_2^2m_1^3}{8 c^2\mu M^3}B^2=-\frac{1}{2}\beta_M \vec{B}^2,
\end{align}
where we neglected terms of order $B^4$.
   Again, the quantum-mechanical calculation agrees with the classical consideration
of Eqs.~(\ref{betaMdia}) and (\ref{betadiapiest}).
   In particular, since we are neglecting the spin of the constituents and, thus, their
magnetic moments, we only obtain a diamagnetic contribution and no paramagnetic contribution
to the magnetic polarizability.

   Also in the case of a static and uniform magnetic field, the eigenstates and
eigenvalues can be constructed exactly.
   Because the motion in the $z$ direction is not affected by a magnetic field in the $z$ direction,
we only consider the Hamiltonian for the relative motion in the $(x,y)$ plane.
   Introducing creation and annihilation operators for right and left ``circular quanta''
\cite{Cohen-Tannoudji},
\begin{align*}
a_r^\dagger&=\frac{1}{\sqrt{2}}(a_1^\dagger+ia_2^\dagger),&
a_r=\frac{1}{\sqrt{2}}(a_1-ia_2),\\
a_l^\dagger&=\frac{1}{\sqrt{2}}(a_1^\dagger-ia_2^\dagger),&
a_l=\frac{1}{\sqrt{2}}(a_1+ia_2),
\end{align*}
the Hamiltonian for the motion in the $(x,y)$ plane can be written as
\begin{align}
H_{xy}&=\frac{p_x^2+p_y^2}{2\mu}+\left(\frac{\mu\omega_0^2}{2}+
\frac{q_1^2m_2^3+q_2^2m_1^3}{8 c^2\mu M^3}\,B^2\right)(x^2+y^2)
-\frac{q_1m_2^2+q_2m_1^2}{2 c\mu M^2}\,l_z B\nonumber\\
&=\left(\hbar\tilde\omega-\frac{q_1m_2^2+q_2m_1^2}{2 c\mu M^2}\, B\right)
\left(N_r+\frac{1}{2}\right)
+\left(\hbar\tilde\omega+\frac{q_1m_2^2+q_2m_1^2}{2c\mu M^2}\, B\right)
\left(N_l+\frac{1}{2}\right),
\end{align}
where $N_r=a_r^\dagger a_r$ and $N_l=a_l^\dagger a_l$ denote the number operators
of the right and left circular quanta, respectively.
   The modified oscillator frequency is given by
\begin{displaymath}
\tilde\omega=\sqrt{\omega_0^2+\frac{q_1^2m_2^3+q_2^2m_1^3}{4 c^2\mu^2 M^3}\,B^2},
\end{displaymath}
and the eigenstates are denoted by $|n_r,n_l\rangle$, $n_r,n_l\in {\mathbbm N}_0$,
with energy eigenvalues
\begin{displaymath}
\epsilon(n_r,n_l)=
\hbar\tilde\omega(n_r+n_l+1)
-\frac{q_1m_2^2+q_2m_1^2}{2 c\mu M^2}\, B(n_r-n_l).
\end{displaymath}
   In particular, we find for the ground state $(n_r=n_l=0)$ a shift in energy
by
\begin{align*}
\Delta\epsilon&=\hbar\tilde\omega+\frac{1}{2}\hbar\omega_0-\frac{3}{2}\hbar\omega_0\\
&=\hbar\omega_0\sqrt{1+\frac{q_1^2m_2^3+q_2^2m_1^3}{4\omega_0^2c^2\mu^2 M^3}\,B^2}-\hbar\omega_0
\approx \hbar\frac{q_1^2m_2^3+q_2^2m_1^3}{8\omega_0 c^2\mu^2 M^3}\,B^2.
\end{align*}
   Using
\begin{displaymath}
\langle x^2\rangle=\langle y^2\rangle=\langle z^2\rangle=\frac{\hbar}{2\mu\omega_0},
\end{displaymath}
we re-express
\begin{displaymath}
\mu\omega_0=\frac{3\hbar}{2r_E^2}
\end{displaymath}
to obtain
\begin{equation}
\Delta\epsilon=-\frac{1}{2}\left(-r_E^2 \frac{q_1^2m_2^3+q_2^2m_1^3}{6 c^2\mu M^3}\right)\vec{B}^2,
\end{equation}
which agrees with Eq.~(\ref{deltae2m}).

   Comparing the results for the polarizabilities derived from Eqs.~(\ref{deltae2e})
and (\ref{deltae2m}) with the Compton polarizabilities $\bar{\alpha}_E$ and
$\bar\beta_M$ of Eqs.~(\ref{alphab}) and (\ref{betab}), we see that they differ
by the terms $Q^2r_E^2/(3M c^2)$ and $-\langle \vec D\,^2)/(2M c^2)$,
respectively.
   In this context, the quantities $\alpha_E$ and $\beta_E$ are sometimes referred to
as static polarizabilities, because they characterize the response to static and uniform
external fields.
   However, as we will see in the next section, in a covariant description, there is
no difference between static polarizabilities and Compton polarizabilities \cite{Lvov:1993fp}.
   Moreover, Eq.~(\ref{alphab}) seems to suggest that electric polarizabilities
of ground-state particles are positive quantities, given that $Q^2 r_E^2\geq 0$.
   However, as we will see in Sec.~\ref{subsection_neutral_pion}, a covariant calculation of the $\pi^0$
electric polarizability in the framework of chiral perturbation theory will give
rise to a negative value for $\bar{\alpha}_E^{\pi^0}$.

\section{Compton Scattering in Quantum Field Theory}
   Now that we have discussed Compton scattering in the classical framework
and also using nonrelativistic quantum mechanics, we will move on to a
discussion within relativistic quantum field theory.
   In particular, we will work out similarities such as the Thomson limit
but also differences as, for example, in the interpretation of the polarizabilities.
   From now on, we make use of (rationalized) natural units, $c=1=\hbar$, \footnote{
The conversion constant is $\hbar c=197.3269788(12)$ MeV fm \cite{Tanabashi:2018oca}.}
such that Maxwell's equations in the vacuum are given by
\begin{align*}
\vec\nabla\cdot\vec B&=0,&\vec\nabla\times\vec E+\frac{\partial\vec B}{\partial t}=0,\\
\vec\nabla\cdot\vec E&=\rho,&\vec\nabla\times\vec B-\frac{\partial\vec E}{\partial t}=\vec J.
\end{align*}
   To define the notation and normalization used, let us discuss
the case of three Cartesian free spin-0 fields,
\begin{equation}
\phi_i(x)=\int_{{\mathbbm R}^3} \frac{d^3 k}{(2\pi)^3 2\omega(\vec k)}\left(a_i(\vec k)e^{-ik\cdot x}
+a_i^\dagger(\vec k)e^{ik\cdot x}\right),\quad i=1,2,3,
\end{equation}
where $k\cdot x=k_0 t-\vec k\cdot\vec x$ with $k_0=\omega(\vec k)=\sqrt{M^2_\pi+\vec k\,^2}$.
   The annihilation and creation operators satisfy the canonical commutation relations,
\begin{equation}
[a_i(\vec k),a_j^\dagger(\vec k')]=(2\pi)^3 2\omega(\vec k)\delta_{ij}\delta^3(\vec k-\vec k'),\quad
[a_i(\vec k),a_j(\vec k')]=0=[a_i^\dagger(\vec k),a_j^\dagger(\vec k')].
\end{equation}
   The physical pion fields are introduced as
\begin{equation}
\pi^+(x)=\frac{1}{\sqrt{2}}\big(\phi_1(x)-i\phi_2(x)\big),\quad\pi^0(x)=\phi_3(x),\quad
\pi^-(x)=\frac{1}{\sqrt{2}}\big(\phi_1(x)+i\phi_2(x)\big),
\end{equation}
such that, e.g.,
\begin{displaymath}
\langle 0|\pi^+(x)|\pi^+(\vec p)\rangle=\frac{1}{\sqrt{2}}\langle 0|\pi^+(x)\left(a_1^\dagger(\vec p)
+ia_2^\dagger(\vec p)\right)|0\rangle=e^{-ip\cdot x}.
\end{displaymath}

\subsection{Electromagnetic vertex of a charged pion}
   In the discussion of the low-energy theorem for the Compton tensor in Sec.~\ref{subsectionLET},
we need the interaction vertex of a single photon with a charged pion.
   In the following, we discuss the properties of this vertex which are derived entirely
based on symmetry considerations.

   Let $\pi^+_0$ and $\pi^-_0$ denote bare, i.e., unrenormalized field operators of some
quantum field theory describing interacting charged pions.
   Let us define the three-point Green's function involving the
electromagnetic current operator $J^\mu_0(z)$ as
\begin{equation}
\label{tpgf} G^\mu(x,y,z)=\langle 0|T\left[\pi^+_0(x)
\pi^-_0(y)J^{\mu}_0(z)\right]|0\rangle.
\end{equation}
   The physical interpretation of Eq.~(\ref{tpgf}) is that $G^\mu$ describes
the transition amplitude for creating a $\pi^+$ at $y$, propagation to $z$,
interaction via the current operator at $z$, propagation to $x$ and annihilation
of the $\pi^+$ at $x$.\footnote{
Alternatively, $G^\mu$ describes the creation of a $\pi^-$ at $x$, propagation to $z$,
interaction via the current operator at $z$, propagation to $y$ and annihilation
of the $\pi^-$ at $y$.
   Moreover, $G^\mu$ also describes the creation of a $\pi^+$ at $y$ and of a
$\pi^-$ at $x$, both propagating to $z$, with an annihilation at $z$ through the
current operator.}
   The corresponding momentum-space Green's function reads
\begin{equation}
\label{gmu} (2\pi)^4 \delta^4(p_f-p_i-q) G^{\mu}(p_f,p_i)= \int
d^4x\, d^4y\, d^4z\, e^{i(p_f \cdot x - p_i \cdot y-q\cdot z )}
G^\mu(x,y,z),
\end{equation}
where $p_i$ and $p_f$ are the four-momenta corresponding to lines of a positively
charged pion entering and leaving the vertex, respectively, and $q=p_f-p_i$
is the momentum transfer at the vertex.
   Defining the renormalized three-point Green's function $G^\mu_R$ as
\begin{equation}
\label{gmur} G^{\mu}_R(p_f,p_i) = Z^{-1}_{\pi} Z^{-1}_J
G^{\mu}(p_f,p_i),
\end{equation}
where $Z_{\pi}$ and $Z_J$ are renormalization
constants,\footnote{In fact, $Z_J=1$ due to gauge invariance.}
   we obtain the one-particle irreducible, renormalized three-point Green's
function by removing the propagators at the external lines,\footnote{Note that we
extracted the proton charge $e$.}
\begin{equation}
\label{gammamuirr}
e\Gamma^{\mu,\rm irr}_R(p_f,p_i) = [i
\Delta_R(p_f)]^{-1} G^{\mu}_R(p_f,p_i)[i\Delta_R(p_i)]^{-1},
\end{equation}
where $\Delta_R(p)$ is the full, renormalized propagator.
   From a perturbative point of view, $\Gamma^{\mu,\rm irr}_R$ is made up of
those Feynman diagrams which cannot be disconnected by cutting any
one single internal line.

   In the following we discuss a few model-independent properties of
$\Gamma^{\mu,\rm irr}_R(p_f,p_i)$ \cite{Rudy:1994qb}.
\begin{enumerate}
\item Imposing Lorentz covariance, the most general parametrization of
$\Gamma^{\mu,\rm irr}_R$ can be written in terms of two independent
four-momenta, $P^\mu=p_f^\mu+p_i^\mu$ and $q^\mu=p_f^\mu-p_i^\mu$,
respectively, multiplied by Lorentz-scalar form functions $F$ and
$G$ depending on three scalars, e.g., $q^2$, $p_i^2$, and
$p_f^2$,
\begin{equation}
\label{par} \Gamma^{\mu,\rm irr}_R(p_f,p_i) = (p_f+p_i)^{\mu}
F(q^2,p_f^2,p_i^2) + (p_f-p_i)^{\mu} G(q^2,p_f^2,p_i^2).
\end{equation}
\item Time-reversal symmetry results in
\begin{equation}
\label{trs} F(q^2,p_f^2,p_i^2)=F(q^2,p_i^2,p_f^2), \quad
G(q^2,p_f^2,p_i^2)=-G(q^2,p_i^2,p_f^2).
\end{equation}
   In particular, from Eq.\ (\ref{trs}) we conclude that
$G(q^2,M^2_\pi,M^2_\pi)=0$.
   This, of course, corresponds to the well-known fact that a spin-0 particle
has only one electromagnetic form factor, $F(q^2)$.
\item Using the charge-conjugation properties $J^\mu\mapsto-J^\mu$ and
$\pi^+\leftrightarrow\pi^-$, it is straightforward to see that form
functions of particles are just the negative of form functions of
antiparticles.
   In particular, the $\pi^0$ does not have any electromagnetic form
functions even off shell, since it is its own antiparticle.
\item Due to the hermiticity of the electromagnetic current operator,
$F(q^2)$ is real in the spacelike region $q^2\leq 0$:
\begin{align*}
(p_f+p_i)^\mu e F^\ast(q^2)&=\langle p_f|J^\mu(0)|p_i\rangle^\ast
=\langle p_i|{J^\mu}^\dagger(0)|p_f\rangle
=\langle p_i|J^\mu(0)|p_f\rangle\\
&=(p_i+p_f)^\mu e F(q^2)\,\,\mbox{for}\,\, q^2\leq 0.
\end{align*}
\item After writing out the various time orderings in Eq.\ (\ref{tpgf}),
let us consider the divergence\footnote{For notational convenience, in Eqs.~(\ref{wtstart})--(\ref{partialg}) we omit the
index 0 denoting bare fields.}
\begin{align}
\label{wtstart}
\partial_\mu^z G^\mu(x,y,z)&=
\langle 0|T[\pi^+(x)\pi^-(y)\partial_\mu J^\mu(z)]|0\rangle\nonumber\\
&\quad+\delta(z^0-x^0)\langle 0|T\{[J^0(z),\pi^+(x)]\pi^-(y)\}|0\rangle\nonumber\\
&\quad+\delta(z^0-y^0)\langle 0|T\{\pi^+(x)[J^0(z),\pi^-(y)]\}|0\rangle.
\end{align}
   Current conservation at the operator level,
$\partial_\mu J^\mu(z)=0$, together with the equal-time commutation
relations of the electromagnetic charge-density operator with the
pion field operators,\footnote{Note that both equations are related
by taking the adjoint.}
\begin{align}
\label{comrel}
[J^0(x),\pi^-(y)] \delta (x^0-y^0) & = e\delta^4(x-y) \pi^-(y), \nonumber \\
{[}J^0(x),\pi^+(y)]  \delta (x^0-y^0) & =  -e\delta^4(x-y) \pi^+(y),
\end{align}
are the basic ingredients for obtaining Ward-Takahashi identities
\cite{Ward:1950xp,Takahashi:1957xn} for electromagnetic processes.
   For example, we obtain from Eq.\ (\ref{wtstart})
\begin{equation}
\label{partialg}
\partial_\mu^z G^\mu(x,y,z)=e\left[\delta^4(z-y)-\delta^4(z-x)\right]
\langle 0|T[\pi^+(x)\pi^-(y)]|0\rangle.
\end{equation}
   Taking the Fourier transformation of Eq.\ (\ref{partialg}), performing
a partial integration, and repeating the same steps which lead from
Eq.\ (\ref{gmu}) to (\ref{gammamuirr}), one obtains the celebrated
Ward-Takahashi identity for the electromagnetic vertex
\begin{equation}
\label{wti} q_{\mu} \Gamma^{\mu,\rm irr}_R(p_f,p_i) =
\Delta_R^{-1}(p_f)-\Delta_R^{-1}(p_i).
\end{equation}
   In general, this technique can be applied to obtain Ward-Takahashi
identities relating Green's functions which differ by insertions of
the electromagnetic current operator.

   Inserting the parametrization of the irreducible vertex, Eq.\ (\ref{par}),
into the Ward-Takahashi identity, Eq.\ (\ref{wti}), the form
functions $F$ and $G$ are constrained to satisfy
\begin{equation}
\label{constraint} (p_f^2-p_i^2) F(q^2,p_f^2,p_i^2)+q^2
G(q^2,p_f^2,p_i^2) = \Delta^{-1}_R(p_f)-\Delta^{-1}_R(p_i).
\end{equation}
   From Eq.\ (\ref{constraint}), it can be shown that, given a consistent
calculation of $F$, the propagator of the particle, $\Delta_R$, as
well as the form function $G$ are completely determined (see
Appendix A of Ref.~\cite{Rudy:1994qb} for details).
   The Ward-Takahashi identity thus provides an important consistency check
for microscopic calculations.
\item As the simplest example, one may consider a structureless
``point pion,''
\begin{displaymath}
\Gamma^\mu(p_f,p_i)=(p_f+p_i)^\mu,\quad q_\mu
\Gamma^\mu=p_f^2-p_i^2=(p_f^2-M_\pi^2)-(p^2_i-M_\pi^2),
\end{displaymath}
i.e., $F(q^2,p_f^2,p_i^2)=1$ and $G(q^2,p_f^2,p_i^2)=0$.
\end{enumerate}
   Finally, it is important to emphasize that the off-shell behavior of form
functions is representation dependent, i.e., form functions
are, in general, not observable.
   In the context of a Lagrangian formulation, this can be understood
as a result of field transformations
 \cite{Chisholm:1961tha,Kamefuchi:1961sb,Scherer:1994aq}.
   This does not render the previous discussion useless, rather the
Ward-Takahashi identities provide important consistency relations
between the building blocks of a quantum-field-theoretical
description.

\subsection{Compton tensor}
   The problem of finding a suitable set of amplitudes parametrizing the Compton
tensor was already addressed and solved by Tarrach \cite{Tarrach:1975tu}
through a projection technique originally proposed by Bardeen and Tung \cite{Bardeen:1969aw}.
   Here, we follow a very simple and powerful
alternative method developed by L'vov {\it et al.}~\cite{Lvov:2001zdg} which avoids projections.
   The general goal is to construct a tensor basis and a related
set of Lorentz-invariant amplitudes $B_i$, which are free from kinematical
singularities and constraints.
   This simplifies the classification of the low-energy characteristics of the pion, and also
provides technical advantages, for instance, when discussing
dispersion relations.
   At the end, the Compton tensor is written
in a manifestly gauge-invariant form and divided into structures contributing to real Compton scattering,
virtual Compton scattering (VCS) with one photon virtual and, finally, VCS with both photons virtual.

   To begin with, we define the amplitude $T_{\rm VCS}$ of
virtual Compton scattering,
\begin{equation}
\label{VCSreaction}
   \gamma^{(\ast)}(\epsilon,q) + \pi(p_i)
        \to \gamma^{(\ast)}(\epsilon',q') + \pi(p_f),
\end{equation}
as
\begin{equation}
  T_{\rm VCS} = \epsilon_\mu \epsilon_\nu^{\prime *} T^{\mu\nu},
\end{equation}
where $\epsilon$ and $\epsilon'$ denote the polarization vectors of real or
virtual photons.
   The Compton tensor $T^{\mu\nu}$ is given in terms of the
covariant $\hat{T}$ product \cite{Itzykson:1980rh} of the electromagnetic
current operators,
\begin{equation}
     T^{\mu\nu} =i \int d^4x\, e^{-iq\cdot x}\, \langle\pi(p_f)|
        \hat T[J^\mu(x) J^\nu(0)] |\pi(p_i)\rangle.
\label{T*}
\end{equation}
   Starting from the $S$ matrix $S=I+iT$, we define the invariant scattering amplitude ${\cal T}_{fi}$ for
the transition $|i\rangle$ to $|f\rangle$ as
\begin{equation}
\langle f|T|i\rangle=(2\pi)^4 \delta^4(P_f-P_i){\cal T}_{fi},
\end{equation}
where $P_i$ and $P_f$ denote the total four-momenta of the initial and final states, respectively.
   In the case of real photons, the RCS invariant scattering amplitude is given
by
\begin{equation}
   {\cal T}_{fi} = \epsilon_\mu \epsilon_\nu^{\prime *} T^{\mu\nu}.
\label{T_fi}
\end{equation}
  Due to four-momentum conservation, $p_i+q=p_f+q'$, the tensor
$T^{\mu\nu}$ depends on three linearly independent four-vectors
$P$, $Q$, and $R$,
\begin{equation}
\label{pqr}
    P=\frac{1}{2}(p_i+p_f),\quad
    Q=\frac{1}{2}(q+q'),\quad
    R=\frac{1}{2}(p_f-p_i) = \frac{1}{2}(q-q').
\end{equation}
   Since we only consider the case of initial and
final pions on the mass shell, the vectors $P$, $Q$, and $R$
are constrained by
\begin{equation}
  P^2 = M^2_\pi - R^2,  \quad P\cdot R=0,
\end{equation}
where $M_\pi$ is the pion mass.
   Hence, we can choose four independent kinematical invariants
which, for the moment, we take to be $Q^2$, $R^2$,
$P\cdot Q$, and $Q\cdot R$.

   The discrete symmetries as well as gauge invariance impose
restrictions on the general form of the Compton tensor.
   For charged pions the combination of pion crossing with
charge-conjugation symmetry results in\footnote{For the neutral pion (but not for the neutral kaon),
which is its own antiparticle, only pion crossing
is required to obtain Eq.\ (\ref{crossing1}).}
\begin{equation}
\label{crossing1}
   T^{\mu\nu}(P,Q,R) = T^{\mu\nu}(-P,Q,R),
\end{equation}
   whereas photon crossing yields
\begin{equation}
\label{crossing2}
   T^{\mu\nu}(P,Q,R) = T^{\nu\mu}(P,-Q,R).
\end{equation}
   Gauge invariance requires
\begin{align}
\label{gauge-invariance}
   q _\mu T^{\mu\nu} & =  (Q+R)_\mu T^{\mu\nu} =0, \nonumber\\
   q'_\nu T^{\mu\nu} & =  (Q-R)_\nu T^{\mu\nu} =0,
\end{align}
   where the first and second equation are not independent
once photon-crossing symmetry is imposed.
   Finding a solution to Eqs.\ (\ref{gauge-invariance}) without
introducing kinematical singularities or constraints
is the main challenge in constructing appropriate amplitudes.
   Because of parity conservation, the Compton tensor transforms
as a proper second-rank Lorentz tensor.
   The most general $T^{\mu\nu}$ satisfying the
crossing-symmetry conditions of Eqs.\ (\ref{crossing1})
and (\ref{crossing2}) can be written as
a linear combination of ten basis tensors
which include the metric tensor $g^{\mu\nu}$ and 9 bi-linear
products of $P$, $Q$, and $R$.
   Due to parity, structures containing a single fully antisymmetric tensor
$\epsilon^{\mu\nu\alpha\beta}$ are excluded.

   In order to parametrize $T^{\mu\nu}$, let us introduce gauge-invariant combinations of
photon polarizations and momenta,
\begin{equation}
   {\cal F}_{\mu\nu}  = -i
      (q_\mu \epsilon_\nu - q_\nu \epsilon_\mu ), \quad
   {\cal F}_{\mu\nu}' =  i
      (q_\mu'\epsilon_\nu^{\prime *} - q_\nu'\epsilon_\mu^{\prime *}).
\end{equation}
   These second-rank tensors represent the Fourier components
of the electromagnetic field-strength tensor
$F_{\mu\nu}(x) = \partial_\mu A_\nu(x) - \partial_\nu A_\mu(x)$
associated with plane-wave initial and final photons
described by vector potentials $A_\mu(x) = \epsilon_\mu \exp(-iq\cdot x)$ and
$A_\mu'(x) = \epsilon_\mu^{\prime *} \exp(iq'\cdot x)$, respectively.
   A gauge transformation $A_\mu\mapsto A_\mu+\partial_\mu\chi$ may be represented as
$\epsilon_\mu\mapsto\epsilon_\mu-i\lambda q_\mu$, such that
\begin{displaymath}
{\cal F}_{\mu\nu}\mapsto-i\left[q_\mu(\epsilon_\nu-i\lambda q_\nu)-q_\nu(\epsilon_\mu-i\lambda q_\mu)\right]
={\cal F}_{\mu\nu}-\lambda q_\mu q_\nu+\lambda q_\nu q_\mu={\cal F}_{\mu\nu},
\end{displaymath}
and similarly ${\cal F}'_{\mu\nu}\mapsto {\cal F}'_{\mu\nu}$.
   The most general VCS amplitude can be written in the following manifestly gauge-invariant form:
\begin{align}
\label{ampl}
    T_{\rm VCS} &= \frac{1}{2}{\cal F}^{\mu\nu} {\cal F}_{\mu\nu}' B_1
      + (P_\mu {\cal F}^{\mu\nu})(P^\rho {\cal F}_{\rho\nu}') B_2 \nonumber\\
   &\quad
    + [(P^\nu q^{\prime \mu} {\cal F}_{\mu\nu})
       (P^\sigma q^{\prime \rho} {\cal F}_{\rho\sigma}')
   +  (P^\nu q^\mu {\cal F}_{\mu\nu})
      (P^\sigma q^\rho {\cal F}_{\rho\sigma}') ] B_3\nonumber\\
&\quad
      + (q_\mu {\cal F}^{\mu\nu}) (q^{\prime \rho} {\cal F}_{\rho\nu}') B_4
      + (P^\nu q^\mu {\cal F}_{\mu\nu})
        (P^\sigma q^{\prime \rho} {\cal F}_{\rho\sigma}') B_5.
\end{align}
   The invariant amplitudes $B_i$ are functions of four independent kinematical
variables and are free from kinematical singularities and constraints.
   In terms of ${\cal F}_{\mu\nu}$ and ${\cal F}_{\mu\nu}'$ it turns out
to be rather straightforward to identify structures contributing
for real or virtual photons.
   Real photons in the initial (final) state satisfy $q^2=q\cdot\epsilon=0$
($q'^2=q'\cdot\epsilon'^\ast=0$), such that
\begin{equation}
q^\mu {\cal F}_{\mu\nu}=-i(q^\mu q_\mu\epsilon_\nu-q_\nu q^\mu\epsilon_\mu)=0,\quad
q'^\mu {\cal F}'_{\mu\nu}=0.
\label{qfmunu}
\end{equation}
   As a consequence of Eqs.\ (\ref{qfmunu}), only the
amplitudes $B_1$ and $B_2$ are needed to describe real Compton
scattering.
   When only one photon is virtual, one more amplitude ($B_3$)
contributes.
   All five amplitudes $B_i$ enter, when both photons are
virtual.

   Of course, after substituting $q\to q_1$, $q'\to -q_2$, $p_i\to -p_1$, and
$p_f\to p_2$, Eq.~(\ref{ampl}) also describes the
general kinematical structure of the amplitude of the crossed
reaction $\gamma(q_1)\gamma(q_2) \to \pi(p_1)\pi(p_2)$ for
on-shell pions.

   As mentioned before, the functions $B_i$ depend on the four invariants
$Q^2$, $R^2$, $(Q\cdot R)^2$, and $(P\cdot Q)^2$.
   As an alternative, the following combinations of the first three quantities
can be used as independent arguments of the $B_i$:
$q^2+{q'}^2 = 2(R^2+Q^2)$,
$q\cdot q' = R^2-Q^2$, and $q^2 {q'}^2 = (R^2+Q^2)^2 - 4(Q\cdot R)^2$.
   Thus, we may write
\begin{equation}
\label{Bifinal}
  B_i = B_i(\nu^2, q\cdot q', q^2 + q^{\prime 2}, q^2 q^{\prime 2}),
\end{equation}
where $\nu$ is defined as
\begin{equation}
   M_\pi\nu = P\cdot Q = P\cdot q = P\cdot q'.
\end{equation}
   Besides being manifestly crossing symmetric, this form of the $B_i$ has
the advantage of having a simple limit if one or both photons
become real.

   Finally, the Mandelstam invariants of the VCS reaction
read
\begin{align}
\label{mandelstam}
   s &= (p_i+q)^2  = M^2_\pi + 2M_\pi\nu + q\cdot q', \nonumber\\
   u &= (p_i-q')^2 = M^2_\pi - 2M_\pi\nu + q\cdot q', \nonumber\\
   t &= (q-q')^2 = q^2 + q^{\prime 2} - 2q\cdot q'.
\end{align}

\subsection{Low-energy behavior of the Compton tensor}
\label{subsectionLET}
   At low energies, the electromagnetic polarizabilities describe the response of a
composite system to the electromagnetic fields in Compton scattering.
   In the following, we discuss properties of the Compton tensor in
the limit $q,q'\to 0$, which are entirely based on gauge invariance,
Lorentz covariance and the discrete symmetries $P$, $C$, and $T$.
   In this context, it will be useful to divide the full Compton tensor into
two pieces so that
\begin{equation}
T^{\mu \nu} = T^{\mu \nu}_A +  T^{\mu \nu}_B.
\end{equation}
   Here $T^{\mu \nu}_A$ will contain all of the terms in the amplitude which
are singular as either $q \rightarrow 0$ or $q' \rightarrow 0$, together,
perhaps, with some additional non-singular terms.
   $T^{\mu \nu}_B$ will contain everything else.
   We stress that this separation is not unique in the sense that
non-singular terms may be shifted from $T^{\mu\nu}_A$ to $T^{\mu\nu}_B$ and
vice versa.

   Using, e.g., the soft-photon technique of {Ref.~\cite{Low:1958sn}}, one
may {\em define} the generalized Born terms of the virtual Compton scattering amplitude
as \cite{Fearing:1996gs}
\begin{equation}
   T^{\mu\nu}_{\rm Born} = e^2 F(q^2) F(q^{\prime 2})
   \left[2g^{\mu\nu}
   - \frac{(2p_i+q )^\mu (2p_f+q')^\nu} {(p_i+q )^2 - M^2_\pi}
   - \frac{(2p_i-q')^\nu (2p_f-q )^\mu} {(p_i-q')^2 - M^2_\pi} \right],
\label{Born}
\end{equation}
where $F$ denotes the on-shell electromagnetic form factor.
   The $s$- and $u$-channel terms provide the singular
contributions proportional to $1/(p_i\cdot q)$ and $1/(p_i\cdot q')$,
respectively, whereas the term proportional to the metric tensor
$g^{\mu\nu}$ makes the generalized Born terms gauge invariant.
   Moreover, $T^{\mu\nu}_{\rm Born}$ is symmetric under photon crossing,
$q\leftrightarrow -q'$, $\mu\leftrightarrow \nu$.
   Equation (\ref{Born}) provides a natural
generalization of the Born amplitude for a point-like particle
to the case of a finite-size particle.
   As discussed in Ref.\ \cite{Fearing:1996gs} in detail, such a
generalization incorporates all low-energy singularities of the total
VCS amplitude, so that the non-Born part of the
amplitude can be expanded in powers of small photon momenta, giving rise
to (generalized) polarizabilities.

   A particularly elegant way of obtaining Eq.~(\ref{Born}) from an effective
Lagrangian was discussed in Ref.~\cite{Lvov:2001zdg} and is outlined in \ref{appendixeffLag}.

\subsection{Electromagnetic polarizabilities}
   The generalized Born terms of Eq.~(\ref{Born}) possess all the symmetries of the
total amplitude $T_{\rm VCS}$ and contain all singularities of $T_{\rm VCS}
$ at low energies.
   After having defined the generalized Born terms, we decompose the invariant amplitudes
$B_i(\nu^2, q\cdot q', q^2 + q^{\prime 2}, q^2 q^{\prime 2})$
into generalized Born and non-Born contributions,
\begin{equation}
    B_i = B_i^{\rm Born} + B_i^{\rm NB}, \quad i=1,\ldots,5.
\label{BiBiBornBiNB}
\end{equation}
   From Eq.~(\ref{Born}), we obtain the generalized Born parts of the invariant
amplitudes $B_i$,
\begin{equation}
\label{B-born}
   B_1^{\rm Born} = (q\cdot q') C, \quad
   B_2^{\rm Born} = -4C, \quad
   C = \frac{2e^2 F(q^2) F(q^{\prime 2})} {(s-M^2_\pi)(u-M^2_\pi)},
\end{equation}
and $B_i^{\rm Born}=0$ for $i=3,4,5$.
   At energies below inelastic thresholds, the non-Born parts of $B_i$ are
regular functions of the kinematical variables, which may be expanded
in a Taylor series.
   In particular, when the momenta of both photons are small,
$q \approx q' \to 0$, one obtains
\begin{equation}
\label{LET-2}
     T_{\rm VCS} = T_{\rm VCS}^{\rm Born}
    + \frac{1}{2} {\cal F}^{\mu\nu} {\cal F}_{\mu\nu}' b_1(0)
      + (P_\mu{\cal F}^{\mu\nu})(P^\rho {\cal F}_{\rho\nu}') b_2(0) + {\cal O}(q^4),
\end{equation}
where the constants $b_{i}(0) \equiv B_{i}^{\rm NB}(0,0,0,0)$, $i=1,2$,
are related to the electric and magnetic polarizabilities of low-energy real Compton scattering,
\begin{equation}
\label{real-a,b}
     8\pi M_\pi \alpha_E = -b_1(0) - M^2_\pi b_2(0), \quad
     8\pi M_\pi \beta_M =  b_1(0).
\end{equation}
   For the sake of notational convenience, from now on we will omit the bar symbol
from Compton polarizabilities resulting from a covariant calculation.
   If we explicitly specify the particle in question, we will also omit
the subscripts $E$ and $M$, because a confusion of the electric polarizability
with the fine-structure constant is then excluded.
   Equations (\ref{LET-2}) and (\ref{real-a,b}) provide a Lorentz-invariant
form of the low-energy theorem for real and virtual Compton scattering up to
and including second order in the photon momenta.

   Again, the non-Born contributions can be interpreted as the matrix element
resulting from an effective Lagrangian (see \ref{effLagNB}).

\subsection{Dispersion relations and sum rules}

   In this section, we shortly review how certain analytic properties of the
invariant amplitudes $B_i$ give rise to dispersion relations which, in turn,
can be translated into sum rules for the polarizabilities.
   For the sake of simplicity, we restrict ourselves to the case of real Compton
scattering.
   For detailed reviews of the application of dispersion relations to real and virtual
Compton scattering, we refer to Refs.~\cite{Drechsel:2002ar,Pasquini:2018wbl}.

\begin{figure}[t]
\centerline{\includegraphics[width=0.6\textwidth]{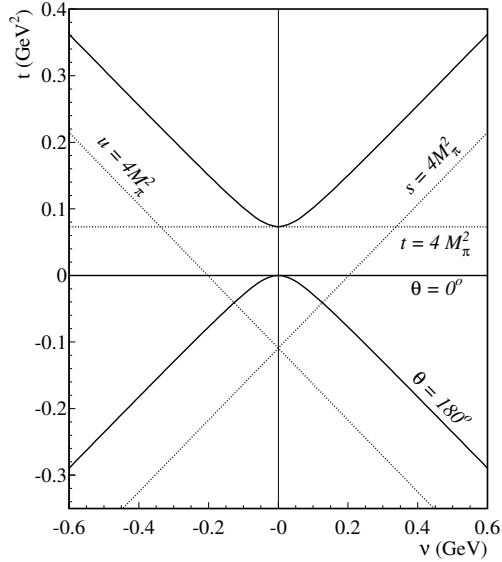}}
\caption{The Mandelstam plane for Compton scattering off the pion.
The functions $A$ and $B$ are real in the interior of a triangle
formed by the dotted lines $s=t=u=4 M^2_{\pi}$.}
\label{fig:mandelstam}
\end{figure}

   Let us consider the kinematics of real Compton scattering, $\gamma(q) +
\pi(p_i) \rightarrow \gamma(q') + \pi(p_f)$.
   In this case, the Mandelstam variables of Eq.~(\ref{mandelstam}) are constrained
by $s+t+u=2M_\pi^2$.
   The crossing-odd variable $\nu$ is defined by
\begin{equation}
\nu=\frac{s-u}{4M_\pi}.
\end{equation}
   The two Lorentz-invariant variables $\nu$ and $t$ span the Mandelstam plane shown in
Fig.~\ref{fig:mandelstam}.
   They are related to the initial ($E_\gamma$) and final ($E'_\gamma$) photon lab energies and to the lab scattering angle $\theta$  by
\begin{align}
\nu&=E_\gamma+\frac{t}{4M_\pi}=\frac{1}{2}(E_\gamma+E'_\gamma),\nonumber\\
t&=-4E_\gamma E'_\gamma  \sin^2 (\theta /2)= -2M_\pi(E_\gamma-E'_\gamma).
\label{eq2.3}
\end{align}
   Starting from Eq.~(\ref{ampl}), the invariant scattering amplitude ${\cal T}_{fi}$ of RCS can be expressed by
2 independent amplitudes $A(\nu, t)$ and $B(\nu,t)$,\footnote{In comparison to {Ref.~\cite{Filkov:1982cx}}, our sign convention for $B$ is opposite.}
\begin{align}
\lefteqn{T_{\rm RCS}=(q\cdot q'\epsilon\cdot\epsilon^{\prime\ast}-\epsilon\cdot q'\epsilon^{\prime\ast}\cdot q)A(\nu,t)}\nonumber\\
&\quad+\left(P\cdot q P\cdot q'\epsilon\cdot\epsilon^{\prime\ast}-P\cdot q\epsilon\cdot q'\epsilon^{\prime\ast}\cdot P
-P\cdot q'\epsilon\cdot P\epsilon^{\prime\ast}\cdot q+q\cdot q'\epsilon\cdot P\epsilon^{\prime\ast}\cdot P\right)B(\nu,t).
\label{TRCS}
\end{align}
   The Lorentz-scalar functions $A$ and $B$ depend on $\nu$ and $t$, they are free of kinematic singularities and
constraints, and, because of the crossing symmetry of $T_{\rm RCS}$, they satisfy the relations $A(\nu, t)=A(-\nu, t)$ and $B(\nu,t)=B(-\nu,t)$.
   We further note that the functions $A$ and $B$ are real
in the interior of a triangle formed by the dotted lines $s=t=u=4 M^2_{\pi}$ in
Fig.~\ref{fig:mandelstam}.

    The unitarity of the $S$ matrix, $S^\dagger S=I$, implies the optical theorem
\cite{Itzykson:1980rh}: the total photoabsorption cross section $\gamma \pi\to \text{hadrons}$,
$\sigma_{\gamma\pi}^{\rm tot}(s)$, and the imaginary part of the elastic forward-scattering transition amplitude $T_{\rm RCS}(s,t=0,\epsilon=\epsilon')$ are related
by
\begin{equation}
\label{optical_theorem}
\text{Im}\left[T_{\rm RCS}(s,t=0,\epsilon=\epsilon')\right]=\sqrt{\lambda(s,M_\pi^2,0)}\,\sigma_{\gamma\pi}^{\rm tot}(s),
\end{equation}
where
\begin{displaymath}
\lambda(x,y,z)\equiv x^2+y^2+z^2-2xy-2yz-2zx
\end{displaymath}
is the K\"all\'en function.
   Using Eq.~(\ref{TRCS}) in combination with Eq.~(\ref{BiBiBornBiNB}),
the forward transition amplitude is given by
\begin{equation}
T_{\rm fw}(\nu)\equiv T_{\rm RCS}(s,t=0,\epsilon=\epsilon')=-2e^2-M_\pi^2\nu^2 B_{\rm fw}^{\rm NB}(\nu),
\label{Tforward}
\end{equation}
where the first term originates from the Born terms of Eq.~(\ref{B-born}) and
$B_{\rm fw}^{\rm NB}(\nu)=B^{\rm NB}(\nu,t=0)$.
   Extending the variable $\nu$ to complex values, $\nu_c=\nu_r+i\nu_i$, the function $T_{\rm fw}(\nu_c)$ is
analytic in the complex $\nu$ plane except for cuts extending along the real axis from $\nu_0=3M_\pi/2$
to $+\infty$ and from $-\infty$ to $-\nu_0$ corresponding to the $s$- and $u$-channel thresholds, respectively
\cite{Itzykson:1980rh}.

\begin{figure}[t]
\centerline{\includegraphics[width=0.6\textwidth]{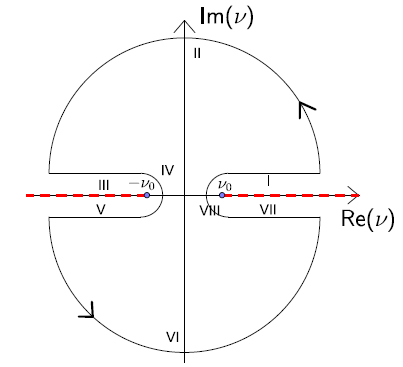}}
\caption{Contour integral in the complex plane.}
\label{fig:complex_nu_plane}
\end{figure}

   Using the contour shown in Fig.~\ref{fig:complex_nu_plane} and applying Cauchy's integral formula to the function
\begin{displaymath}
\tilde{T}_{\rm fw}(\nu)=\frac{T_{\rm fw}(\nu)-T_{\rm fw}(0)}{\nu^2},
\end{displaymath}
   we obtain a (subtracted) forward dispersion relation of the form (see \ref{dispersion_relations})
\begin{equation}
\label{dispersion_relation_ttilde}
\tilde{T}_{\rm fw}(\nu+i\epsilon)=\frac{1}{\pi}\int_{\nu_0}^\infty d\nu'\, \text{Im}\left[\tilde{T}_{\rm fw}(\nu')\right]\left(\frac{1}{\nu'-\nu-i\epsilon}
+\frac{1}{\nu'+\nu+i\epsilon}\right).
\end{equation}
   We rearrange Eq.~(\ref{dispersion_relation_ttilde}) as
\begin{displaymath}
T_{\rm fw}(\nu+i\epsilon)=T_{\rm fw}(0)+\frac{(\nu+i\epsilon)^2}{\pi}\int_{\nu_0}^\infty d\nu'\, \text{Im}\left[\tilde{T}_{\rm fw}(\nu')\right]\left(\frac{1}{\nu'-\nu-i\epsilon}
+\frac{1}{\nu'+\nu+i\epsilon}\right)
\end{displaymath}
and consider the real part for $\nu>0$ in the limit $\epsilon\to 0^+$.
   To this end, we employ Eq.~(\ref{Tforward}) for $\nu =0$, make use of the optical theorem
to rewrite $\text{Im}[\tilde{T}_{\rm fw}(\nu')]=2M_\pi\sigma_{\gamma\pi}^{\rm tot}(\nu')/\nu'$,
and apply the principal value prescription, $\frac{1}{\nu'-\nu-i\epsilon}=P.V.\left(\frac{1}{\nu'-\nu}\right)+i\pi\delta(\nu'-\nu)$,
to obtain
\begin{equation}
\text{Re}\left[T_{\rm fw}(\nu)\right]=-2e^2+4 M_\pi\frac{\nu^2}{\pi}P.V.\int_{\nu_0}^\infty d\nu'\,\frac{\sigma^{\rm tot}_{\gamma\pi}(\nu')}{\nu^{\prime 2}-\nu^2},
\label{dispersion_relation_sigma_tot}
\end{equation}
where $P.V.$ denotes the Cauchy principle value.
   For $0\leq\nu<\nu_0$ and $\nu'\geq\nu_0$, we may expand
\begin{displaymath}
\frac{1}{\nu'^2-\nu^2}=\frac{^1}{\nu'^2}\left(1+\frac{\nu^2}{\nu'^2}+\cdots\right),
\end{displaymath}
such that\footnote{Since the integral starts at $\nu_0$, we can omit the principal value  symbol $P.V.$}
\begin{displaymath}
\int_{\nu_0}^\infty d\nu'\,\frac{\sigma^{\rm tot}_{\gamma\pi}(\nu')}{\nu^{\prime 2}-\nu^2}
=\int_{\nu_0}^\infty d\nu'\,\frac{\sigma^{\rm tot}_{\gamma\pi}(\nu')}{\nu^{\prime 2}}
+\nu^2\int_{\nu_0}^\infty d\nu'\,\frac{\sigma^{\rm tot}_{\gamma\pi}(\nu')}{\nu^{\prime 4}}+\cdots.
\end{displaymath}
   For $|\nu|<\nu_0$, the forward transition amplitude is real and we, therefore, obtain from Eq.~(\ref{dispersion_relation_sigma_tot})
\begin{equation}
T_{\rm fw}(\nu)=-2e^2+4M_\pi\frac{\nu^2}{\pi}\int_{\nu_0}^\infty d\nu'\,\frac{\sigma^{\rm tot}_{\gamma\pi}(\nu')}{\nu^{\prime 2}}
+4M_\pi\frac{\nu^4}{\pi}\int_{\nu_0}^\infty d\nu'\,\frac{\sigma^{\rm tot}_{\gamma\pi}(\nu')}{\nu^{\prime 4}}
+{\cal O}(\nu^6).
\label{Tforwardexpansion}
\end{equation}
   Expanding Eq.~(\ref{Tforward}) in terms of $\nu$,
\begin{displaymath}
T_{\rm fw}(\nu)=-2e^2-M_\pi^2\nu^2 B^{\rm NB}(\nu,0)
=-2e^2-M_\pi^2\nu^2 B^{\rm NB}(0,0)+{\cal O}(\nu^4),
\end{displaymath}
and using the results of Eq.~(\ref{real-a,b}),
\begin{displaymath}
B^{\rm NB}(0,0)=-\frac{8\pi}{M_\pi}(\alpha_E+\beta_M),
\end{displaymath}
we obtain, by comparing the terms proportional to $\nu^2$, the celebrated
Baldin sum rule \cite{Baldin,Petrunkin2},
\begin{equation}
\label{baldin_sum_rule}
\alpha_E+\beta_M=\frac{1}{2\pi^2}\int_{\nu_0}^\infty d\nu\,\frac{\sigma^{\rm tot}_{\gamma\pi}(\nu)}{\nu^2}.
\end{equation}
   In an analogous fashion, the integral multiplying the power $\nu^4$ in Eq.~(\ref{Tforwardexpansion}) provides a sum rule
involving the electric and magnetic quadrupole polarizabilities $\alpha_{E2}$ and $\beta_{M2}$  as well as the so-called dispersive
corrections $\alpha_{E\nu}$ and $\beta_{M\nu}$ to the electric and magnetic dipole polarizabilities \cite{Babusci:1998ww},
\begin{equation}
\alpha_{E\nu}+\beta_{M\nu}+\frac{1}{12}(\alpha_{E2}+\beta_{M2})=
\frac{1}{2\pi^2}\int_{\nu_0}^\infty d\nu\,\frac{\sigma^{\rm tot}_{\gamma\pi}(\nu)}{\nu^4}.
\end{equation}

   Establishing a sum rule for the difference $\alpha_E-\beta_M$ analogous to that for
$\alpha_E+\beta_M$ of Eq.~(\ref{baldin_sum_rule})
turns out to be more complex.
   Various approaches have been suggested, differing by the way how the contour integral is
chosen in the complex plane \cite{Bernabeu:1974zu,Bernabeu:1977hp,Lvov:1979zd,Petrunkin2,Filkov:1982cx}.
   By this, a certain amount of model dependence is introduced to the determination of
$\alpha_E-\beta_M$ \cite{Lvov:1993fp}.
   For a discussion of a dispersive treatment of the crossed-channel reaction $\gamma\gamma\to\pi\pi$
and its relation to the polarizabilities $\alpha_E$ and $\beta_M$,
we refer the interested reader to
Refs.~\cite{Donoghue:1993kw,Filkov:1998rwz,Filkov:2005ccw,Filkov:2005suj,Pasquini:2008ep,GarciaMartin:2010cw,Hoferichter:2011wk,Moussallam:2013una,Dai:2016ytz,Danilkin:2018qfn}.

\section{Compton Scattering off the Pion in Chiral Perturbation Theory}
   In this section, we will focus on the predictions of chiral perturbation theory
(ChPT) \cite{Weinberg:1978kz,Gasser:1983yg,Scherer:2002tk,Bijnens:2014lea} for the electromagnetic
polarizabilities of the pion.
   A short discussion of the chiral Lagrangian can be found in \ref{chiralLagrangian}.
   In fact, within the framework of the partially conserved axial-vector current (PCAC)
hypothesis and current algebra, the electromagnetic
polarizabilities of the charged pion are related to the
radiative charged-pion beta decay $\pi^+\to e^+\nu_e\gamma$~\cite{Terentev:1972ix}.
   Chiral perturbation theory provides a framework for reproducing the predictions
of current algebra and for systematically analyzing corrections to the current-algebra results.

\subsection{Chiral perturbation theory at ${\cal O}(p^4)$}
   For the sake of completeness, we discuss the amplitude $T_{\rm VCS}$ for both
photons virtual.
   The polarizabilities of real Compton scattering are obtained from
the invariant amplitudes evaluated for $q^2=q'^2=0$, whereas the generalized
polarizabilities, discussed in \ref{generalized_polarizabilities}, refer to $q^2<0$ and $q'^2=0$.

\subsubsection{Generalized Born amplitude}
   According to Weinberg's power counting, a calculation of the $s$- and
$u$-channel pole terms at ${\cal O}(p^4)$ involves the renormalized
irreducible vertex at ${\cal O}(p^4)$,
\begin{equation}
\label{gammamu}
\Gamma^{\mu,\text{irr}}_R(p_f,p_i)=(p_f+p_i)^\mu F(q^2)+(p_f-p_i)^\mu \frac{p_f^2-p_i^2}{q^2}
\left[1-F(q^2)\right],\quad q=p_f-p_i,
\end{equation}
   where $F(q^2)$ is the prediction for the electromagnetic form factor
of the pion (see Eq.\ (15.3) of Ref.\ \cite{Gasser:1983yg}).
   At ${\cal O}(p^4)$, the renormalized propagator is simply given by
\begin{equation}
\label{prop}
i\Delta_R(p)=\frac{i}{p^2-M_\pi^2+i0^+},
\end{equation}
with $M^2_\pi$ the ${\cal O}(p^4)$ result for the pion mass squared
(see Eq.\ (12.2) of Ref.\ \cite{Gasser:1983yg}).
   Note that Eqs.\ (\ref{gammamu}) and (\ref{prop}) satisfy the
Ward-Takahashi identity \cite{Ward:1950xp,Takahashi:1957xn}.

\begin{figure}[t]
\centerline{\includegraphics[width=\textwidth]{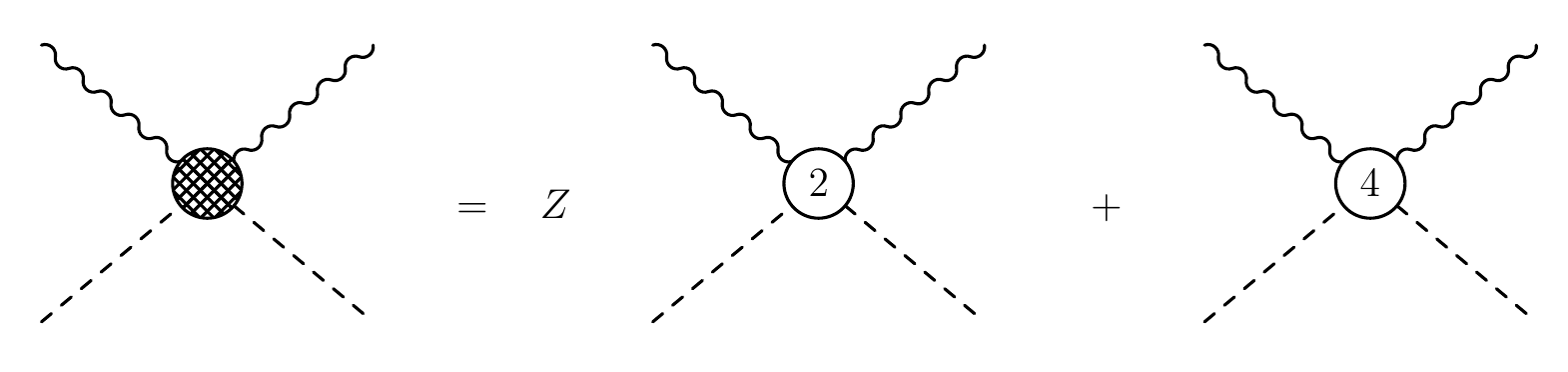}}
\centerline{\includegraphics[width=\textwidth]{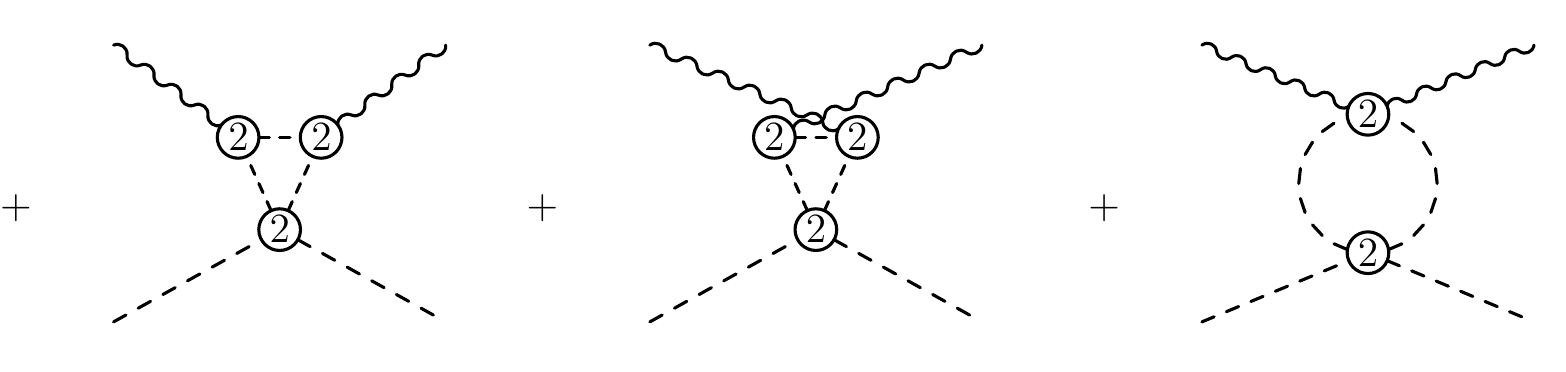}}
\caption{Diagrams related to the one-particle-irreducible residual
amplitude at ${\cal O}(p^4)$. The vertices are denoted by their
chiral order. $Z$ denotes the wave function renormalization
constant. At ${\cal O}(p^4)$, only the tree-level contribution from
${\cal L}_2$ has to be multiplied by $Z$.
\label{figure_chpt_diagrams1}}
\end{figure}

   With these ingredients, the VCS amplitude at ${\cal O}(p^4)$ can be expressed as
\begin{equation}
\label{tvcschpt}
T_{\rm VCS}=T_{\rm VCS}^{\rm pole}+T_{\rm VCS}^{\rm res},
\end{equation}
where, using Eq.~(\ref{gammamu}), the result for the pole terms at ${\cal O}(p^4)$ reads
\begin{align}
\label{bp}
T_{\rm VCS}^{\rm pole}&=-e^2\left(\frac{\epsilon^{\prime\ast}_\nu\Gamma_R^{\nu,\rm irr}(p_f,p_f+q')
\epsilon_\mu\Gamma^{\mu,\rm irr}_R(p_i+q,p_i)}{s-M_\pi^2+i0^+}\right.\nonumber\\
&\left.\quad\quad\quad
+\frac{\epsilon_\mu\Gamma^{\mu,\rm irr}_R(p_f,p_f-q)
\epsilon^{\prime\ast}_\nu\Gamma_R^{\nu,\rm irr}(p_i-q',p_i)}{u-M_\pi^2+i0^+}
\right).
\end{align}
   Here, $s$ and $u$ denote the Mandelstam variables of Eq.~(\ref{mandelstam}).
   Both the pole contribution $T_{\rm VCS}^{\rm pole}$ and the residual contribution
$T_{\rm VCS}^{\rm res}$ are not separately gauge invariant.
   The explicit expression for $T_{\rm VCS}^{\rm pole}$ is given in \ref{explicit_expressions_ChPT}.
   The set of one-particle-irreducible diagrams shown in
Fig.\ \ref{figure_chpt_diagrams1} gives rise to a residual part of the form
\begin{equation}
\label{br}
T_{\rm VCS}^{\rm res}=\Delta T_{\rm VCS}+T_{\rm VCS}^{\rm NB},
\end{equation}
such that
\begin{equation}
T_{\rm VCS}^{\rm Born}=T_{\rm VCS}^{\rm pole}+\Delta T_{\rm VCS}
\end{equation}
corresponds to the generalized Born terms of Eq.~(\ref{Born}).
   The generalized Born amplitudes $T_{\rm VCS}^{\rm Born}$ and
the non-Born amplitude $T_{\rm VCS}^{\rm NB}$
are now separately gauge invariant.
   In particular, if we consider the case of real Compton scattering,
$q^2=q'^2=0$, $\epsilon\cdot q=\epsilon^{\prime\ast}\cdot q'=0$,
the generalized Born terms reduce to the result of scalar QED,
\begin{displaymath}
T_{\rm RCS}^{\rm Born}=e^2\left(2\epsilon'^\ast\cdot\epsilon
-4 \frac{p_f\cdot\epsilon'^\ast \, p_i\cdot\epsilon}{s-M_\pi^2+i0^+}
-4\frac{p_f\cdot\epsilon\, p_i\cdot\epsilon'^\ast}{u-M_\pi^2+i0^+}\right).
\end{displaymath}

\subsubsection{Non-Born amplitude}
   Here, we concentrate on the results for real Compton scattering \cite{Bijnens:1987dc}.
   The expressions for one photon or both photons off shell are
given in Ref.~\cite{Unkmeir:1999md}.
   The concept of generalized polarizabilities is discussed in \ref{generalized_polarizabilities}.
   At ${\cal O}(p^4)$, the expression for the non-Born amplitude reads
\begin{equation}
\label{tbrcharged}
T_{\rm RCS}^{\rm NB}
=e^2\left(q\cdot q'\epsilon\cdot\epsilon'^\ast
-\epsilon\cdot q'\, \epsilon'^\ast\cdot q\right)
\frac{1}{8\pi^2 F_\pi^2}
\left(-\frac{\bar{l}_\Delta}{3}+\text{loops}\right),
\end{equation}
where
\begin{displaymath}
\text{loops}=1-\frac{M_\pi^2}{q\cdot q'}J^{(-1)}\left(-\frac{2q\cdot q'}{M_\pi^2}\right)
\end{displaymath}
originates from the loop diagrams in the second row of Fig.~\ref{figure_chpt_diagrams1}.
   Furthermore, $F_\pi=92.2$ MeV denotes the pion-decay constant \cite{Tanabashi:2018oca} and
$\bar l_\Delta\equiv(\bar l_6-\bar l_5)$ is a linear combination of
scale-independent parameters of the Lagrangian of Gasser and Leutwyler
\cite{Gasser:1983yg}.
   The one-loop function $J^{(-1)}$ is given by \cite{Unkmeir:1999md}
\begin{align}
J^{(-1)}(x)&=\int_0^1 dy\, y^{-1}\ln[1+x(y^2-y)-i0^+]\nonumber\\
&=\begin{cases}
\frac{1}{2}\ln^2\left(\frac{\sigma-1}{\sigma+1}\right)\quad(x< 0),\\
-\frac{1}{2}\arccos^2\left(1-\frac{x}{2}\right)\quad (0\le x <4),\\
\frac{1}{2}\ln^2\left(\frac{1-\sigma}{1+\sigma}\right)-\frac{\pi^2}{2}
+i\pi\ln\left(\frac{1-\sigma}{1+\sigma}\right)\quad (4< x),
\end{cases}
\end{align}
with
\begin{displaymath}
\sigma(x)=\sqrt{1-\frac{4}{x}},\quad x\notin [0,4].
\end{displaymath}
   Comparing Eq.~(\ref{tbrcharged}) with Eq.~(\ref{ampl}) and using
\begin{displaymath}
q\cdot q'\epsilon\cdot\epsilon'^\ast
-\epsilon\cdot q'\, \epsilon'^\ast\cdot q
=\frac{1}{2}{\cal F}^{\mu\nu} {\cal F}_{\mu\nu}',
\end{displaymath}
   we see that, at ${\cal O}(p^4)$, the non-Born contribution
to the amplitude $B_2(\nu^2,q\cdot q',0,0)$ vanishes.
   Moreover, at this order, the amplitude $B_1(\nu^2,q\cdot q',0,0)$ is
not dependent on $\nu$.
   Using the Taylor expansion
\begin{displaymath}
J^{(-1)}(x)=-\frac{1}{2}x+\frac{1}{24}x^2+\cdots,
\end{displaymath}
we obtain
\begin{align*}
b_1(0)&=B_1^{\rm NB}(0,0,0,0)=-e^2\frac{\bar{l}_\Delta}{24\pi^2 F_\pi^2},\\
b_2(0)&=B_2^{\rm NB}(0,0,0,0)=0,
\end{align*}
and, thus, from Eq.~(\ref{real-a,b}) the following prediction for the charged-pion polarizabilities,
\begin{equation}
\alpha_{\pi^\pm}=-\beta_{\pi^\pm}=\frac{e^2}{4\pi}\frac{\bar{l}_\Delta}{48 F_\pi^2 M_\pi}.
\end{equation}
   First of all, we notice the degeneracy $\beta_{\pi^\pm}=-\alpha_{\pi^\pm}$ at ${\cal O}(p^4)$.
   Secondly, the polarizabilities diverge as $1/M_\pi$ in the chiral limit.
   Finally, the loop contribution to the charged-pion polarizabilities is zero at this order,
such that the polarizabilities are entirely predicted in terms of the combination $\bar{l}_\Delta=\bar l_6-\bar l_5$.
   At ${\cal O}(p^4)$, this difference is related to
the ratio $\gamma=F_A/F_V$ of the pion axial-vector form factor
$F_A$ and the vector form factor $F_V$ of radiative pion beta decay
\cite{Gasser:1983yg}, $\gamma=l_\Delta/6$.
   Once this ratio is known, chiral symmetry makes an {\em absolute}
prediction for the polarizabilities.
   This observation was already made in Ref.~\cite{Terentev:1972ix} within the framework of the
PCAC hypothesis and current algebra.
   The situation is similar to the case of $\pi\pi$ scattering \cite{Weinberg:1966kf},
where the $s$-wave $\pi\pi$-scattering lengths are predicted once $F_\pi$ has been
determined from pion decay.
   In terms of the results of the PIBETA Collaboration
for $F_A$ and $F_V$, the ${\cal O}(p^4)$ prediction reads \cite{Bychkov:2008ws}
\begin{equation}
\label{PIBETAprediction}
\alpha_{\pi^\pm}=2.78(2)_{\rm expt}(10)_{F_V}\times 10^{-4}\, \text{fm}^3.
\end{equation}
   The uncertainties originate from the fit to the experimental data and the uncertainty in the
vector form factor $F_V$.
   Furthermore, the prediction does not include effects from
higher orders in the quark-mass expansion.

\subsection{Results at ${\cal O}(p^6)$}
\label{subsection_results_p6}

\begin{figure}[t]
\centerline{\includegraphics[width=0.3\textwidth]{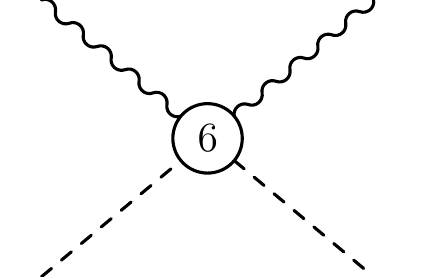}}
\caption{Contact diagram at ${\cal O}(p^6)$
\label{figure_contact_diagram_p6}}
\end{figure}

   The first calculation of the charged-pion polarizabilities at ${\cal O}(p^6)$ was performed in
Refs.~\cite{Burgi:1996mm,Burgi:1996qi}.
   At this order, the calculation involves far more than 100 diagrams.
   The contact diagram at ${\cal O}(p^6)$ (see Fig.~\ref{figure_contact_diagram_p6}) involves two independent contributions
in terms of linear combinations of renormalized, scale-dependent low-energy
couplings from ${\cal L}_6$.
   These contributions were estimated using resonance saturation with vector and axial-vector
mesons ($J^{PC}=1^{--},1^{+-},1^{++}$).
   In units of $10^{-4}$ fm$^3$, the result for the charged-pion polarizabilities as reported
in Refs.~\cite{Burgi:1996mm,Burgi:1996qi}
reads
\begin{align}
\label{alphaEp6}
\alpha_{\pi^\pm}&=\underbrace{2.68}_{{\cal O}(p^4)}+\underbrace{0.08+0.33-0.70}_{{\cal O}(p^6)}=\underbrace{2.39}_{\text{total}},\\
\label{betaMp6}
\beta_{\pi^\pm}&=\underbrace{-2.68}_{{\cal O}(p^4)}+\underbrace{0.07-0.01+0.75}_{{\cal O}(p^6)}=\underbrace{-1.87}_{\text{total}}.
\end{align}
   In Eqs.~(\ref{alphaEp6}) and (\ref{betaMp6}), the corrections at ${\cal O}(p^6)$ are split into
contact contributions estimated using resonance saturation (first term), contributions which are pure numbers
independent of low-energy constants (LECs) and quark masses (second term), and, finally, chiral logarithms (third term).
   The total corrections at ${\cal O}(p^6)$ amount to 11\% and 24\% of the ${\cal O}(p^4)$ predictions
for $\alpha_{\pi^\pm}$ and $\beta_{\pi^\pm}$,
respectively.\footnote{The ${\cal O}(p^4)$ result is given in terms of the LECs available in 1996.}
   The convergence behaviour is similar to that of the $s$-wave $\pi\pi$ scattering lengths \cite{Bijnens:1995yn}.
   Note that the degeneracy $\alpha_{\pi^\pm}=-\beta_{\pi^\pm}$ has been lifted at ${\cal O}(p^6)$.
   The result for the sum of the polarizabilities, $\alpha_{\pi^\pm}+\beta_{\pi^\pm}=0.52\times 10^{-4}$~fm$^3$ is within 25\%
of the Baldin sum rule estimate of {Ref.~\cite{Petrunkin2}},
\begin{equation}
\label{baldincharged}
\alpha_{\pi^\pm}+\beta_{\pi^\pm}=(0.39\pm 0.04)\times 10^{-4}\,\text{fm}^3.
\end{equation}

   In the meantime, the calculation was repeated in Ref.~\cite{Gasser:2006qa}, essentially confirming
the results of Refs.~\cite{Burgi:1996mm,Burgi:1996qi}.
   Using updated values of the LECs, the present status of the charged-pion polarizabilities at
${\cal O}(p^6)$ is
\begin{equation}
\alpha_{\pi^\pm}-\beta_{\pi^\pm}=(5.7\pm 1.0)\times 10^{-4}\,\text{fm}^3,\quad
\alpha_{\pi^\pm}+\beta_{\pi^\pm}=0.16 \times 10^{-4}\,\text{fm}^3.
\end{equation}
   Note that the chiral expansion of the sum, $\alpha_{\pi^\pm}+\beta_{\pi^\pm}$, only starts at ${\cal O}(p ^6)$
and has been quoted without an error \cite{Gasser:2006qa}.
   A dispersive calculation by Kaloshin and Serebryakov \cite{Kaloshin:1993wj} gives $\alpha_{\pi^\pm}-\beta_{\pi^\pm}
=(6.6\pm 1.2)\times 10^{-4}\,\text{fm}^3$, in good agreement with this result
(see also the next section).

\subsection{Polarizabilities of the neutral pion}
\label{subsection_neutral_pion}

   The case of the neutral pion is rather different from the charged pion.
   First, because the $\pi^0$ is its own antiparticle, the $s$- and $u$-channel
pole terms are identically zero.
   Moreover, current algebra does not provide a relation between the Compton tensor
on the one hand and a weak decay on the other hand.

   At ${\cal O}(p^4)$, the polarizabilities entirely originate from loop
contributions (see second row of Fig.~\ref{figure_chpt_diagrams1}), and chiral perturbation theory thus makes a
parameter-free prediction \cite{Bijnens:1987dc,Donoghue:1988eea}
\begin{equation}
\alpha_{\pi^0}=-\beta_{\pi^0}=-\frac{e^2}{4\pi}\frac{1}{96\pi^2 F_\pi^2 M_\pi}
=-0.5 \times 10^{-4}\, \mbox{fm}^3.
\end{equation}
   Note in particular that the electric polarizability is negative.
   Two-loop calculations of the $\gamma\gamma\to\pi^0\pi^0$ reaction have been performed in
Refs.\ \cite{Bellucci:1994eb,Gasser:2005ud} and,
using updated values for the LECs, the two-loop results for the sum and
the difference of polarizabilities are given by \cite{Gasser:2005ud}
\begin{equation}
\label{alphabetapi0}
\begin{split}
(\alpha_{\pi^0}+\beta_{\pi^0})_\text{two-loop}&=(1.1\pm 0.3) \times 10^{-4}\, \mbox{fm}^3,\\
(\alpha_{\pi^0}-\beta_{\pi^0})_\text{two-loop}&=-(1.9\pm 0.2) \times 10^{-4}\, \mbox{fm}^3.
\end{split}
\end{equation}
   As in the case of the charged pions,
the degeneracy $\alpha_{\pi^0}+\beta_{\pi^0}=0$ is lifted at the two-loop
level.
   The Baldin sum rule estimate is $\alpha_{\pi^0}+\beta_{\pi^0}=(1.04\pm0.07)\times 10^{-4}\, \mbox{fm}^3$
\cite{Petrunkin2}, in very good agreement with Eq.~(\ref{alphabetapi0}).

   A first determination of the experimental value of the electric polarizability of the neutral pion ($\pi^0$) was made by Babusci {\it et al.}~\cite{Babusci:1991sk}
   They did so by comparing ChPT at the one-loop level with DESY Crystal Ball (CB) $\gamma\gamma\to\pi^0\pi^0$ total cross section data \cite{Marsiske:1990hx} for
invariant pion-pair masses $M_{\pi^0\pi^0} < 0.5$ GeV/$c^2$ in the angular range $|\cos(\theta)| < 0.8$.
   Assuming $\alpha_{\pi^0}+\beta_{\pi^0}=0$, keeping $\alpha_{\pi^0}$ as a free parameter, and assuming that their one-loop expression is fully justified,
they found $|\alpha_{\pi^0}|=(0.69 \pm 0.07_{\rm stat}\pm 0.04_{\rm syst})\times 10^{-4}\, \mbox{fm}^3$.\footnote{In the given approximation, the total
cross section is proportional to $|\alpha_{\pi^0}|^2$ such that only the absolute value of
$\alpha_{\pi^0}$ could be extracted.}
   Babusci {\it et al.}~cautioned that the $\pi^0$ polarizability value would likely change if multi-loop contributions were found to be large.
   In fact, Bellucci, Gasser, and Sainio later concluded that the two-loop ChPT calculation for the $\gamma\gamma\to\pi^0\pi^0$ cross section
agrees well with the CB data \cite{Marsiske:1990hx} in the low-energy region (see Fig.~5 of Ref.~\cite{Bellucci:1994eb}).
   Subsequently, Donoghue and Holstein (DH) \cite{Donoghue:1993kw} also showed via a dispersive analysis that higher-loop corrections to ChPT are significant.
   They found that CB data agree with their dispersive calculation including the constraints due to chiral symmetry for $\pi^0$ polarizabilities
$\alpha_{\pi^0}=(-1.3, -0.5, 0.3)\times 10^{-4}\, \mbox{fm}^3$, demonstrating thereby that the data have little
sensitivity to the polarizability.
   DH conclude that the CB data are consistent with predicted ChPT $\pi^0$ polarizabilities.
   Kaloshin and Serebryakov \cite{Kaloshin:1993wj} performed a simultaneous analysis of charged and neutral pion polarizabilities
in terms of the Mark II \cite{Boyer:1990vu} and CB data \cite{Marsiske:1990hx}.
   In their $S$-matrix approach, via a combined three-parameter fit, they found $\alpha_{\pi^\pm}-\beta_{\pi^\pm}=(6.6\pm 1.2)\times 10^{-4}\,\text{fm}^3$ and
$\alpha_{\pi^0}-\beta_{\pi^0}=(-0.7\pm 2.3)\times 10^{-4}\,\text{fm}^3$, consistent with ChPT predictions.
   Fil'kov and Kashevarov \cite{Filkov:1998rwz} constructed dispersion relations at fixed $t$ with one subtraction
for the invariant helicity amplitudes of $\gamma\pi$ scattering.
   Using these DRs for the description of the process $\gamma\gamma\to\pi^0\pi^0$,
from a fit to the CB data they obtained the values $\alpha_{\pi^0}+\beta_{\pi^0}=(0.98\pm 0.03)\times 10^{-4}\,\text{fm}^3$
and $\alpha_{\pi^0}-\beta_{\pi^0}=(-1.6\pm 2.2)\times 10^{-4}\,\text{fm}^3$.

   In the meantime, additional data of the $\gamma\gamma\to\pi^0\pi^0$ process, measured in the kinematic range $0.6\, \text{GeV}/c^2 < M_{\pi^0\pi^0} < 4.1\, \text{GeV}/c^2$,
$|\cos(\theta^\ast)|<0.8$, were provided by the Belle collaboration \cite{Uehara:2008ep,Uehara:2009cka}.
   In a very recent dispersive analysis of $\gamma^\ast\gamma\to\pi^0\pi^0$, Danilkin and Vanderhaeghen \cite{Danilkin:2018qfn}
obtain from a coupled-channel approach, as a byproduct, $\alpha_{\pi^0}-\beta_{\pi^0} =9.5\times 10^{-4}\, \mbox{fm}^3$;
a similar number, namely, $8.9 \times 10^{-4}\, \mbox{fm}^3$, was reported in Ref.~\cite{Colangelo:2017fiz} using a single-channel approach.
   The DV predicted low-energy cross sections agree well with data (see Fig.~2 of Ref.~\cite{Danilkin:2018qfn}).
   It should be noted, however, that both approaches make use of unsubtracted dispersion relations such that the value for
$\alpha_{\pi^0}-\beta_{\pi^0}$ is a prediction rather than a free parameter for fitting the cross section data.
   Both analyses emphasize that the inclusion of the $\omega$ meson in the $t$ channel, contributing to the left-hand cuts, most likely
will produce a large correction to the predicted value \cite{Danilkin:2018qfn,Colangelo:2017fiz}.
   Garc\'{\i}a-Mart\'{\i}n and Moussallam \cite{GarciaMartin:2010cw}, using a subtracted dispersion relations analysis of
CB and Belle data, found $\alpha_{\pi^0}-\beta_{\pi^0}=(-1.25\pm 0.17) \times 10^{-4}\, \mbox{fm}^3$ (see also Ref.~\cite{Moussallam:2013una}).
   This result is an indirect determination that includes fitting to the quadrupole $\pi^0$ polarizabilities, plus input
based on a chiral formula [see Eq.~(78) of Ref.~\cite{GarciaMartin:2010cw}] which relates the
dipole polarizabilties $(\alpha_1-\beta_1)_{\pi^0}$, the quadrupole polarizabilities $(\alpha_2-\beta_2)_{\pi^0}$,
and the chiral coupling $c_{34}$ of the ${\cal O}(p^6)$ chiral Lagrangian \cite{Bijnens:1999sh},
\begin{equation}
6(\alpha_1-\beta_1)_{\pi^0} + M_{\pi^0}^2 (\alpha_2-\beta_2)_{\pi^0}= [6.20 +0.25\, c^\text{eff}_{34}(m_\rho)]\times 10^{-4}\,\text{fm}^3,
\end{equation}
with $c_{34}^{\text{eff}}(m_\rho)=4.75\pm 1.71$.
   They include the theory constraint
that the charged-pion dipole polarizability difference should lie in the 67\% confidence interval of the ChPT calculation.
   Their low-energy cross section is very similar to the unitarized ChPT result (see Fig.~11 of Ref.~\cite{GarciaMartin:2010cw}).
   Finally, using various model scenarios, Dai and Pennington \cite{Dai:2016ytz} obtain central values for $\alpha_{\pi^0}-\beta_{\pi^0}$
ranging from $-1.9$ to $-0.8 \times 10^{-4}\, \mbox{fm}^3$.

   From the experimental side, the BESIII collaboration has taken high-statistics data covering for the first time the full
$M_{\pi\pi}$ mass range down to the threshold region for both the $\gamma\gamma\to\pi^0\pi^0$ and $\gamma\gamma\to\pi^+\pi^-$
channels in the low-mass region \cite{Redmer:2018uew,Guo:2019gjf}, which are presently being analyzed.

\subsection{Kaon polarizabilities}
   In terms of the LECs of SU(3) ChPT \cite{Gasser:1984gg}, the one-loop predictions for the charged-kaon polarizabilities
are determined by the same linear combination of low-energy constants as those of the charged-pion polarizabilities.
   One simply needs to replace the pion mass and pion-decay constant by the kaon mass
and kaon-decay constant, respectively \cite{Donoghue:1989si,Guerrero:1997rd}:
\begin{equation}
\alpha_{K^\pm}=-\beta_{K^\pm}=\frac{e^2}{4\pi}\frac{4\left(L_9^r(\mu)+L_{10}^r(\mu)\right)}{F_K^2 M_K}
=0.58 \times 10^{-4}\, \mbox{fm}^3,
\end{equation}
while the neutral-kaon polarizabilities vanish at this order \cite{Guerrero:1997rd}.
  Even though the renormalized coupling constants $L_9^r(\mu)$ and $L_{10}^r(\mu)$ depend on the scale $\mu$, their sum is {\it not} scale dependent.
   Unfortunately, except for an upper limit $|\alpha_{K^-}|<200 \times 10^{-4}\, \mbox{fm}^3$
from kaonic atoms \cite{Backenstoss:1973jx}, no experimental information on kaon polarizabilities
is available, though, in principle, the charged-kaon polarizabilities could also be investigated by the
COMPASS collaboration via the Primakoff reaction with a kaon beam \cite{Moinester:2003rb,Denisov:2018unj}.
   The BESIII collaboration is taking data for $\gamma^\ast\gamma\to K^+ K^-$, which may be analyzed
to determine kaon polarizabilities.

\section{Pion Polarizabilities from Experiment}
   We now turn to the extraction of the charged-pion polarizabilities from experiment.
   As there is no stable pion target, empirical information about the pion
polarizabilities is not easy to obtain.
   For this purpose, one has to consider reactions which contain
the Compton scattering amplitude as a building block,
such as, e.g., the Primakoff effect in high-energy pion-nucleus bremsstrahlung, $\pi^-Z\to \pi^-Z \gamma$,
radiative pion photoproduction on the nucleon, $\gamma p\to \gamma \pi^+n$, and pion pair
production in $e^+e^-$ scattering, $e^+e^-\to e^+e^-\pi^+\pi^-$.

\subsection{COMPASS}

   The COMPASS collaboration at CERN determined the difference $\alpha_{\pi^\pm}-\beta_{\pi^\pm}$ by investigating
pion Compton scattering $\gamma\pi\to\gamma\pi$ at center-of-mass energies below $3.5$ pion masses \cite{Adolph:2014kgj}.
   Compton scattering was measured via radiative pion Primakoff scattering (bremsstrahlung of 190 GeV/$c$ negative pions)
in the nuclear Coulomb field of the Ni nucleus: $\pi^-\text{Ni}\to\pi^-\text{Ni}\gamma$.
   Exchanged quasi-real photons are selected by isolating the sharp Coulomb peak observed at lowest four-momentum transfers to the target nucleus,
$Q^2 < 0.0015\,\text{GeV}^2/c^2$.
   The resulting data are equivalent to $\gamma\pi\to\gamma\pi$ Compton scattering for laboratory $\gamma$'s
having momenta of the order 1~GeV/$c$, incident on a target pion at rest.
   In the reference frame of this target pion, the cross section is sensitive to $\alpha_{\pi^\pm}-\beta_{\pi^\pm}$
at backward angles of the scattered $\gamma$'s [see Eq.~(\ref{dsdopol}) and Ref.~\cite{Buenerd:1995dd}].
   This corresponds to the most forward angles in the laboratory frame for the highest-energy Primakoff $\gamma$'s.
   Historically, pion Compton scattering was first observed via Primakoff scattering by Kowalewski {\it et al.}~in 1984 \cite{Kowalewski:1984it}.

   Pion Primakoff scattering at COMPASS is an ultra-peripheral reaction on a virtual-photon target.
   The initial and final-state pions are at a distance (impact parameter $b$) more than 50 fm from the target nucleus,
significantly reducing meson-exchange and final-state interactions.
   This follows from the extremely small four-momentum transfer $Q_{\rm min}$ to the target nucleus in a Primakoff reaction.
   For COMPASS radiative pion Primakoff scattering, the four-momentum transfer $Q$ to the target nucleus is in the range up to
3~$Q_{\rm min}$, with average value $\approx$~2 $Q_{\rm min}$.
   By the uncertainty principle, with $Q_{\rm min}\approx 1\, \text{MeV}/c$ and $\Delta Q_{\rm min}\approx 2\, \text{MeV}/c$,  the impact
parameter is $\Delta b \sim \hbar c/(2c\Delta Q_{\rm min})\approx 50\,\text{fm}$.

   COMPASS used a 190 GeV/$c$ beam of negative hadrons (96.8\%~$\pi^-$, 2.4\%~$K^-$, 0.8\%~$\bar{p}$).
   The COMPASS spectrometer has a silicon tracker to measure precise meson scattering angles,
   electromagnetic calorimeters for $\gamma$ detection and for triggering, and Cherenkov threshold detectors
for $K/\pi$ separation \cite{Abbon:2007pq}.
   From a 2009 data sample of 63,000 events, the extracted pion polarizabilities were determined.

\begin{figure}[t]
\centerline{\includegraphics[width=0.8\textwidth]{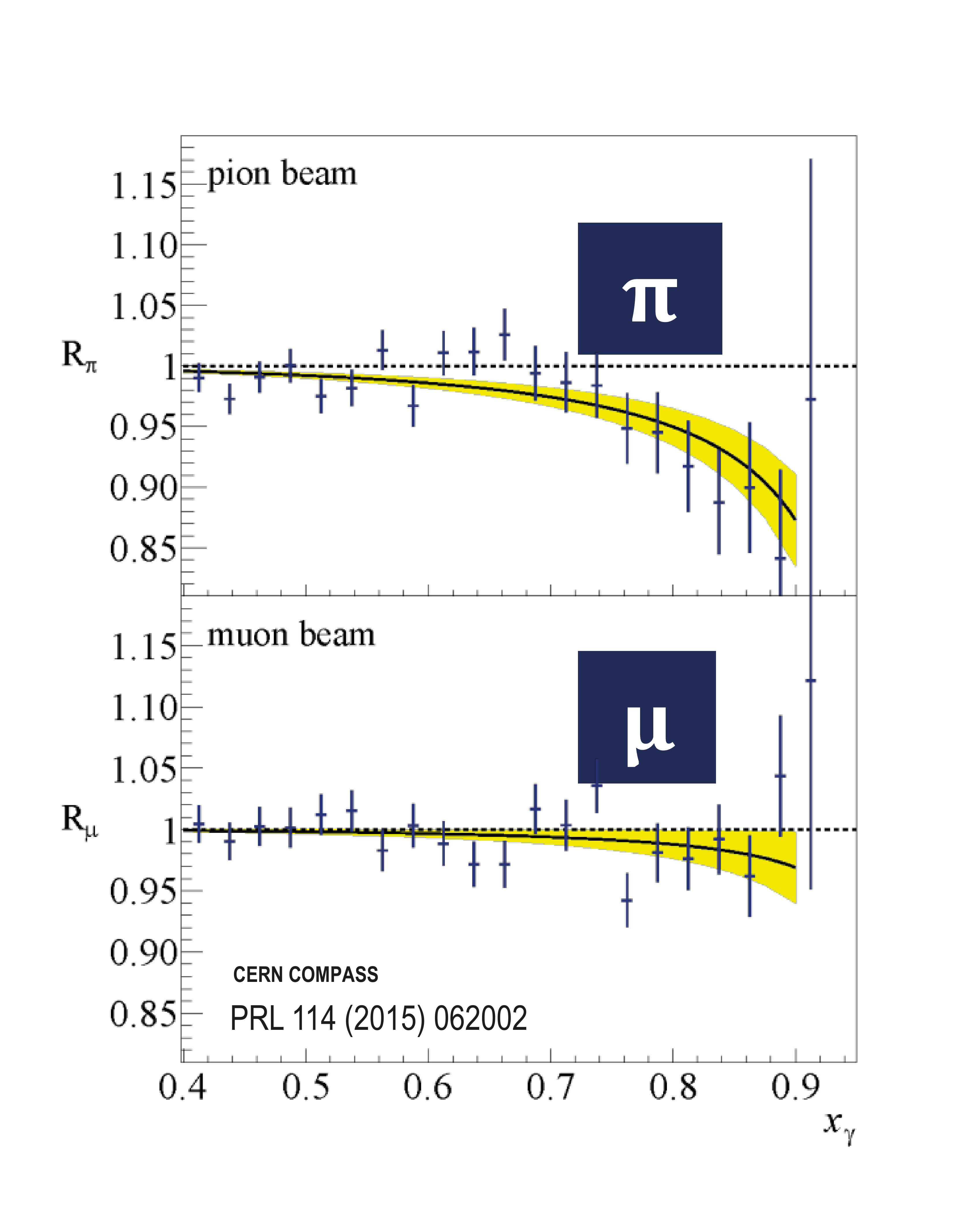}}
\caption{Upper panel: Determination of the pion polarizability by fitting the $x_\gamma$ distribution of the experimental
ratios $R_\pi$ (data points) to the theoretical ratio $R_T$ (solid line) [see Eq.~(\ref{Rpi})];
   lower panel: Measurement with a muon beam. (From Ref.~\cite{Adolph:2014kgj})
\label{figure_ratios_R}}
\end{figure}

   Assuming $\alpha_{\pi^\pm}+\beta_{\pi^\pm}=0$, the dependence of the laboratory differential cross section on
$x_\gamma=E_\gamma/E_\pi$ is used to determine $\alpha_{\pi^\pm}$, where $x_\gamma$ is the fraction of the beam energy
carried by the final-state $\gamma$.
   The variable $x_\gamma$ is related to the $\gamma$ scattering angle for $\gamma\pi\to\gamma\pi$,
so that the selected range in $x_\gamma$ corresponds to backward scattering, where the sensitivity to
$\alpha_{\pi^\pm}-\beta_{\pi^\pm}$ is largest.
   Let $\sigma_{\rm E}(x_\gamma)$ denote the experimental laboratory frame differential cross section
as a function of $x_\gamma$.
   Furthermore, let $\sigma_{\rm MC}(x_\gamma,\alpha_{\pi^\pm})$ denote the calculated cross section for
polarizability $\alpha_{\pi^\pm}$, using a Monte Carlo simulation, such that $\sigma_{\rm MC}(x_\gamma,\alpha_{\pi^\pm}=0)$ denotes the cross section
for a point-like pion having zero polarizability.
   The $\sigma_{\rm E}(x_\gamma)$ data are obtained after subtracting backgrounds from the
$\pi^-\text{Ni}\to\pi^-\text{Ni}\gamma$ diffractive channel and the $\pi^-\text{Ni}\to\pi^-\text{Ni}\pi^0$ diffractive and Primakoff channels.
   The ratios $R_\pi=\sigma_{\rm E}(x_\gamma)/\sigma_{\rm MC}(x_\gamma,\alpha_{\pi^\pm}=0)$ are the experimental data points
shown in the upper panel of Fig.~\ref{figure_ratios_R}.
   The polarizability $\alpha_{\pi^\pm}$ and its statistical error are extracted by fitting $R_\pi$ to the theoretical expression
\begin{equation}
\label{Rpi}
R_\pi=1-\frac{3}{2}M_\pi^3\frac{x^2_\gamma}{1-x_\gamma} \frac{4\pi\alpha_{\pi^\pm}}{e^2}=1-72.73\times 10^{-4}  \frac{x^2_\gamma}{1-x_\gamma}\alpha_{\pi^\pm},
\end{equation}
where $\alpha_{\pi^\pm}$ is given in units of $10^{-4}$~fm$^3$.
   The ratio $R_\pi$ with best fit $\alpha_{\pi^\pm}$ is shown in the upper panel of Fig.~\ref{figure_ratios_R} as the solid curve \cite{Adolph:2014kgj}.
   Systematic uncertainties were controlled by many tests, including measuring $\mu^-\text{Ni}\to\mu^-\text{Ni}\gamma$
Primakoff cross sections by replacing pions by muons while keeping the same beam momentum.
   The muon Compton scattering cross section is precisely known, since muons have zero polarizabilities.
   The lower panel of Fig.~\ref{figure_ratios_R} shows the analogous ratio $R_\mu$ for the muon measurement.
   The main contribution to the systematic uncertainties comes from the Monte Carlo description of the COMPASS setup.
   Comparing experimental and theoretical $x_\gamma$ dependences of $R_\pi$ yields:
\begin{equation}
\alpha_{\pi^\pm}=-\beta_{\pi^\pm}=(2.0 \pm 0.6_{\rm stat} \pm 0.7_{\rm syst})\times 10^{-4}\,\text{fm}^3
\end{equation}
or equivalently $\alpha_{\pi^\pm}-\beta_{\pi^\pm}=(4.0 \pm 1.2_{\rm stat}\pm 1.4_{\rm syst})\times 10^{-4}\,\text{fm}^3$.
   The COMPASS data analysis \cite{Adolph:2014kgj,Guskov:2010zna,Nagel:2012cla,Friedrich:2012ybb} included corrections
for one-photon-loop radiative effects \cite{Akhundov:1994uv,Kaiser:2008hj}, chiral loop effects \cite{Nagel:2012cla,Kaiser:2008ss},
and the electromagnetic form factor of the Nickel nucleus.
   The corrections were modest, increasing the extracted polarizability values by $\approx 0.6\times {10}^{-4}\,\text{fm}^3$
after they are applied \cite{Adolph:2014kgj}.
   The ChPT one-loop calculation and dispersion relation calculations of Pasquini agree with one another on the permille level
\cite{Friedrich:2016gqb}.

   Antipov {\it et al.}~\cite{Antipov:1982kz,Antipov:1984ez} previously carried out a Primakoff polarizability experiment at
Serpukhov using a 40~GeV/$c$ beam of negative pions, and reported
$\alpha_{\pi^\pm}-\beta_{\pi^\pm} = (13.6 \pm 2.8_{\rm stat} \pm 2.4_{\rm syst})\times 10^{-4}\,\text{fm}^3$,
higher than the COMPASS result.
   However, since this low-statistics experiment ($\approx$ 7000 events) did not allow complete precision studies
of systematic errors, their result is not considered further in the present review.
   Higher statistics data ($\approx$ 5 times) taken by COMPASS in 2012 are expected to provide
an independent and improved determination of  $\alpha_{\pi^\pm} - \beta_{\pi^\pm}$.
   Measurement of the kaon polarizability would become possible if and when COMPASS achieves a radio-frequency-separated kaon beam.

\subsection{Mainz}

\begin{figure}[t]
\centerline{\includegraphics[width=0.3\textwidth]{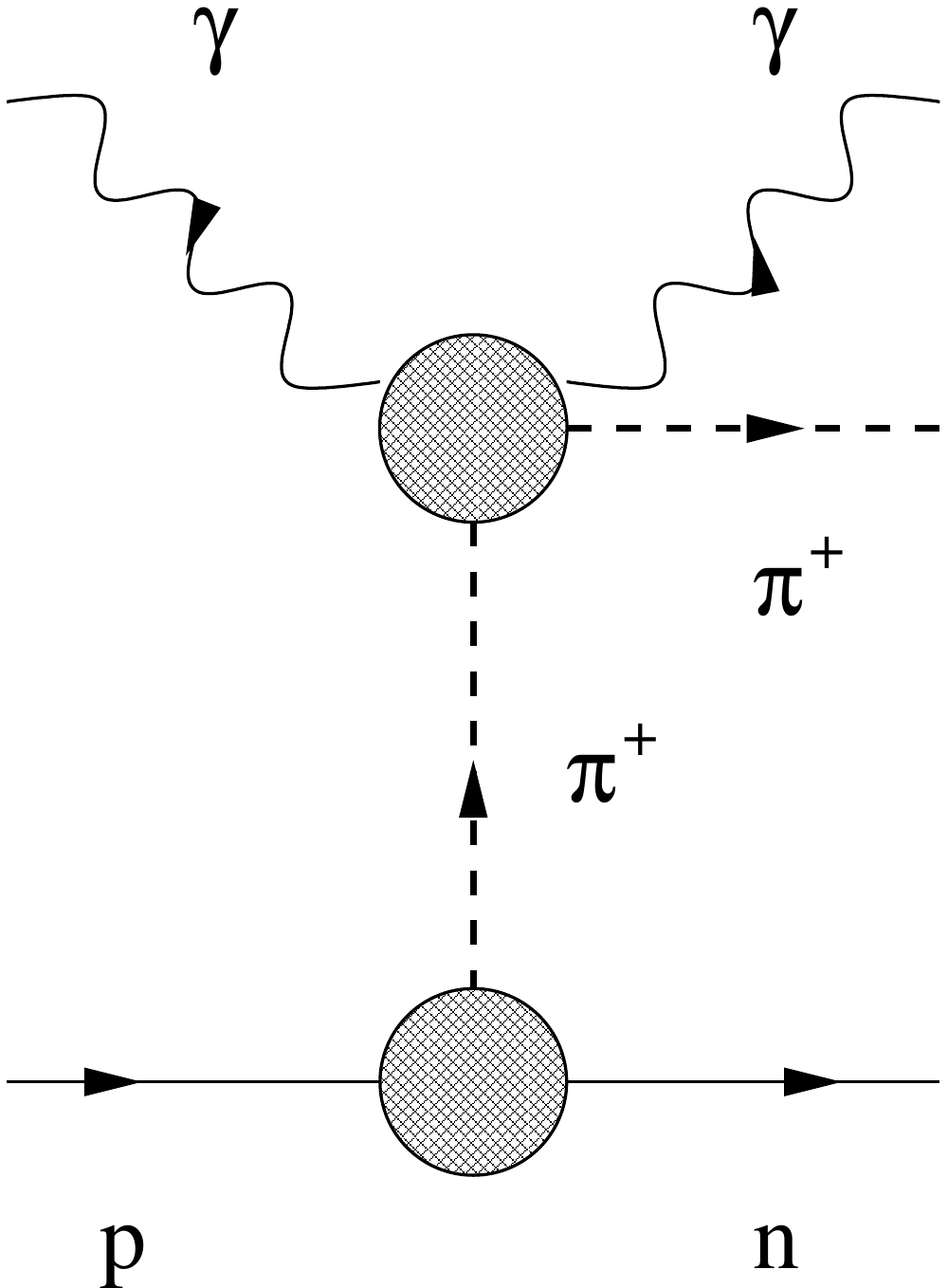}}
\caption{The reaction $\gamma p\to\gamma\pi^+n$ contains Compton scattering off
a pion as a sub diagram in the $t$ channel, where $t=(p_n-p_p)^2$.}
\label{fig:tchannel}
\end{figure}

\begin{figure}
\centerline{\includegraphics[width=0.6\textwidth]{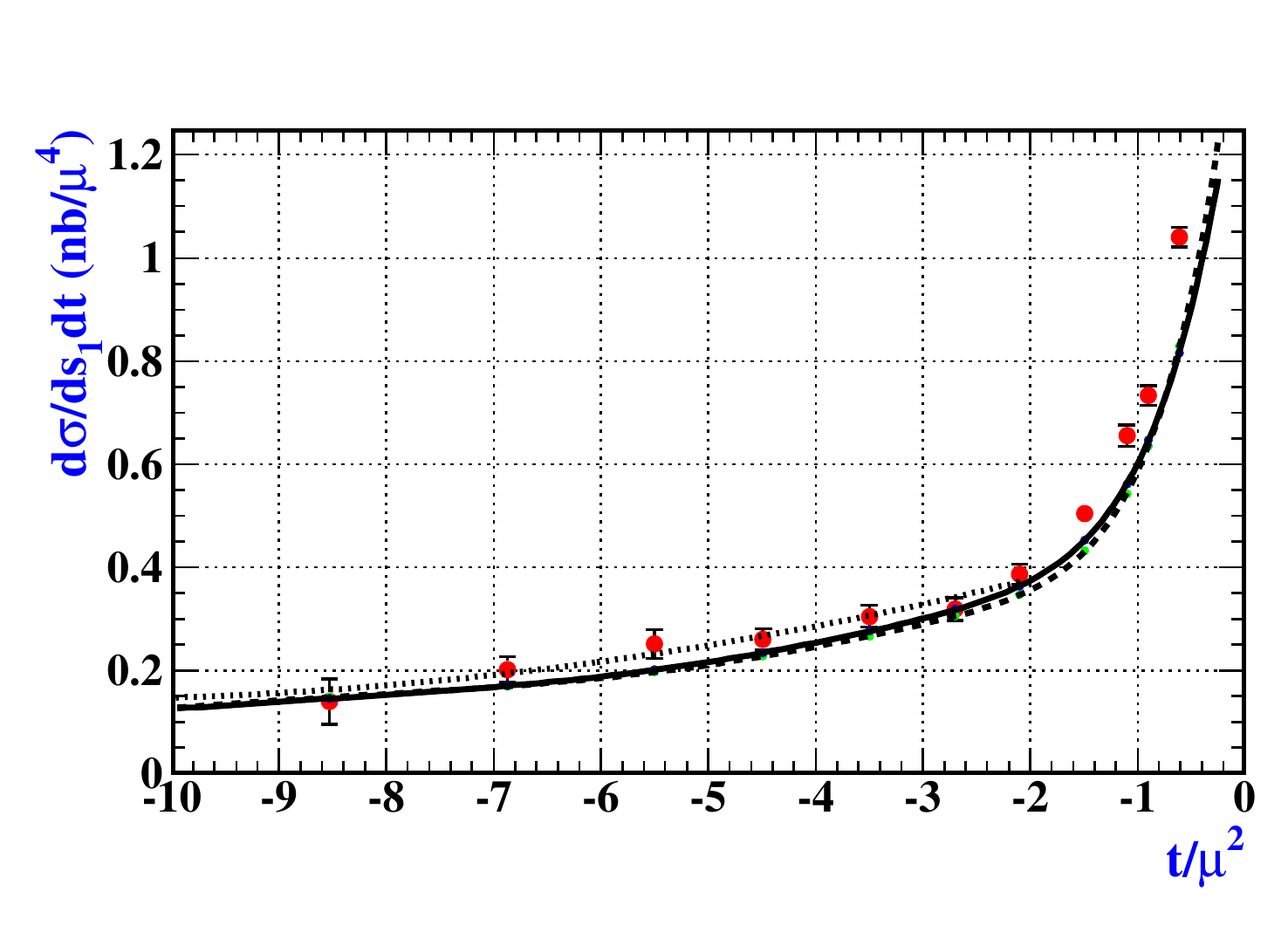}}
\caption{Differential cross section averaged over 537 MeV $< E_\gamma <$ 817 MeV
and 1.5 $M_\pi^2<s_1<5 M_\pi^2$. Solid line: model 1; dashed line: model 2;
dotted line: fit to experimental data. Note that $\mu^2=M_\pi^2$. From Ref.~\cite{Ahrens:2004mg}.} \label{fig:cros1c}
\end{figure}

\begin{figure}
\centerline{\includegraphics[width=0.6\textwidth]{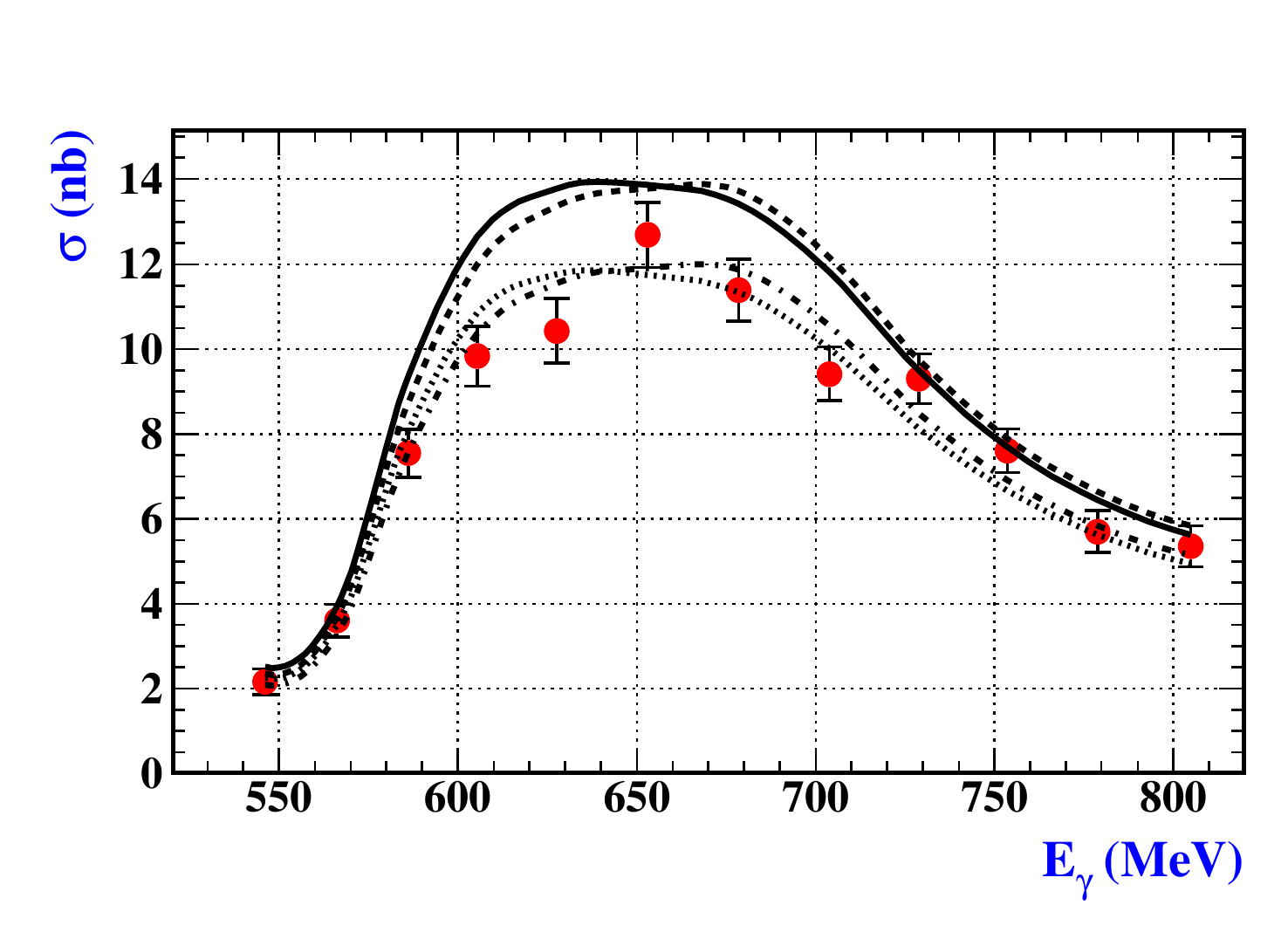}}
\caption{
The cross section of the process $\gamma p\to\gamma\pi^+n$ integrated over $s_1$
and $t$ in the region where the contribution of the pion polarizability is
biggest and the difference between the predictions of the theoretical models
under consideration does not exceed 3\%. The dashed and dashed-dotted lines are
predictions of model 1 and the solid and dotted lines of model 2 for
$(\alpha-\beta)_{\pi^\pm}=0$ and $(\alpha-\beta)_{\pi^\pm}=14\times 10^{-4}\,
\mbox{fm}^3$, respectively. From Ref.~\cite{Ahrens:2004mg}.} \label{fig:cros2c}
\end{figure}

   The potential of studying the influence of the pion polarizabilities on radiative pion
photoproduction from the proton was extensively studied in Ref.~\cite{Drechsel:1994kh}.
   In terms of Feynman diagrams, the reaction $\gamma p\to\gamma\pi^+n$ contains
real Compton scattering off a charged pion as a pion-pole diagram (see
Fig.~\ref{fig:tchannel}).
   Radiative $\pi^+$-meson photoproduction from the proton ($\gamma p\to \gamma\pi^+n$)
was studied at the Mainz Microtron in the kinematic region  537 MeV $< E_\gamma <$ 817 MeV,
$140^{\circ}\le\theta^{\rm cm}_{\gamma\gamma'}\leq 180^{\circ}$,
where $\theta^{\rm cm}_{\gamma\gamma'}$ is the polar angle between the initial and final photon in the c.m.~system
of the outgoing $\gamma$ and pion \cite{Ahrens:2004mg}.
   The experimental challenge is that the incident $\gamma$ ray is scattered from an off-shell pion, and the polarizability contribution to
the Compton cross section from the pion pole diagrams is only a small fraction of the measured cross section.
   The $\pi^+$-meson polarizability was determined from a comparison of the data with the predictions of two theoretical models.
   Model 1 includes eleven pion and nucleon pole diagrams, using the pseudoscalar pion-nucleon interaction and including the anomalous magnetic
moments of protons and neutrons.
   Model 2 consists of five nucleon and pion pole diagrams without the anomalous magnetic moments but including contributions from the
resonances $\Delta (1232)$, $P_{11}(1440)$, $D_{13}(1520)$, $S_{11}(1535)$, and the $\sigma$ meson.
   The validity of these two models was studied by comparing the predictions with the experimental data in the kinematic region where the
pion polarizability contribution is negligible ($s_1 < 5 M_\pi^2$),
where $s_1$ is the square of the total energy in the $\gamma p\to\gamma\pi^+n$ c.m.~system,
and where the difference between the predictions of the two models does not exceed 3\% (see Fig.~\ref{fig:cros1c}).
   In the region where the pion polarizability contribution is substantial
($5 < s_1/M_\pi^2 <15$; $-12 < t/M_\pi^2 < -2$),
$\alpha_{\pi^\pm}-\beta_{\pi^\pm}$ was determined from a fit of the calculated cross section to the data, as illustrated in
Fig.~\ref{fig:cros2c}.
   The deduced polarizabilities are \cite{Ahrens:2004mg}
\begin{equation}
\alpha_{\pi^\pm}-\beta_{\pi^\pm} = (11.6 \pm 1.5_{\rm stat} \pm 3.0_{\rm syst} \pm 0.5_{\rm model})\times 10^{-4}\,\text{fm}^3.
\end{equation}
   Combining statistical and systematic errors, the Mainz polarizability is $\alpha_{\pi^\pm}-\beta_{\pi^\pm} = (11.6 \pm 3.4)\times 10^{-4}\,\text{fm}^3$.
   The large uncertainty is due mainly to the calculated efficiency of the neutron detectors used in the experiment.
   The 95\% confidence interval for the polarizabilities is then approximately 5 to $18 \times 10^{-4}\,\text{fm}^3$, too large
to constrain models.

   The quoted model uncertainty $0.5_{\rm model}\times 10^{-4}\,\text{fm}^3$ denotes the uncertainty associated with using the two chosen theoretical models.
   It was estimated as half the difference between the model-1 and model-2 polarizability values.
   However, it does not take into account that comparisons with other possible models may significantly increase the model error.
   A larger model uncertainty could help explain the difference between COMPASS and Mainz polarizabilities.

   It would be of interest to improve the estimate of the model uncertainty by using an independent model to extract the polarizability.
   A step towards a third model was taken by Kao, Norum, and Wang \cite{Kao:2008pf} who studied the $\gamma p\to \gamma\pi^+n$ reaction
within the framework of heavy-baryon chiral perturbation theory.
   They found that the contributions from two unknown low-energy constants in the $\pi N$ chiral Lagrangian are comparable
with the contributions of the charged pion polarizabilities.
   Their model therefore suggests that higher-order contributions could substantially increase the model error.

\subsection{Mark II}

   Charged-pion polarizabilities were determined by comparing Mark II total cross section data ($\gamma\gamma\to\pi^+\pi^-$) \cite{Boyer:1990vu}
for invariant pion-pair masses $M_{\pi^+\pi^-}\leq 0.5\,\text{GeV}/c^2$ with a ChPT one-loop calculation \cite{Babusci:1991sk,Alexander:1993rz}.
   The Mark II experiment was carried out via the reaction $e^+e^-\to e^+ e^-\pi^+\pi^-$ at a center-of-mass energy of
29~GeV for invariant pion-pair masses $M_{\pi^+\pi^-}$ between 350~MeV/$c^2$ and 1.6~GeV/$c^2$ \cite{Boyer:1990vu}.
   Only the region below $M_{\pi^+\pi^-}=0.5\,\text{GeV}/c^2$ is considered within the domain of validity of ChPT.

   The most important problem in studying the $e^+e^-\to e^+ e^-\pi ^+\pi ^-$ reaction is the elimination of the dominant two-prong
QED reactions $e^+e^-\to e^+e^-e^+e^-$ and $e^+e^-\to e^+e^-\mu^+\mu^-$.
   These leptonic backgrounds below $M_{\pi^+\pi^-}=0.5\,\text{GeV}/c^2$ are expected each to be more than 10 times larger than the expected signal.
   For the critical $M_{\pi^+\pi^-}$ region between 350 and 400 MeV/$c^2$,
Mark II eliminated these backgrounds by identifying pion pairs using time of flight (TOF),
by requiring both tracks to hit an active region of the liquid-argon calorimeter, and by requiring both tracks to have a
summed transverse momentum with respect to the $e^+e^-$ axis of less than 150 MeV/$c$ \cite{Boyer:1990vu}.
   Summarizing, Mark II at SLAC has the highest statistics and lowest systematic error data for
$\gamma\gamma\to\pi^+\pi^-$ for $M_{\pi^+\pi^-}\leq 0.5\,\text{GeV}/c^2$.

\begin{figure}
\centerline{\includegraphics[width=\textwidth]{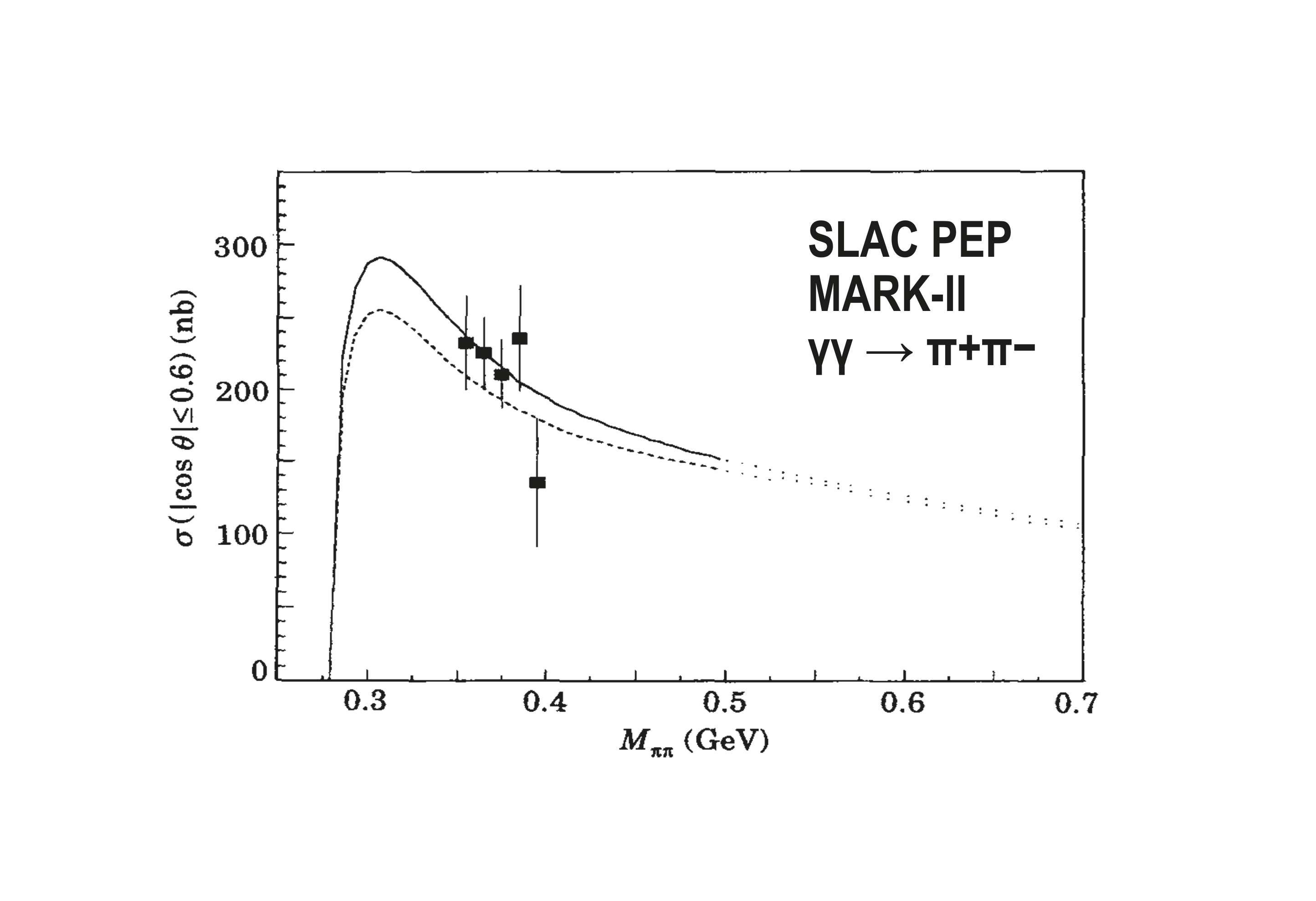}}
\caption{Mark II total cross section data ($\gamma\gamma\to\pi^+\pi^-$) for $M_{\pi^+\pi^-}\leq 0.5\,\text{GeV}/c^2$.
The theoretical curves are: Born terms (dashed line); ChPT with $\alpha_{\pi^\pm}-\beta_{\pi^\pm} = 5.4\times 10^{-4}\,\text{fm}^3$ (full line).
The region above $M_{\pi^+\pi^-}=0.5\,\text{GeV}/c^2$  is considered outside the domain of validity of ChPT. From Ref.~\cite{Alexander:1993rz}.}
\label{fig:MarkII}
\end{figure}

   A number of theoretical papers subsequently made use of the Mark II data to deduce pion polarizabilities.
   For example, theoretical curves from Ref.~\cite{Alexander:1993rz} are shown in Fig.~\ref{fig:MarkII} for the Born cross section (dashed line)
and the ChPT cross section with $\alpha_{\pi^\pm}-\beta_{\pi^\pm} = 5.4\times 10^{-4}\,\text{fm}^3$ (full line).
   The cross section excess below $M_{\pi^+\pi^-} = 0.5\,\text{GeV}/c^2$ compared to the Born calculation was interpreted as due to pion polarizabilities,
with best fit value $\alpha_{\pi^\pm}-\beta_{\pi^\pm} = (4.4\pm 3.2_{\rm stat+syst})\times 10^{-4}\,\text{fm}^3$ \cite{Babusci:1991sk}.
   A similar analysis from Ref.~\cite{Donoghue:1993kw} gave $\alpha_{\pi^\pm}-\beta_{\pi^\pm} \approx 5.3 \times 10^{-4}\,\text{fm}^3$, consistent with this result.
   The 95\% confidence interval from these analyses is then approximately 0 to $11\times 10^{-4}\,\text{fm}^3$.
   Using updated low-energy constants, the most recent two-loop analysis gave $\alpha_{\pi^\pm}-\beta_{\pi^\pm}=(5.7\pm 1.0)\times 10^{-4}\,\text{fm}^3$
\cite{Gasser:2006qa}.

\subsection{Present and future experiments}

   The JLab pion polarizability experiment E12-13-008 \cite{Aleksejevs:2013}, beginning about 2021, plans to measure $\gamma\gamma\to\pi^+\pi^-$ cross sections and asymmetries
via the Primakoff  $\gamma\gamma\to\pi^+\pi^-$ reaction.
   They will use a 6 GeV tagged linearly polarized photon beam produced via
coherent bremsstrahlung, an Sn ''virtual-photon target,'' the GlueX detector in Hall-D, and auxiliary detectors.
   GlueX is based on a solenoidal hermetic detector optimized for tracking of charged particles and detection of gamma rays.
   An important problem in studying the $\gamma\gamma\to\pi^+\pi^-$ reaction is the elimination of the two-prong
QED reactions $\gamma\gamma\to e^+e^-$ and $\gamma\gamma\to\mu^+\mu^-$.
   The muon pair background below $M_{\pi^+\pi^-} = 0.5$~GeV/$c^2$
is expected to be roughly 5 times larger than the expected signal;
while the electron-positron background is expected to be negligible.
   TOF will not be useful for particle identification (PID) in the JLab measurement because of the extreme relativistic velocities of the pions.
   For $e^+/e^-$, they will use the GlueX FCAL lead-glass calorimeter along with tracking information to determine energy and momentum.
   For pion/muon PID, they will use a combination of the responses of the FCAL and two sets of downstream wire chambers
separated by a passive iron hadronic absorber to distinguish pions from muons.
   The JLab experiment plans to use linearly polarized incident photons and to use the asymmetry and
the azimuthal dependences of the $\pi^+\pi^-$ and $\mu^+\mu^-$ systems to help distinguish between
signal and background.
   The advantages of measuring azimuthal correlations for both $\pi^+\pi^-$ and $\pi^0\pi^0$ channels
was discussed in Refs.~\cite{Alexander:1993rz,Ong:1988tr}.
   Since the $\pi^+\pi^-$ channel will have contributions from both coherent $\rho^0$ photo-production on the nuclear target and Primakoff production of pion pairs,
the asymmetry dependence will allow separating these channels.
   Dispersion relation calculations \cite{Pasquini:2008ep,Dai:2016ytz} show that the total cross section for $|\cos(\theta)| < 0.6$  for $\gamma\gamma\to\pi^+\pi^-$
at $M_{\pi^+\pi^-}=0.4$~GeV/$c^2$ equals approximately 170 nb and 210 nb for $\alpha_{\pi^\pm}-\beta_{\pi^\pm}$ equal to
$5.7\times 10^{-4}\,\text{fm}^3$  and $13.0\times 10^{-4}\,\text{fm}^3$, respectively.
   These calculations provide important guidance in planning the JLab experiment
that if $\alpha_{\pi^\pm}-\beta_{\pi^\pm}\approx 5.7\times 10^{-4}\,\text{fm}^3$, the cross section must be measured to about 3 nb uncertainty to achieve an accuracy of
approximately 10\%.

   Pion polarizability studies are also being carried out at the Beijing Spectrometer III (BESIII)
at the Beijing Electron-Positron Collider II (BEPC II).
   BESIII is a cylindrically symmetric detector surrounding the interaction point of the
$e^+e^-$ colliding beams.
   The experiment has collected high-statistics data samples for untagged and single-tagged
events, $\gamma\gamma^\ast\to\pi^+\pi^-$ and $\gamma\gamma^\ast\to\pi^0\pi^0$ \cite{Redmer:2018uew,Guo:2019gjf}.
   Currently, an analysis of single-tagged events is in an advanced state,
for which the virtuality of one of the photons is in the range $0.1\,\text{GeV}^2/c^2$ to $4.0\,\text{GeV}^2/c^2$.
   QED $e^+e^-$ or $\mu^+\mu^-$ backgrounds are removed via high-quality particle identification and
Monte Carlo simulations.
   Backgrounds with two pions in the final state are subtracted by fitting to the two-pion invariant mass
spectrum.

\subsection{Dispersive calculations of $\gamma\gamma\to\pi\pi$ and pion polarizabilities}
   Pion polarizabilities are determined by how the $\gamma\pi\to\gamma\pi$ Compton scattering amplitudes approach threshold.
   According to Eqs.~(\ref{LET-2}) and (\ref{real-a,b}), the polarizabilities are given in
terms of the non-Born invariant amplitudes $B_1^{\rm NB}$ and $B_2^{\rm NB}$  [see Eq.~(\ref{Bifinal})
evaluated at $\nu=q\cdot q'=q^2=q'^2=0$].
   In terms of the Mandelstam variables defined for the Compton process, Eqs.~(\ref{mandelstam}),
this corresponds to $s=u=M_\pi^2$ and $t=0$.
   By crossing symmetry, the $\gamma(q)+\pi(p_i)\to\gamma(q')+\pi(p_f)$ amplitudes also describe the $\gamma(q_1)+\gamma(q_2)\to\pi(p_1)+\pi(p_2)$ process.
   Denoting the Mandelstam variables of the two-photon-fusion process by
\begin{align}
\label{Mandelstamfusion}
\bar{s}&=(q_1+q_2)^2=(p_1+p_2)^2,\nonumber\\
\bar{t}&=(q_1-p_1)^2=(p_2-q_2)^2,\nonumber\\
\bar{u}&=(q_1-p_2)^2=(p_1-q_2)^2,
\end{align}
and identifying $q_1=q$, $q_2=-q'$, $p_1=p_f$, and $p_2=-p_i$, the Mandelstam variables of the
Compton and the two-photon-fusion processes are related by $s=\bar{u}$, $t=\bar{s}$, and $u=\bar{t}$.
   In particular, since the physical two-photon-fusion process requires $\bar{s}\geq 4M_\pi^2$, the kinematical
point, at which the polarizabilities are defined, is, with respect to the two-photon-fusion reaction,
in the unphysical (sub-threshold) region.
   The polarizability contribution to the low-energy region of the $\gamma\gamma\to\pi\pi$
cross section will compete with other mechanisms in this channel such as the important $\pi\pi$
final-state rescattering effects.

   A dispersive approach to the $\gamma\gamma\to\pi^+\pi^-$ process provides the possibility
of constructing amplitudes with analytic properties which are consistent with respect to the constraints of unitarity
and causality.
   In terms of the obtained amplitudes, the polarizabilities can either be predicted or, indirectly, be used
as parameters for fitting the available data, depending on whether one uses unsubtracted or subtracted dispersion relations,
respectively.
   Implementing a coupled-channel unitarity, the most recent DR calculation of Danilkin and Vanderhaeghen \cite{Danilkin:2018qfn}
makes use of an unsubtracted DR to predict $\alpha_{\pi^\pm}-\beta_{\pi^\pm}=6.1\times 10^{-4}\,\text{fm}^3$.
   Their results are very similar to the DR prediction of Ref.~\cite{Colangelo:2017fiz} in terms of a single-channel
approach, $\alpha_{\pi^\pm}-\beta_{\pi^\pm}=5.7\times 10^{-4}\,\text{fm}^3$, in perfect agreement with ChPT.\footnote{
Note that due to the SU(3) relations for the radiative couplings \cite{Danilkin:2018qfn,Colangelo:2017fiz},
in the charged-pion case, the inclusion of the $\rho^0$ left-hand cut is expected to produce
only a small contribution in comparison with the neutral-pion case.}
   Recently, Dai and Pennington (DP) carried out DR calculations \cite{Dai:2016ytz} for $\gamma\gamma\to\pi\pi$.
   In their formalism, the $\pi^0$ and $\pi^\pm$ polarizability values are correlated, so that knowing one allows calculating the other.
   Using the COMPASS result $\alpha_{\pi^\pm}-\beta_{\pi^\pm}= 4.0\times 10^{-4}\,\text{fm}^3$ and the Mainz result
$\alpha_{\pi^\pm}-\beta_{\pi^\pm}= 11.6\times 10^{-4}\,\text{fm}^3$ as input, DP calculate both $\gamma\gamma\to\pi^+\pi^-$ and $\gamma\gamma\to\pi^0\pi^0$ cross sections.
   They compare these with MARK II $\gamma\gamma\to\pi^+\pi^-$ data \cite{Boyer:1990vu} and DESY Crystal Ball $\gamma\gamma\to\pi^0\pi^0$ data \cite{Marsiske:1990hx}.
   With the COMPASS value, they find excellent agreement for $\gamma\gamma\to\pi^+\pi^-$ and reasonable agreement for $\gamma\gamma\to\pi^0\pi^0$.
   With the Mainz value, their DR calculations and Crystal Ball data do not agree at all.
   The differences are too large to be explained by uncertainties in the DP calculation.
   DP conclude that $\alpha_{\pi^\pm}-\beta_{\pi^\pm} = 11.6\times 10^{-4}\,\text{fm}^3$
is excluded by the Crystal Ball $\gamma\gamma\to\pi^0\pi^0$ data.

   Corrections for higher-order effects via a one-pion-loop low-energy expansion in ChPT
increased the COMPASS polarizability values by $0.6\times 10^{-4}\,\text{fm}^3$ \cite{Adolph:2014kgj}.
   Pasquini showed using subtracted DRs for the pion Compton amplitude \cite{Friedrich:2016gqb,Pasquini:2017} that
the yet higher-order energy contributions neglected in the COMPASS analysis are very small.
   DRs take into account the full energy dependence, while ChPT uses a low-energy expansion.
   In the COMPASS kinematic region of interest, the ChPT one-loop calculation and subtracted DRs agree in the mass range
up to $4M_\pi$ at the two per mille level \cite{Friedrich:2016gqb,Pasquini:2017}.
   Furthermore, the DR predictions $\alpha_{\pi^\pm}-\beta_{\pi^\pm}= 5.70\times 10^{-4}\,\text{fm}^3$ of Pasquini, Drechsel, and
Scherer \cite{Pasquini:2008ep}, using unsubtracted DRs for the $\gamma\gamma\to\pi^+\pi^-$ amplitudes, agree with the results of ChPT.

   By contrast, Fil'kov and Kashevarov (FK) claim that there is significant disagreement between the ChPT one-loop calculation and their DR calculation
\cite{Filkov:2005ccw,Filkov:2005suj,Filkov:2008vak,Filkov:2017,Filkov:2018}.
   They claim therefore that the higher-order corrections cannot be made via a one-pion-loop calculation in ChPT.
   In their DR calculation, the contribution of the $\sigma$ meson to the COMPASS pion Compton scattering cross section is very substantial.
   FK claim that the COMPASS deduced $\alpha_{\pi^\pm}-\beta_{\pi^\pm}$ is very sensitive to $\sigma$-meson contributions.
   They find via their DR calculations, taking into account the contribution of the $\sigma$-meson, that $\alpha_{\pi^\pm}-\beta_{\pi^\pm}\approx 11\times 10^{-4}\,\text{fm}^3$
 for the COMPASS experiment.

   Pasquini, Drechsel, and Scherer (PDS) claim, however, that the FK discrepancies arise due to the way that they implement the dispersion relations
\cite{Pasquini:2008ep,Pasquini:2009ze}.
   DRs may be based on specific forms for the absorptive part of the Compton amplitudes.
   The scalar $\sigma$ meson has low (Breit-Wigner) mass ($M_\sigma=(400-550)$~MeV) and a large (Breit-Wigner) width ($\Gamma_\sigma=(400-700)$~MeV) \cite{Tanabashi:2018oca}.
   In the DR calculation of FK, this resonance is modeled by an amplitude characterized by a pole, width, and coupling constant.
   They choose an analytic form (small-width approximation, and an energy-dependent coupling constant) that leads to a dispersion integral that
diverges like $1/\sqrt{t}$  for $t\to 0$.
   PDS examined the analytic properties of different analytic forms, and showed that the strong enhancement by the $\sigma$ meson, as found by FK,
is connected with spurious (unphysical) singularities of this nonanalytic function.
   PDS explain that the FK resonance model gives large (unstable) results for the pion polarizability, because it diverges at $t = 0$, precisely
where the polarizability is determined.
   That is, via DRs, the imaginary part of the Compton amplitudes serves as input
to determine the polarizabilities at the Compton threshold ($s = M_\pi^2$, $t = 0$).
   PDS found that if the basic requirements of dispersion relations are taken into account, DR results and effective field theory are consistent.
   FK disagree with these PDS conclusions, claiming that their DR results are not due to spurious singularities.

\section{Summary}

   In this article, we reviewed experimental and theoretical pion polarizability studies.
   The electric and magnetic polarizabilities $\alpha_\pi$ and $\beta_\pi$ describe the induced dipole moments of the pion during
$\gamma\pi\to\gamma\pi$ Compton scattering.
   We started out with a classical description of Thomson scattering and provided a classical interpretation
of the polarizabilities in terms of a system of two harmonically bound point particles, yielding an
order-of-magnitude estimate of $10^{-4}\,\text{fm}^3$ for the charged-pion polarizabilities.
   Within the framework of nonrelativistic quantum mechanics, we derived the differential cross section of
Compton scattering of real photons off pions in terms of their polarizabilities.
   We proceeded to a discussion of the (virtual) Compton tensor within relativistic quantum field theory
with particular emphasis on the consequences of gauge invariance, crossing symmetry, and the discrete symmetries.
   We derived the low-energy theorem, a dispersion relation for the forward-scattering amplitude, and sum rules
for the polarizabilities.
   We then discussed Compton scattering off the pion within the framework of chiral perturbation theory,
resulting in the two-loop predictions $\alpha_{\pi^\pm}-\beta_{\pi^\pm}=(5.7\pm 1.0)\times 10^{-4}\,\text{fm}^3$
and $\alpha_{\pi^\pm}+\beta_{\pi^\pm}=0.16\times 10^{-4}\,\text{fm}^3$ for the charged pion \cite{Gasser:2006qa}, and
$\alpha_{\pi^0}-\beta_{\pi^0}=(1.1\pm 0.3)\times 10^{-4}\,\text{fm}^3$
and $\alpha_{\pi^0}+\beta_{\pi^0}=-(1.9\pm 0.2)\times 10^{-4}\,\text{fm}^3$ for the neutral pion \cite{Gasser:2005ud}.
   We reviewed the determination of neutral pion polarizabilities from $\gamma\gamma\to\pi^0\pi^0$ data and of
charged-pion polarizabilities from various experiments.
   The combination $\alpha_{\pi^\pm}-\beta_{\pi^\pm}$ for charged pions was most recently measured by: (1) CERN COMPASS via radiative pion Primakoff scattering
in the nuclear Coulomb field, $\pi^-Z\to\pi^- Z\gamma$ \cite{Adolph:2014kgj},
(2) SLAC PEP Mark II via two-photon production of
pion pairs, $\gamma\gamma\to\pi^+\pi^-$, via the $e^+e^-\to e^+e^-\pi^+\pi^-$ reaction \cite{Boyer:1990vu}, and (3)
Mainz Microtron via radiative pion photoproduction from the proton,
$\gamma p\to\gamma\pi^+n$ \cite{Ahrens:2004mg}.
   We also described a planned JLab polarizability experiment via Primakoff scattering of high-energy $\gamma$'s in the nuclear Coulomb field leading to
two-photon fusion production of pion pairs, $\gamma\gamma\to\pi\pi$ \cite{Aleksejevs:2013}.
   To date, only the COMPASS polarizability measurement has acceptably small uncertainties.
   Its value $\alpha_{\pi^\pm}-\beta_{\pi^\pm} = (4.0\pm 1.8)\times 10^{-4}\,\text{fm}^3$ agrees well
with the two-loop ChPT prediction $\alpha_{\pi^\pm}-\beta_{\pi^\pm} = (5.7\pm 1.0)\times 10^{-4}\,\text{fm}^3$,
strengthening the identification of the pion with the Goldstone boson of QCD.

\appendix

\section{Low-energy effective Lagrangian and generalized Born terms}
\label{appendixeffLag}
   In the following we discuss an effective Lagrangian giving rise to
the generalized Born terms of Eq.~(\ref{Born}).
   In order to describe the electromagnetic, non-pointlike structure of,
say, the $\pi^+$, let us define
\begin{equation}
f(q^2)=\frac{1}{q^2}\left[F(q^2)-1\right],
\label{functionf}
\end{equation}
   where $F$ denotes the on-shell electromagnetic form factor.
   We introduce the noncanonical covariant derivative
\begin{equation}
D_\mu^f \pi^+=\{\partial_\mu+ieA_\mu+ie[f(-\Box)\partial^\nu F_{\mu\nu}]\}\pi^+,
\end{equation}
where $F_{\mu\nu}=\partial_\mu A_\nu-\partial_\nu A_\mu$, and the function $f$
of Eq.~(\ref{functionf}) acts on the field-strenghth tensor.\footnote{To be specific,
$f(-\Box)$ acting on a plane wave $e^{iq\cdot x}$ results in $f(-\Box)e^{iq\cdot x}
=f(q^2)e^{iq\cdot x}$.}
   Note that the noncanonical covariant derivative differs from the ordinary covariant
derivative of a point particle by the additional term proportional to the function
$f$.
   Under a gauge transformation of the second kind we have
\begin{align*}
\pi^+(x)&\mapsto\exp(-ie\chi(x))\pi^+(x),\\
A^\mu(x)&\mapsto A^\mu(x)+\partial^\mu\chi(x),\\
D_\mu^f\pi^+(x)&\mapsto\exp(-ie\chi(x))D_\mu^f\pi^+(x).
\end{align*}
   The effective Lagrangian taking the finite size of the charged
pion into account is given by
\begin{equation}
{\cal L}_{\rm eff}^{\rm Born}=(D_\mu^f \pi^+)^\dagger D^\mu_f \pi^+-M_\pi^2\pi^-\pi^+.
\end{equation}
   Organized in powers of the elementary charge, the Lagrangian  can
be expressed in terms of the canonical covariant derivative
$D_\mu\pi^+=(\partial_\mu+ieA_\mu)\pi^+$
as
\begin{equation}
{\cal L}_{\rm eff}^{\rm Born}={\cal L}_0+{\cal L}_1+{\cal L}_2,
\end{equation}
where
\begin{align*}
{\cal L}_0&=(D_\mu\pi^+)^\dagger D^\mu\pi^+-M_\pi^2\pi^-\pi^+,\\
{\cal L}_1&=ie (D_\mu\pi^+)^\dagger\pi^+ [f(-\Box)\partial_\nu F^{\mu\nu}]-ie\pi^-D_\mu\pi^+[f(-\Box)\partial_\nu F^{\mu\nu}]^\dagger,\\
{\cal L}_2&=e^2\pi^-\pi^+[f(-\Box)\partial^\nu F_{\mu\nu}][f(-\Box)\partial_\rho F^{\mu\rho}]^\dagger.
\end{align*}
   The first term is just the Lagrangian of scalar QED, wheres
the second and third terms encode the finite-size effects.
   In particular, ${\cal L}_0$ and ${\cal L}_1$ give rise to the
electromagnetic vertex
\begin{equation}
  \Gamma^\mu(p_f,p_i) = (p_f+p_i)^\mu F(q^2)+ q^\mu (p_i^2 - p_f^2) f(q^2), \quad q=p_f-p_i,
\end{equation}
which satisfies the Ward-Takahashi identity
\begin{equation}
  q_\mu \Gamma^\mu(p_f,p_i) = p_f^2-p_i^2=
    \Delta^{-1}(p_f) - \Delta^{-1}(p_i)
\end{equation}
in combination with the free propagator.
The last term, ${\cal L}_2$, vanishes when at least one of the photons is real, thus
satisfying $\partial_\nu F^{\mu\nu} = 0$.

\section{Effective Lagrangian of the non-Born terms}
\label{effLagNB}
   In \ref{appendixeffLag} we have discussed the effective Lagrangian
giving rise to the generalized Born terms.
   The non-Born terms can be thought to originate from the following effective Lagrangian,
\begin{align}
\label{L-gen}
  {\cal L}_{\rm eff}^{\rm NB}&= \frac{1}{4}\left(\hat{B}_1 F_{\mu\nu}F^{\mu\nu}
      + 2\hat{B}_4 \partial_\mu F^{\mu\nu}\partial^\rho F_{\rho\nu}\right) \pi^-\pi^+\nonumber\\
&\quad
    + \frac{1}{2} \left[\hat{B}_2 F^{\alpha\mu} F_{\beta\mu}
   + \hat{B}_5 (\partial_\mu F^{\alpha\mu})(\partial^\nu F_{\beta\nu})
   -  2\hat{B}_3 F^{\alpha\mu} (\partial_\mu\partial^\nu F_{\beta\nu})\right]
       \hat{P}_\alpha \hat{P}^\beta \pi^-\pi^+,
\end{align}
where
\begin{displaymath}
\hat P_\mu(\pi^-\pi^+)=\frac{i}{2}(\pi^-\partial_\mu\pi^+-\partial_\mu\pi^-\pi^+)-eA_\mu\pi^-\pi^+.
\end{displaymath}
   The $\hat{B}_i$ are differential operators acting on all the
fields and are determined by their Fourier components $B_i$.
   In practice, one expands the general VCS amplitude of Eq.~(\ref{ampl}) to a given order.
   The effective Lagrangian corresponding to that order is then constructed in terms of
the appropriate (covariant) derivatives.

\section{Dispersion relations}
\label{dispersion_relations}
   The starting point for deriving dispersion relations \cite{Drechsel:2002ar} is Cauchy's integral formula,
\begin{displaymath}
f(z)=\frac{1}{2\pi i}\oint_{\partial D}d\zeta\, \frac{f(\zeta)}{\zeta-z} ,
\end{displaymath}
where $D\subset {\mathbbm C}$ is a domain with boundary $\partial D$ and $f$ is holomorphic in $D$.
   In order to obtain dispersion sum rules, we consider the forward transition amplitude
$T_{\rm fw}(\nu)$ of Eq.~(\ref{Tforward}).
   Because $T_{\rm fw}(\nu)$ does not fall off sufficiently fast as $\nu\to\infty$, we
introduce the subtracted function\footnote{Since $T_{\rm fw}(\nu)$ is an even function of $\nu$, $\tilde{T}_{\rm fw}(\nu)$ is not singular at $\nu=0$.}
\begin{displaymath}
\tilde{T}_{\rm fw}(\nu)=\frac{T_{\rm fw}(\nu)-T_{\rm fw}(0)}{\nu^2},
\end{displaymath}
and consider the extension of $\tilde{T}_{\rm fw}$ to a complex variable $\nu_c=\nu_r+i\nu_i$.
   We assume that $\tilde{T}_{\rm fw}(\nu_c)$ is analytic in both the upper half plane, $\nu_i>0$, and
the lower half plane, $\nu_i<0$, and real analytic on the interval $]-\nu_0,\nu_0[$ of the real axis.
   Furthermore, we have two branch cuts along the real axis, extending from $-\infty$ to $-\nu_0$ and
from $\nu_0$ to $\infty$, respectively.
   Choosing for $\epsilon>0$ the domain and contour as shown in Fig.~{\ref{fig:complex_nu_plane}}, we write
\begin{displaymath}
\tilde{T}_{\rm fw}(\nu+i\epsilon)
=\frac{1}{2\pi i}\sum_{i=\rm I}^{\rm VIII}\int_{\gamma_i}d\zeta\,\frac{\tilde{T}_{\rm fw}(\zeta)}{\zeta-\nu-i\epsilon},
\end{displaymath}
i.e., we join eight paths to form the closed path $\gamma$.
   The integrals over $\gamma_{\rm II}$ and $\gamma_{\rm VI}$ vanish as the radius of the
semicircles approaches infinity.
   The integrals over $\gamma_{\rm IV}$ and $\gamma_{\rm VIII}$ vanish as the radius of the
semicircles approaches zero.
   We are thus left with
\begin{align}
\tilde{T}_{\rm fw}(\nu+i\epsilon)
&=\frac{1}{2\pi i}\lim_{\epsilon'\to 0^+}\left[\int_{\nu_0}^\infty d\nu'\frac{\tilde{T}_{\rm fw}(\nu'+i\epsilon')}{\nu'+i\epsilon'-\nu-i\epsilon}
+\int_{\infty}^{\nu_0}d\nu'\frac{\tilde{T}_{\rm fw}(\nu'-i\epsilon')}{\nu'-i\epsilon'-\nu-i\epsilon}\right.\nonumber\\
&\quad\left.+\int_{-\infty}^{-\nu_0}d\nu' \frac{\tilde{T}_{\rm fw}(\nu'+i\epsilon')}{\nu'+i\epsilon'-\nu-i\epsilon}
+\int_{-\nu_0}^{-\infty}d\nu' \frac{\tilde{T}_{\rm fw}(\nu'-i\epsilon')}{\nu'-i\epsilon'-\nu-i\epsilon}\right].
\label{TildeTfwint}
\end{align}
   Appealing to the Schwarz reflection principle in the form
\begin{align}
\lim_{\epsilon'\to 0^+}\left[\tilde{T}_{\rm fw}(\nu'+i\epsilon')-\tilde{T}_{\rm fw}(\nu'-i\epsilon')\right]
&=\lim_{\epsilon'\to 0^+}\left[\tilde{T}_{\rm fw}(\nu'+i\epsilon')-\tilde{T}_{\rm fw}^\ast(\nu'+i\epsilon')\right]\nonumber\\
&=2i\text{Im}\left[\tilde{T}_{\rm fw}(\nu')\right],
\label{schwarz}
\end{align}
we can, after reversing the direction of integration in the second integral, combine the first two integrals
of Eq.~(\ref{TildeTfwint}) to obtain
\begin{displaymath}
2i\int_{\nu_0}^\infty d\nu' \frac{\text{Im}\left[\tilde{T}_{\rm fw}(\nu')\right]}{\nu'-\nu-i\epsilon}.
\end{displaymath}
   Using
$$\int_{-b}^{-a}dx f(x)=\int_a^b dx f(-x)\quad\text{and}\quad \int_{-a}^{-b} dx f(x)=-\int_{a}^bdxf(-x),$$
the third and fourth integral of Eq.~(\ref{TildeTfwint}) may be written as
\begin{displaymath}
\lim_{\epsilon'\to 0^+}\left[-\int_{\nu_0}^{\infty}d\nu' \frac{\tilde{T}_{\rm fw}(-\nu'+i\epsilon')}{\nu'-i\epsilon'+\nu+i\epsilon}
+\int_{\nu_0}^{\infty}d\nu' \frac{\tilde{T}_{\rm fw}(-\nu'-i\epsilon')}{\nu'+i\epsilon'+\nu+i\epsilon}\right].
\end{displaymath}
   Making use of crossing symmetry, $\tilde{T}_{\rm fw}(-\nu'\pm i\epsilon')=\tilde{T}_{\rm fw}(\nu'\mp i\epsilon')$,
in combination with Eq.~(\ref{schwarz}), we then find
\begin{displaymath}
2i\int_{\nu_0}^\infty d\nu' \frac{\text{Im}\left[\tilde{T}_{\rm fw}(\nu')\right]}{\nu'+\nu+i\epsilon}.
\end{displaymath}
   Combining the two terms, we obtain the (subtracted) forward dispersion relation
\begin{equation}
\tilde{T}_{\rm fw}(\nu+i\epsilon)=\frac{1}{\pi}\int_{\nu_0}^\infty d\nu'\, \text{Im}\left[\tilde{T}_{\rm fw}(\nu')\right]\left(\frac{1}{\nu'-\nu-i\epsilon}
+\frac{1}{\nu'+\nu+i\epsilon}\right).
\end{equation}

\section{The chiral Lagrangian}
\label{chiralLagrangian}
  Chiral perturbation theory \cite{Weinberg:1978kz,Gasser:1983yg,Scherer:2002tk,Bijnens:2014lea}
is based on the chiral $\mbox{SU(2)}_L\times \mbox{SU(2)}_R$ symmetry
of QCD in the limit of vanishing $u$- and $d$-quark masses.
   The assumption of spontaneous symmetry breaking down to $\mbox{SU(2)}_V$
gives rise to three massless pseudoscalar Goldstone bosons with vanishing
interactions in the limit of zero energies.
   These Goldstone bosons are identif\/ied with the physical pion triplet,
the nonzero pion masses resulting from an explicit symmetry breaking
in QCD through the quark masses.
   The effective Lagrangian of the pion interaction is organized in a
so-called momentum expansion,
\begin{equation}
  {\cal L}_{\mbox{\footnotesize eff}}
={\cal L}_2+{\cal L}_4+\cdots,
\end{equation}
   where the subscripts refer to the order in the expansion.
   Interactions with external f\/ields, such as the electromagnetic
f\/ield, as well as explicit symmetry breaking due to the f\/inite
quark masses, are systematically incorporated into the effective
Lagrangian.
   Covariant derivatives and quark-mass terms count as ${\cal O}(p)$
and ${\cal O}(p^2)$, respectively.
   Weinberg's power counting scheme \cite{Weinberg:1978kz} allows for a
classif\/ication of the Feynman diagrams by establishing a relation between
the momentum expansion and the loop expansion.
    The most general chiral Lagrangian at ${\cal O}(p^2)$ is given by
\begin{equation}
\label{l2}
{\cal L}_2 = \frac{F^2}{4} \mbox{Tr} \left[ D_{\mu} U (D^{\mu}U)^{\dagger}
+\chi U^{\dagger}+ U \chi^{\dagger} \right],
\end{equation}
where $F$ denotes the pion-decay constant in the chiral limit:
$F_\pi=F[1+{\cal O}(\hat{m})]=92.2$ MeV.
   The pion fields are contained in the unimodular, unitary, $(2\times 2)$ matrix $U$,
\begin{equation}
\begin{split}
U(x)&=\textnormal{exp}\left(i\frac{\Phi(x)}{F}\right),\\
\Phi(x)&=\sum_{i=1}^3\tau_i \phi_i(x)=
\left(\begin{array}{cc}\pi^0(x) & \sqrt{2}\pi^+(x)\\ \sqrt{2}\pi^-(x)&-\pi^0(x)\end{array}\right).
\end{split}
\label{eqn:pionmatrix}
\end{equation}
   We will work in the isospin-symmetric limit $m_u=m_d=\hat{m}$.
   Furthermore, $\chi=2B(s+ip)$ includes the quark masses as
$\chi=2B\hat{m}=M^2$, where $M^2$ is the squared pion mass at leading order in the
quark-mass expansion, and $B$ is related to the scalar singlet
quark condensate $\left\langle \bar{q}q\right\rangle_0$
in the chiral limit \cite{Gasser:1983yg,Colangelo:2001sp}.
   Finally, the interaction with an external electromagnetic four-vector potential $A_\mu$ is generated
through the covariant derivative
\begin{displaymath}
D_\mu U=\partial_\mu U +i\frac{e}{2}A_\mu[\tau_3, U].
\end{displaymath}
   The most general structure of ${\cal L}_4$, f\/irst obtained by
Gasser and Leutwyler, reads,
in the standard trace notation,
\begin{align}
\label{l4gl}
{\cal L}^{GL}_4 &=
\frac{l_1}{4} \left\{\mbox{Tr}[D_{\mu}U (D^{\mu}U)^{\dagger}] \right\}^2
+\frac{l_2}{4}\mbox{Tr}[D_{\mu}U (D_{\nu}U)^{\dagger}]
\mbox{Tr}[D^{\mu}U (D^{\nu}U)^{\dagger}]\nonumber\\
&\quad
+\frac{l_3}{16}\left[\mbox{Tr}(\chi U^\dagger+ U\chi^\dagger)\right]^2
+\frac{l_4}{4}\mbox{Tr}[D_\mu U(D^\mu\chi)^\dagger
+D_\mu\chi(D^\mu U)^\dagger]\nonumber\\
&\quad
+l_5\left[\mbox{Tr}(F^R_{\mu\nu}U F^{\mu\nu}_LU^\dagger)
-\frac{1}{2}\mbox{Tr}(F_{\mu\nu}^L F^{\mu\nu}_L
+F_{\mu\nu}^R F^{\mu\nu}_R)\right]\nonumber\\
&\quad+i\frac{l_6}{2}\mbox{Tr}[ F^R_{\mu\nu} D^{\mu} U (D^{\nu} U)^{\dagger}
+ F^L_{\mu\nu} (D^{\mu} U)^{\dagger} D^{\nu} U]
-\frac{l_7}{16}\left[\mbox{Tr}(\chi U^\dagger-U\chi^\dagger)\right]^2
+\cdots,
\end{align}
   where three terms containing only external f\/ields have been omitted.
   For the electromagnetic interaction, the f\/ield-strength tensors
are given by $F^{\mu\nu}_L=F^{\mu\nu}_R=-\frac{e}{2}\tau_3
(\partial^\mu A^\nu-\partial^\nu A^\mu)$.
   Finally, we have omitted the Wess-Zumino-Witten effective
action \cite{Wess:1971yu,Witten:1983tw} describing the effects of the
anomaly, because it does not contribute to Compton scattering at
${\cal O}(p^4)$ \cite{Bijnens:1987dc,Donoghue:1988eea}.

\section{Explicit expressions in chiral perturbation theory}
\label{explicit_expressions_ChPT}
   The explicit expression for the pole terms of Eq.~(\ref{bp}) reads
\begin{align}
\label{bpappendix}
T_{\rm VCS}^{\rm pole}&=-e^2\left\{F(q'^2)F(q^2)\left[\frac{\epsilon'^\ast\cdot(2p_f+q')\epsilon\cdot(2p_i+q)}{s-M_\pi^2+i0^+}
+\frac{\epsilon\cdot(2p_f-q)\epsilon'^\ast\cdot(2p_i-q')}{u-M_\pi^2+i0^+}\right]\right.\nonumber\\
&\quad+2 F(q'^2)\frac{1-F(q^2)}{q^2}\epsilon'^\ast\cdot q\epsilon\cdot q
+2F(q^2)\frac{1-F(q'^2)}{q'^2}\epsilon\cdot q'\epsilon'^\ast\cdot q'
\nonumber\\
&\quad\left.+\epsilon'^\ast\cdot q'\epsilon\cdot q \frac{1-F(q'^2)}{q'^2}\frac{1-F(q^2)}{q^2}(q^2+q'^2-t)
+2\epsilon'^\ast\cdot\epsilon F(q'^2)F(q^2)\right\}.
\end{align}
   The first line of Eq.~(\ref{bpappendix}) corresponds to the pole contribution of the generalized
pole terms of Eq.~(\ref{Born}).

\section{Generalized polarizabilities}
\label{generalized_polarizabilities}
\begin{figure}[t]
\centerline{\includegraphics[width=0.9\textwidth]{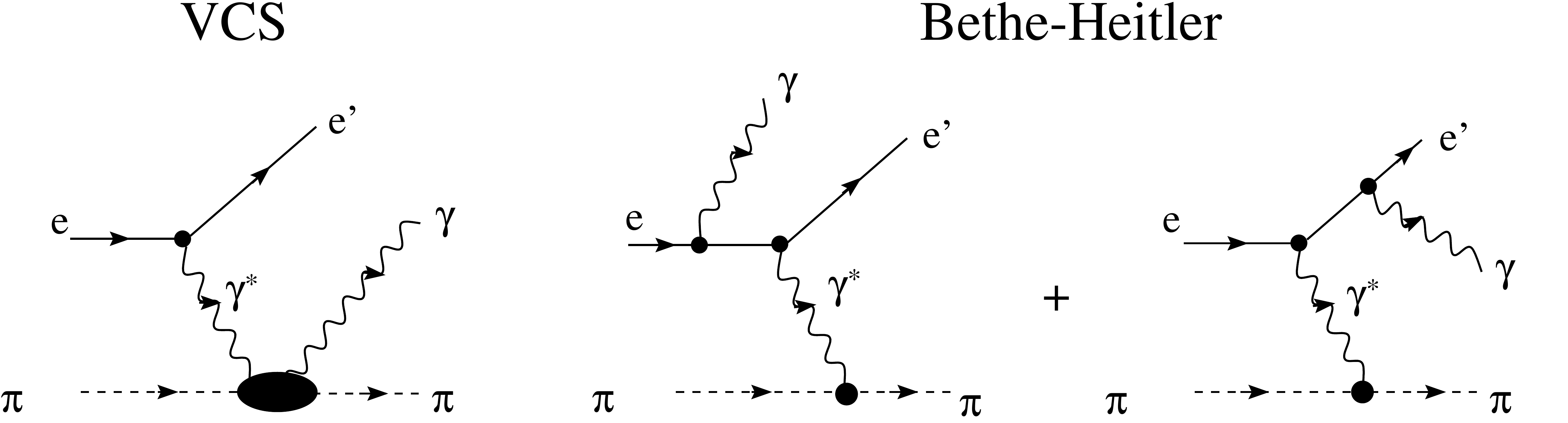}}
\caption{\label{diagrams:fig} The reaction
$\pi/K(p_i)+e^-(k)\to \pi/K(p_f) + e^-(k')+\gamma(q')$
to lowest order in the electromagnetic coupling: VCS and Bethe-Heitler
diagrams.}
\label{fig_bhvcs}
\end{figure}

    A more general case of Eq.\ (\ref{T*}), namely
$q^2<0$ and $q'^2=0$, can be studied in the electron-pion bremsstrahlung reaction
$\pi^\pm(p_i)+e^-(k)\to \pi^\pm(p_f) + e^-(k')+\gamma(q')$.
   We will refer to this situation as virtual Compton scattering (VCS)
because at least one photon is off shell.
   At lowest order in the electromagnetic coupling,
the amplitude for this reaction is given by the sum of the virtual Compton scattering (VCS)
and the Bethe-Heitler (BH) contributions,
$T = T_{\rm VCS}+ T_{\rm BH}$ (see Fig. \ref{fig_bhvcs}).
   The BH diagrams correspond to the emission of the final photon
from the electron in the initial or final states and involves only on-shell
information of the target, like the mass, the charge, and the electromagnetic form factor.
   Compared to RCS, the VCS reaction contains more information because of the independent variation of
energy and momentum transfer to the target and the additional longitudinal
polarization of the virtual photon.

   Since the invariant amplitude is given by the sum of the two contributions
$T_{\rm BH}$ and $T_{\rm VCS}$, the differential cross
section is more complex than in RCS or even in standard electron
scattering in the one-photon-exchange approximation:
\begin{displaymath}
d\sigma\sim|T_{\rm BH}+T_{\rm VCS}|^2.
\end{displaymath}
   In particular, the four-momenta exchanged by the virtual photons
in the BH and VCS diagrams read $r\equiv p_f-p_i$ and $q\equiv k-k'=
r+q'$, respectively.
   Unfortunately, the structure-dependent part of the VCS amplitude is
only a small contribution of the total amplitude.
   However, the different behavior of $T_{\rm BH}$ and
$T_{\rm VCS}$ under the substitution $\pi^-\to\pi^+$ ,
\begin{equation}
T_{\rm BH}(\pi^-)= -T_{\rm BH}(\pi^+),\quad
T_{\rm VCS}(\pi^-)= T_{\rm VCS}(\pi^+),
\end{equation}
   may be used to identify this contribution
by comparing the reactions involving a $\pi^-$ and a $\pi^+$ beam
for the same kinematics:\footnote{This argument works for any particle
which is not its own antiparticle such as the $K^+$ or $K^0$.
Of course, one could also employ the substitution $e^-\to e^+$.}
\begin{equation}
d\sigma(\pi^+)-d\sigma(\pi^-)\sim 4 \text{Re}\left[
T_{\rm BH}(\pi^+)
T^\ast_{\rm VCS}(\pi^+)\right].
\end{equation}

   We will now discuss a generalization of the RCS polarizabilities of Eq.~(\ref{real-a,b})
to the case $q^2\leq 0$ and $q'^2=0$.
   The corresponding invariant amplitude can be parametrized in terms
of three functions $B_1$, $B_2$, and $B_3$ which depend on three scalar variables
[see Eq.~(\ref{ampl}) for $q'^2=0$],
\begin{equation}
\label{ampvcs}
T_{\rm VCS}=\frac{1}{2}{\cal F}^{\mu\nu}{\cal F}'_{\mu\nu}B_1
+(P_\mu{\cal F}^{\mu\nu})(P^\rho{\cal F}'_{\rho\nu})B_2
+(P^\nu q^\mu{\cal F}_{\mu\nu})(P^\sigma q^\rho{\cal F}'_{\rho\sigma})B_3.
\end{equation}
   Equation (\ref{ampvcs}) has a particularly simple form in the pion Breit frame
(PBF) defined by $\vec P=0$, i.e., $\vec{p}_i=-\vec{p}_f$,
in which the initial and final pion are treated on a symmetrical footing.
   Introducing the Fourier components of the electric and magnetic
fields as
\begin{equation}
\label{fields}
   \vec{E} = i(q_0\vec{\epsilon} - \vec{q}\epsilon_0),
\quad
    \vec{B} = i\vec{q}\times\vec\epsilon,
\quad
   \vec{E}' = -i(q'_0\vec{\epsilon}^{\,\prime *}
        - \vec{q}^{\,\prime}\epsilon_0^{\prime *}),
\quad
    \vec{B}' = -i\vec{q}^{\,\prime}\times\vec{\epsilon}^{\,\prime *},
\end{equation}
we can rewrite Eq.~(\ref{ampvcs}) in the PBF as
\begin{equation}
\label{T-Breit}
  T_{\rm VCS} = \left[(\vec{B}\cdot \vec{B}') B_1
      - (\vec{E}\cdot \vec{E}') (B_1 + P^2 B_2)
      + (\vec{E}\cdot \vec{q}) (\vec{E}'\cdot \vec{q}) P^2 B_3\right]_{\rm PBF}.
\end{equation}
   Finally, decomposing $\vec{E}=\vec{E}_T+\vec{E}_L$ into components
which are orthogonal and parallel to $\hat{q}$, the parametrization of
the invariant amplitude in the PBF reads
\begin{eqnarray}
\label{mvcspbf2}
T_{\rm VCS}&=&\left\{
(\vec{B}\cdot\vec{B}')B_1-(\vec{E}_T\cdot\vec{E}')\left(B_1+P^2 B_2\right)
\right.\nonumber\\
&&\left.
+(\vec{E}_L\cdot\vec{E}')\left[P^2|\vec{q}|^2B_3-\left(B_1+P^2 B_2\right)\right]
\right\}_{\rm PBF}.
\end{eqnarray}
  Equation (\ref{mvcspbf2}) serves as the basis of taking the low-energy limit
$\omega'\to 0$.
   Discussing only the non-Born amplitudes,
\begin{equation}
b_i(q^2)=B_i^{\rm NB}(0,0,q^2,0),
\end{equation}
it is natural to def\/ine the following three generalized dipole
polarizabilities
\begin{align}
\label{beta}
8\pi M_\pi\beta(q^2)&\equiv b_i(q^2),\\
\label{alphat}
8\pi M_\pi
\alpha_T(q^2)&\equiv-b_1(q^2)-\left(M^2_\pi-\frac{q^2}{4}\right)b_2(q^2),\\
\label{alphal}
8\pi M_\pi\alpha_L(q^2)&\equiv-b_1(q^2)-\left(M^2_\pi-\frac{q^2}{4}
\right)\left[b_2(q^2)+q^2 b_3(q^2)\right],
\end{align}
where $P^2$ has been taken in the limit $q'=0$ as well,
i.e.,  $P^2= M^2_\pi - q^2/4$.
   In general, the transverse and longitudinal electric polarizabilities
$\alpha_T$ and $\alpha_L$ will differ by a term, vanishing however in the
RCS limit $q^2=0$.
   At $q^2=0$, the usual RCS polarizabilities are recovered,
\begin{equation}
\label{rcslimit}
\beta(0)=\beta_M,\quad
\alpha_L(0)=\alpha_T(0)=\alpha_E.
\end{equation}
   Observe that $[\vec{B}\cdot\vec{B}']_{\rm PBF}$ and $
[\vec{E}_L\cdot \vec{E}']_{\rm PBF}$ are of ${\cal O}(\omega')$
whereas $[\vec{E}_T\cdot\vec{E}']_{\rm PBF.}= {\cal O}(\omega'^2)$,
since $[\vec{E}_T]_{\rm PBF}=iq_0(\vec{\epsilon}-\vec{\epsilon}\cdot
\hat{q}\hat{q})={\cal O}(\omega')$.
   In other words, different powers of $\omega'$ have been kept.

\begin{figure}[t]
\centerline{\includegraphics[width=0.9\textwidth]{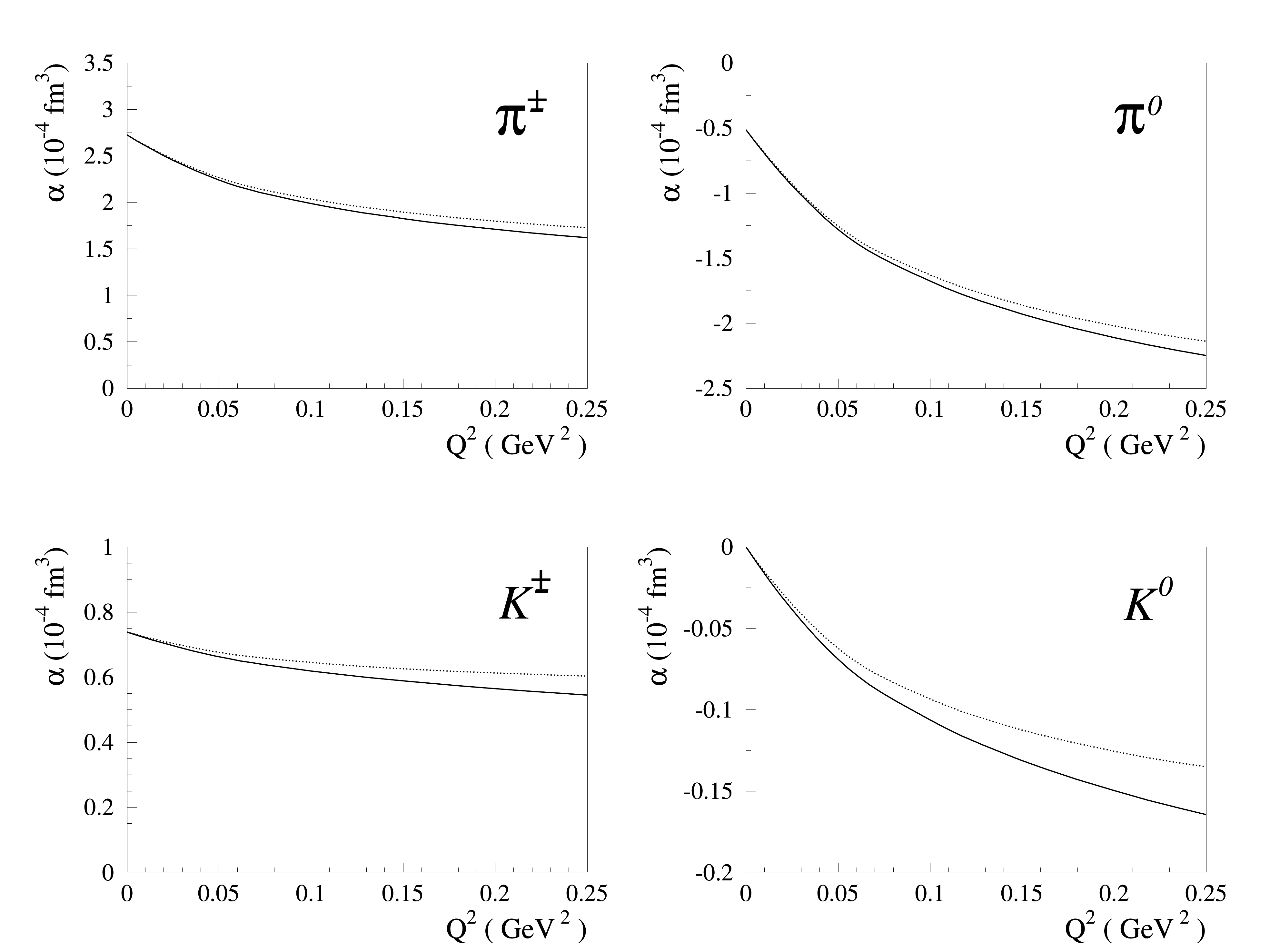}}
\caption{\label{pol} ${\cal O}(p^4)$ predictions for the generalized
pion and kaon polarizabilities $\alpha_L(q^2=-Q^2)$.
The solid lines correspond to the total results, while
the dotted lines result from pion-loop contributions, only.}
\end{figure}

   The results for the generalized polarizabilities at ${\cal O}(p^4)$ can be
found in Refs.~\cite{Lvov:2001zdg,Unkmeir:1999md,Fuchs:2000pn}.
   In particular, at ${\cal O}(p^4)$ one finds $b_2(q^2) = b_3(q^2)=0$,
which leads to the relation
  \begin{equation}
    \alpha_L(q^2) = \alpha_T(q^2) = -\beta(q^2).
  \end{equation}
   The results for the generalized polarizabilities as function of $Q^2=-q^2$
are shown in Fig. \ref{pol}.\cite{Fuchs:2000pn}

   A first measurement of pion VCS was reported by the Fermilab E781 SELEX experiment \cite{Ocherashvili:2001aj}.
   SELEX used a 600 GeV/$c$ $\pi^-$ beam incident on target atomic electrons, detecting the incident $\pi^-$ and the final-state $\pi^-$, electron, and
$\gamma$.
   Theoretical predictions based on ChPT \cite{Unkmeir:2001gw} were incorporated into a Monte Carlo simulation of the experiment,
and agreed reasonably well with the data.
   The two-photon physics program at BESIII \cite{Redmer:2018uew,Guo:2019gjf} involves, among other processes,
the study of the reaction $\gamma\gamma^\ast\to\pi^+\pi^-$ at spacelike momentum
transfers $0.2\,\text{GeV}^2/c^2<Q^2<2.0\,\text{GeV}^2/c^2$, invariant masses between
$2M_\pi<M_{\pi\pi}<2.0\,\text{GeV}/c^2$, at a full coverage of the pion helicity angle
$\cos(\theta^\ast)$, with a similar analysis for the $\pi^0\pi^0$ and $\pi^0\eta$ final
states in preparation.
   It will be interesting to see, to what extent information on the generalized
polarizabilities can be extracted from the forthcoming data.

\section*{Acknowledgments}

Thanks are due to I.~Danilkin, B.~Moussallam, and B.~Pasquini for helpful suggestions, to
D.~Lawrence for information concerning the planned JLab experiment, and to A.~Denig, Y.~Guo, and C.~F.~Redmer
for discussions on the two-photon physics program at BESIII.
S.~S.~was supported by the Deutsche Forschungsgemeinschaft (SFB 1044).


\begin{thebibliography}{000} %for 3 digits

%\cite{Tanabashi:2018oca}
\bibitem{Tanabashi:2018oca}
  M.~Tanabashi {\it et al.} (Particle Data Group),
  %``Review of Particle Physics,''
  Phys.\ Rev.\ D {\bf 98}, 030001 (2018).
%  doi:10.1103/PhysRevD.98.030001
  %%CITATION = doi:10.1103/PhysRevD.98.030001;%%

%\cite{Klein:1955zz}
\bibitem{Klein:1955zz}
  A.~Klein,
  %``Low-Energy Theorems for Renormalizable Field Theories,''
  Phys.\ Rev.\ {\bf 99}, 998 (1955).
%  doi:10.1103/PhysRev.99.998
  %%CITATION = doi:10.1103/PhysRev.99.998;%%

%\cite{Holstein:1990qy}
\bibitem{Holstein:1990qy}
  B.~R.~Holstein,
  %``Pion polarizability and chiral symmetry,''
  Comments Nucl.\ Part.\ Phys.\  {\bf 19}, 221 (1990).
  %%CITATION = CNPPA,19,221;%%

%\cite{Lvov:1993fp}
\bibitem{Lvov:1993fp}
  A.~I.~L'vov,
  %``Theoretical aspects of the polarizability of the nucleon,''
  Int.\ J.\ Mod.\ Phys.\ A {\bf 8}, 5267 (1993).
%  doi:10.1142/S0217751X93002095
  %%CITATION = doi:10.1142/S0217751X93002095;%%

%\cite{Moinester:1997um}
\bibitem{Moinester:1997um}
  M.~A.~Moinester and V.~Steiner,
  %``Pion and kaon polarizabilities and radiative transitions,''
  Lect.\ Notes Phys.\  {\bf 513}, 247 (1998).
%  doi:10.1007/BFb0104910
%  [hep-ex/9801008].
  %%CITATION = doi:10.1007/BFb0104910;%%

%\cite{Scherer:1999yw}
\bibitem{Scherer:1999yw}
  S.~Scherer,
  %``Real and virtual Compton scattering at low-energies,''
  Czech.\ J.\ Phys.\  {\bf 49}, 1307 (1999).
%  doi:10.1023/A:1022872211333
%  [nucl-th/9901056].
  %%CITATION = doi:10.1023/A:1022872211333;%%

%\cite{Holstein:2013kia}
\bibitem{Holstein:2013kia}
  B.~R.~Holstein and S.~Scherer,
  %``Hadron Polarizabilities,''
  Ann.\ Rev.\ Nucl.\ Part.\ Sci.\  {\bf 64}, 51 (2014).
%  doi:10.1146/annurev-nucl-102313-025555
%  [arXiv:1401.0140 [hep-ph]].
  %%CITATION = doi:10.1146/annurev-nucl-102313-025555;%%

%\cite{Gasser:2006qa}
\bibitem{Gasser:2006qa}
  J.~Gasser, M.~A.~Ivanov, and M.~E.~Sainio,
  %``Revisiting gamma gamma ---> pi+ pi- at low energies,''
  Nucl.\ Phys.\ B {\bf 745}, 84 (2006).
%  doi:10.1016/j.nuclphysb.2006.03.022
%  [hep-ph/0602234].
  %%CITATION = doi:10.1016/j.nuclphysb.2006.03.022;%%

%\cite{Adolph:2014kgj}
\bibitem{Adolph:2014kgj}
  C.~Adolph {\it et al.} (COMPASS Collaboration),
  %``Measurement of the charged-pion polarizability,''
  Phys.\ Rev.\ Lett.\  {\bf 114}, 062002 (2015).
%  doi:10.1103/PhysRevLett.114.062002
%  [arXiv:1405.6377 [hep-ex]].
  %%CITATION = doi:10.1103/PhysRevLett.114.062002;%%

%\cite{Boyer:1990vu}
\bibitem{Boyer:1990vu}
  J.~Boyer {\it et al.},
  %``Two photon production of pion pairs,''
  Phys.\ Rev.\ D {\bf 42}, 1350 (1990).
%  doi:10.1103/PhysRevD.42.1350
  %%CITATION = doi:10.1103/PhysRevD.42.1350;%%

%\cite{Ahrens:2004mg}
\bibitem{Ahrens:2004mg}
  J.~Ahrens {\it et al.},
  %``Measurement of the pi+ meson polarizabilities via the gamma p ---> gamma pi+ n reaction,''
  Eur.\ Phys.\ J.\ A {\bf 23}, 113 (2005).
%  doi:10.1140/epja/i2004-10056-2
%  [nucl-ex/0407011].
  %%CITATION = doi:10.1140/epja/i2004-10056-2;%%

\bibitem{Aleksejevs:2013}
A.~Aleksejevs {\it et al.}, Jefferson Lab E12-13-008 (2013), https://www.jlab.org/exp\_prog/proposals/13/PR12-13-008.pdf;
and  David Lawrence (private communication).

\bibitem{Landau} L.~D.~Landau and E.~M.~Lifshitz, {\it The Classical Theory of
Fields}, Vol.~2, 4th ed. (Elsevier Butterworth Heinemann, Amsterdam, 1995).

%\cite{Thirring:1950cy}
\bibitem{Thirring:1950cy}
  W.~E.~Thirring,
  %``Radiative corrections in the nonrelativistic limit,''
  Phil.\ Mag.\ Ser.\ 7 {\bf 41}, 1193 (1950).

%\cite{Low:1954kd}
\bibitem{Low:1954kd}
  F.~E.~Low,
  %``Scattering of light of very low frequency by systems of spin 1/2,''
  Phys.\ Rev.\  {\bf 96}, 1428 (1954).
%  doi:10.1103/PhysRev.96.1428
  %%CITATION = doi:10.1103/PhysRev.96.1428;%%

%\cite{GellMann:1954kc}
\bibitem{GellMann:1954kc}
  M.~Gell-Mann and M.~L.~Goldberger,
  %``Scattering of low-energy photons by particles of spin 1/2,''
  Phys.\ Rev.\  {\bf 96}, 1433 (1954).
%  doi:10.1103/PhysRev.96.1433
  %%CITATION = doi:10.1103/PhysRev.96.1433;%%

\bibitem{bd} S.~Brandt and H.~D.~Dahmen, {\it Elektrodynamik}, 4th ed.
(Springer, Berlin, 2005).

\bibitem{fnm}
  R.~P.~Feynman, R.~Leighton, M.~Sands, {\it The Feynman Lectures on Physics}, Vol.~2
  (Addison-Wesley, Reading, MA, 1964) Sec.~34-4.

\bibitem{fnm2}
  R.~P.~Feynman, R.~Leighton, M.~Sands, {\it The Feynman Lectures on Physics}, Vol.~1
  (Addison-Wesley, Reading, MA, 1963) Sec.~32-5.

%\cite{Mandl:1985bg}
\bibitem{Mandl:1985bg}
  F.~Mandl and G.~Shaw, {\it Quantum Field Theory}, 2nd ed.\ (Wiley, Chichester, UK, 2010).

\bibitem{Petrunkin}
V.~A.~Petrunkin, Nucl.~Phys.~{\bf 55}, 197 (1964).

%\cite{Pasquini:2000ue}
\bibitem{Pasquini:2000ue}
  B.~Pasquini, S.~Scherer, and D.~Drechsel,
  %``Generalized polarizabilities of the proton in a constituent quark model revisited,''
  Phys.\ Rev.\ C {\bf 63}, 025205 (2001).
%  doi:10.1103/PhysRevC.63.025205
%  [nucl-th/0008046].
  %%CITATION = doi:10.1103/PhysRevC.63.025205;%%

%\cite{Ericson:1973dtc}
\bibitem{Ericson:1973dtc}
  T.~E.~O.~Ericson and J.~H\"ufner,
  %``Low-frequency photon scattering by nuclei,''
  Nucl.\ Phys.\ B {\bf 57}, 604 (1973).
%  doi:10.1016/0550-3213(73)90120-X
  %%CITATION = doi:10.1016/0550-3213(73)90120-X;%%

\bibitem{Dirac}
P.~A.~M.~Dirac, {\it The Principles of Quantum Mechanics}, 4th ed.\
(Oxford University Press, Oxford, 1989).

\bibitem{Cohen-Tannoudji}
C.~Cohen-Tannoudji, B.~Diu, F.~Lalo$\ddot{\rm e}$, {\it Quantum Mechanics}, Vol.~1
(Wiley, New York, 1977).

%\cite{Rudy:1994qb}
\bibitem{Rudy:1994qb}
  T.~E.~Rudy, H.~W.~Fearing, and S.~Scherer,
  %``The Off-shell electromagnetic form-factors of pions and kaons in chiral perturbation theory,''
  Phys.\ Rev.\ C {\bf 50}, 447 (1994).
%  doi:10.1103/PhysRevC.50.447
%  [hep-ph/9401302].
  %%CITATION = doi:10.1103/PhysRevC.50.447;%%

%\cite{Ward:1950xp}
\bibitem{Ward:1950xp}
  J.~C.~Ward,
  %``An Identity in Quantum Electrodynamics,''
  Phys.\ Rev.\  {\bf 78}, 182 (1950).
%  doi:10.1103/PhysRev.78.182
  %%CITATION = doi:10.1103/PhysRev.78.182;%%

%\cite{Takahashi:1957xn}
\bibitem{Takahashi:1957xn}
  Y.~Takahashi,
  %``On the generalized Ward identity,''
  Nuovo Cim.\  {\bf 6}, 371 (1957).
%  doi:10.1007/BF02832514
  %%CITATION = doi:10.1007/BF02832514;%%

%\cite{Chisholm:1961tha}
\bibitem{Chisholm:1961tha}
  J.~S.~R.~Chisholm,
  %``Change of variables in quantum field theories,''
  Nucl.\ Phys.\  {\bf 26}, no. 3, 469 (1961).
%  doi:10.1016/0029-5582(61)90106-7
  %%CITATION = doi:10.1016/0029-5582(61)90106-7;%%

%\cite{Kamefuchi:1961sb}
\bibitem{Kamefuchi:1961sb}
  S.~Kamefuchi, L.~O'Raifeartaigh, and A.~Salam,
  %``Change of variables and equivalence theorems in quantum field theories,''
  Nucl.\ Phys.\  {\bf 28}, 529 (1961).
  %%CITATION = doi:10.1016/0029-5582(61)90056-6, 10.1016/0029-5582(61)91075-6;%%

%\cite{Scherer:1994aq}
\bibitem{Scherer:1994aq}
  S.~Scherer and H.~W.~Fearing,
  %``Compton scattering by a pion and off-shell effects,''
  Phys.\ Rev.\ C {\bf 51}, 359 (1995).
%  doi:10.1103/PhysRevC.51.359
%  [hep-ph/9408312].
  %%CITATION = doi:10.1103/PhysRevC.51.359;%%

%\cite{Tarrach:1975tu}
\bibitem{Tarrach:1975tu}
  R.~Tarrach,
  %``Invariant Amplitudes for Virtual Compton Scattering Off Polarized Nucleons Free from Kinematical Singularities, Zeros and Constraints,''
  Nuovo Cim.\ A {\bf 28}, 409 (1975).
%  doi:10.1007/BF02894857
  %%CITATION = doi:10.1007/BF02894857;%%

%\cite{Bardeen:1969aw}
\bibitem{Bardeen:1969aw}
  W.~A.~Bardeen and W.~K.~Tung,
  %``Invariant amplitudes for photon processes,''
  Phys.\ Rev.\  {\bf 173}, 1423 (1968).
  Erratum: [Phys.\ Rev.\ D {\bf 4}, 3229 (1971)].
%  doi:10.1103/physrevd.4.3229.2, 10.1103/PhysRev.173.1423
  %%CITATION = doi:10.1103/physrevd.4.3229.2, 10.1103/PhysRev.173.1423;%%

%\cite{Lvov:2001zdg}
\bibitem{Lvov:2001zdg}
  A.~I.~L'vov, S.~Scherer, B.~Pasquini, C.~Unkmeir, and D.~Drechsel,
  %``Generalized dipole polarizabilities and the spatial structure of hadrons,''
  Phys.\ Rev.\ C {\bf 64}, 015203 (2001).
%  doi:10.1103/PhysRevC.64.015203
%  [hep-ph/0103172].
  %%CITATION = doi:10.1103/PhysRevC.64.015203;%%

%\cite{Itzykson:1980rh}
\bibitem{Itzykson:1980rh}
  C.~Itzykson and J.~B.~Zuber, {\it Quantum Field Theory} (McGraw-Hill, New York, 1980).

%\cite{Low:1958sn}
\bibitem{Low:1958sn}
  F.~E.~Low,
  %``Bremsstrahlung of very low-energy quanta in elementary particle collisions,''
  Phys.\ Rev.\  {\bf 110}, 974 (1958).
%  doi:10.1103/PhysRev.110.974
  %%CITATION = doi:10.1103/PhysRev.110.974;%%

%\cite{Fearing:1996gs}
\bibitem{Fearing:1996gs}
  H.~W.~Fearing and S.~Scherer,
  %``Virtual Compton scattering off spin zero particles at low-energies,''
  Few Body Syst.\  {\bf 23}, 111 (1998).
%  doi:10.1007/s006010050067
%  [nucl-th/9607056].
  %%CITATION = doi:10.1007/s006010050067;%%

%\cite{Drechsel:2002ar}
\bibitem{Drechsel:2002ar}
  D.~Drechsel, B.~Pasquini, and M.~Vanderhaeghen,
  %``Dispersion relations in real and virtual Compton scattering,''
  Phys.\ Rep.\  {\bf 378}, 99 (2003).
%  doi:10.1016/S0370-1573(02)00636-1
%  [hep-ph/0212124].
  %%CITATION = doi:10.1016/S0370-1573(02)00636-1;%%

%\cite{Pasquini:2018wbl}
\bibitem{Pasquini:2018wbl}
  B.~Pasquini and M.~Vanderhaeghen,
  %``Dispersion Theory in Electromagnetic Interactions,''
  Ann.\ Rev.\ Nucl.\ Part.\ Sci.\  {\bf 68}, 75 (2018).
%  doi:10.1146/annurev-nucl-101917-020843
%  [arXiv:1805.10482 [hep-ph]].
  %%CITATION = doi:10.1146/annurev-nucl-101917-020843;%%

%\cite{Filkov:1982cx}
\bibitem{Filkov:1982cx}
  L.~V.~Filkov, I.~Guiasu, and E.~E.~Radescu,
  %``Pion Polarizabilities From Backward and Fixed $u$ Sum Rules,''
  Phys.\ Rev.\ D {\bf 26}, 3146 (1982).
%  doi:10.1103/PhysRevD.26.3146
  %%CITATION = doi:10.1103/PhysRevD.26.3146;%%

\bibitem{Baldin}
A.~M.~Baldin, Nucl.\ Phys.\ {\bf 18}, 310 (1960).

\bibitem{Petrunkin2}
V.~A.~Petrun'kin, Sov.\ J.\ Part.\ Nucl. {\bf 12}, 278 (1981).

%\cite{Babusci:1998ww}
\bibitem{Babusci:1998ww}
  D.~Babusci, G.~Giordano, A.~I.~L'vov, G.~Matone, and A.~M.~Nathan,
  %``Low-energy Compton scattering of polarized photons on polarized nucleons,''
  Phys.\ Rev.\ C {\bf 58}, 1013 (1998).
%  doi:10.1103/PhysRevC.58.1013
%  [hep-ph/9803347].
  %%CITATION = doi:10.1103/PhysRevC.58.1013;%%

%\cite{Bernabeu:1974zu}
\bibitem{Bernabeu:1974zu}
  J.~Bernabeu, T.~E.~O.~Ericson, and C.~Ferro Fontan,
  %``The Nucleon Electromagnetic Polarizabilities,''
  Phys.\ Lett.\  {\bf 49B}, 381 (1974).
%  doi:10.1016/0370-2693(74)90186-5
  %%CITATION = doi:10.1016/0370-2693(74)90186-5;%%

%\cite{Bernabeu:1977hp}
\bibitem{Bernabeu:1977hp}
  J.~Bernabeu and R.~Tarrach,
  %``The Electromagnetic Polarizabilities of the Proton and the Scalar-Isoscalar gamma gamma --> pi pi Amplitude,''
  Phys.\ Lett.\  {\bf 69B}, 484 (1977).
%  doi:10.1016/0370-2693(77)90851-6
  %%CITATION = doi:10.1016/0370-2693(77)90851-6;%%

%\cite{Lvov:1979zd}
\bibitem{Lvov:1979zd}
  A.~I.~L'vov, V.~A.~Petrunkin, and S.~A.~Startsev,
  %``Finite Energy Sum Rule For Nucleon Polarizabilities Difference. (in Russian),''
  Yad.\ Fiz.\ {\bf 29}, 1265 (1979); Sov.\ J.\ Nucl.\ Phys.\ {\bf 29}, 651 (1979).
  %%CITATION = YAFIA,29,1265;%%

%\cite{Donoghue:1993kw}
\bibitem{Donoghue:1993kw}
  J.~F.~Donoghue and B.~R.~Holstein,
  %``Photon-photon scattering, pion polarizability and chiral symmetry,''
  Phys.\ Rev.\ D {\bf 48}, 137 (1993).
%  doi:10.1103/PhysRevD.48.137
%  [hep-ph/9302203].
  %%CITATION = doi:10.1103/PhysRevD.48.137;%%

%\cite{Filkov:1998rwz}
\bibitem{Filkov:1998rwz}
  L.~V.~Fil'kov and V.~L.~Kashevarov,
  %``Compton scattering on the charged pion and the process gamma gamma ---> pi0 pi0,''
  Eur.\ Phys.\ J.\ A {\bf 5}, 285 (1999).
%  doi:10.1007/s100500050287
%  [nucl-th/9810074].
  %%CITATION = doi:10.1007/s100500050287;%%

%\cite{Filkov:2005ccw}
\bibitem{Filkov:2005ccw}
  L.~V.~Fil'kov and V.~L.~Kashevarov,
  %``Determination of pi0 meson quadrupole polarizabilities from the process gamma gamma ---> pi0 pi0,''
  Phys.\ Rev.\ C {\bf 72}, 035211 (2005).
%  doi:10.1103/PhysRevC.72.035211
%  [nucl-th/0505058].
  %%CITATION = doi:10.1103/PhysRevC.72.035211;%%

%\cite{Filkov:2005suj}
\bibitem{Filkov:2005suj}
  L.~V.~Fil'kov and V.~L.~Kashevarov,
  %``Determination of pi+- meson polarizabilities from the gamma gamma ---> pi+ pi- process,''
  Phys.\ Rev.\ C {\bf 73}, 035210 (2006).
%  doi:10.1103/PhysRevC.73.035210
%  [nucl-th/0512047].
  %%CITATION = doi:10.1103/PhysRevC.73.035210;%%

%\cite{Pasquini:2008ep}
\bibitem{Pasquini:2008ep}
  B.~Pasquini, D.~Drechsel, and S.~Scherer,
  %``The Polarizability of the pion: No conflict between dispersion theory and chiral perturbation theory,''
  Phys.\ Rev.\ C {\bf 77}, 065211 (2008).
%  doi:10.1103/PhysRevC.77.065211
%  [arXiv:0805.0213 [hep-ph]].
  %%CITATION = doi:10.1103/PhysRevC.77.065211;%%

%\cite{GarciaMartin:2010cw}
\bibitem{GarciaMartin:2010cw}
  R.~Garc\'{\i}a-Mart\'{\i}n and B.~Moussallam,
  %``MO analysis of the high statistics Belle results on $\gamma\gamma\to \pi^+\pi^-,\pi^0\pi^0$ with chiral constraints,''
  Eur.\ Phys.\ J.\ C {\bf 70}, 155 (2010).
%  doi:10.1140/epjc/s10052-010-1471-7
%  [arXiv:1006.5373 [hep-ph]].
  %%CITATION = doi:10.1140/epjc/s10052-010-1471-7;%%

%\cite{Hoferichter:2011wk}
\bibitem{Hoferichter:2011wk}
  M.~Hoferichter, D.~R.~Phillips, and C.~Schat,
  %``Roy-Steiner equations for gamma gamma -> pi pi,''
  Eur.\ Phys.\ J.\ C {\bf 71}, 1743 (2011).
%  doi:10.1140/epjc/s10052-011-1743-x
%  [arXiv:1106.4147 [hep-ph]].
  %%CITATION = doi:10.1140/epjc/s10052-011-1743-x;%%

%\cite{Moussallam:2013una}
\bibitem{Moussallam:2013una}
  B.~Moussallam,
  %``Unified dispersive approach to real and virtual photon-photon scattering at low energy,''
  Eur.\ Phys.\ J.\ C {\bf 73}, 2539 (2013).
%  doi:10.1140/epjc/s10052-013-2539-y
%  [arXiv:1305.3143 [hep-ph]].
  %%CITATION = doi:10.1140/epjc/s10052-013-2539-y;%%

%\cite{Dai:2016ytz}
\bibitem{Dai:2016ytz}
  L.~Y.~Dai and M.~R.~Pennington,
  %``Pion polarizabilities from $\gamma\gamma\to\pi\pi$ analysis,''
  Phys.\ Rev.\ D {\bf 94}, no. 11, 116021 (2016).
%  doi:10.1103/PhysRevD.94.116021
%  [arXiv:1611.04441 [hep-ph]].
  %%CITATION = doi:10.1103/PhysRevD.94.116021;%%

%\cite{Danilkin:2018qfn}
\bibitem{Danilkin:2018qfn}
  I.~Danilkin and M.~Vanderhaeghen,
  %``Dispersive analysis of the $\gamma\gamma^{*} \to \pi \pi$ process,''
  Phys.\ Lett.\ B {\bf 789}, 366 (2019).
%  doi:10.1016/j.physletb.2018.12.047
%  [arXiv:1810.03669 [hep-ph]].
  %%CITATION = doi:10.1016/j.physletb.2018.12.047;%%

%\cite{Weinberg:1978kz}
\bibitem{Weinberg:1978kz}
  S.~Weinberg,
  %``Phenomenological Lagrangians,''
  Physica A {\bf 96}, 327 (1979).
%  doi:10.1016/0378-4371(79)90223-1
  %%CITATION = doi:10.1016/0378-4371(79)90223-1;%%

%\cite{Gasser:1983yg}
\bibitem{Gasser:1983yg}
  J.~Gasser and H.~Leutwyler,
  %``Chiral Perturbation Theory to One Loop,''
  Annals Phys.\  {\bf 158}, 142 (1984).
%  doi:10.1016/0003-4916(84)90242-2
  %%CITATION = doi:10.1016/0003-4916(84)90242-2;%%

%\cite{Scherer:2002tk}
\bibitem{Scherer:2002tk}
  S.~Scherer,
  %``Introduction to chiral perturbation theory,''
  Adv.\ Nucl.\ Phys.\  {\bf 27}, 277 (2003).
%  [hep-ph/0210398].
  %%CITATION = HEP-PH/0210398;%%

%\cite{Bijnens:2014lea}
\bibitem{Bijnens:2014lea}
  J.~Bijnens and G.~Ecker,
  %``Mesonic low-energy constants,''
  Ann.\ Rev.\ Nucl.\ Part.\ Sci.\  {\bf 64}, 149 (2014).
%  doi:10.1146/annurev-nucl-102313-025528
%  [arXiv:1405.6488 [hep-ph]].
  %%CITATION = doi:10.1146/annurev-nucl-102313-025528;%%

%\cite{Terentev:1972ix}
\bibitem{Terentev:1972ix}
  M.~V.~Terentev,
  %``Pion polarizability, virtual compton-effect and pi ---> e nu gamma decay,''
  Sov.\ J.\ Nucl.\ Phys.\  {\bf 16}, 87 (1973)
  [Yad.\ Fiz.\  {\bf 16}, 162 (1972)].
  %%CITATION = SJNCA,16,87;%%

%\cite{Bijnens:1987dc}
\bibitem{Bijnens:1987dc}
  J.~Bijnens and F.~Cornet,
  %``Two Pion Production in Photon-Photon Collisions,''
  Nucl.\ Phys.\ B {\bf 296}, 557 (1988).
%  doi:10.1016/0550-3213(88)90032-6
  %%CITATION = doi:10.1016/0550-3213(88)90032-6;%%

%\cite{Unkmeir:1999md}
\bibitem{Unkmeir:1999md}
  C.~Unkmeir, S.~Scherer, A.~I.~L'vov, and D.~Drechsel,
  %``Generalized polarizabilities of the pion in chiral perturbation theory,''
  Phys.\ Rev.\ D {\bf 61}, 034002 (2000).
%  doi:10.1103/PhysRevD.61.034002
%  [hep-ph/9904442].
  %%CITATION = doi:10.1103/PhysRevD.61.034002;%%

%\cite{Weinberg:1966kf}
\bibitem{Weinberg:1966kf}
  S.~Weinberg,
  %``Pion scattering lengths,''
  Phys.\ Rev.\ Lett.\  {\bf 17}, 616 (1966).
%  doi:10.1103/PhysRevLett.17.616
  %%CITATION = doi:10.1103/PhysRevLett.17.616;%%

%\cite{Bychkov:2008ws}
\bibitem{Bychkov:2008ws}
  M.~Bychkov {\it et al.},
  %``New Precise Measurement of the Pion Weak Form Factors in pi+ ---> e+ nu gamma Decay,''
  Phys.\ Rev.\ Lett.\  {\bf 103}, 051802 (2009).
%  doi:10.1103/PhysRevLett.103.051802
%  [arXiv:0804.1815 [hep-ex]].
  %%CITATION = doi:10.1103/PhysRevLett.103.051802;%%

%\cite{Burgi:1996mm}
\bibitem{Burgi:1996mm}
  U.~B\"urgi,
  %``Charged pion polarizabilities to two loops,''
  Phys.\ Lett.\ B {\bf 377}, 147 (1996).
%  doi:10.1016/0370-2693(96)00304-8
%  [hep-ph/9602421].
  %%CITATION = doi:10.1016/0370-2693(96)00304-8;%%

%\cite{Burgi:1996qi}
\bibitem{Burgi:1996qi}
  U.~B\"urgi,
  %``Charged pion pair production and pion polarizabilities to two loops,''
  Nucl.\ Phys.\ B {\bf 479}, 392 (1996).
%  doi:10.1016/0550-3213(96)00454-3
%  [hep-ph/9602429].
  %%CITATION = doi:10.1016/0550-3213(96)00454-3;%%

%\cite{Bijnens:1995yn}
\bibitem{Bijnens:1995yn}
  J.~Bijnens, G.~Colangelo, G.~Ecker, J.~Gasser, and M.~E.~Sainio,
  %``Elastic pi pi scattering to two loops,''
  Phys.\ Lett.\ B {\bf 374}, 210 (1996).
%  doi:10.1016/0370-2693(96)00165-7
%  [hep-ph/9511397].
  %%CITATION = doi:10.1016/0370-2693(96)00165-7;%%

%\cite{Kaloshin:1993wj}
\bibitem{Kaloshin:1993wj}
  A.~E.~Kaloshin and V.~V.~Serebryakov,
  %``pi+ and pi0 polarizabilities from gamma gamma ---> pi pi data on the base of S matrix approach,''
  Z.\ Phys.\ C {\bf 64}, 689 (1994).
%  doi:10.1007/BF01957778
%  [hep-ph/9306224].

%\cite{Donoghue:1988eea}
\bibitem{Donoghue:1988eea}
  J.~F.~Donoghue, B.~R.~Holstein, and Y.~C.~Lin,
  %``The Reaction gamma Gamma ---> pi0 pi0 and Chiral Loops,''
  Phys.\ Rev.\ D {\bf 37}, 2423 (1988).
%  doi:10.1103/PhysRevD.37.2423
  %%CITATION = doi:10.1103/PhysRevD.37.2423;%%

%\cite{Bellucci:1994eb}
\bibitem{Bellucci:1994eb}
  S.~Bellucci, J.~Gasser, and M.~E.~Sainio,
  %``Low-energy photon-photon collisions to two loop order,''
  Nucl.\ Phys.\ B {\bf 423}, 80 (1994)
  Erratum: [Nucl.\ Phys.\ B {\bf 431}, 413 (1994)].
%  doi:10.1016/0550-3213(94)90566-5, 10.1016/0550-3213(94)90111-2
%  [hep-ph/9401206].
  %%CITATION = doi:10.1016/0550-3213(94)90566-5, 10.1016/0550-3213(94)90111-2;%%

%\cite{Gasser:2005ud}
\bibitem{Gasser:2005ud}
  J.~Gasser, M.~A.~Ivanov, and M.~E.~Sainio,
  %``Low-energy photon-photon collisions to two loops revisited,''
  Nucl.\ Phys.\ B {\bf 728}, 31 (2005).
%  doi:10.1016/j.nuclphysb.2005.09.010
%  [hep-ph/0506265].
  %%CITATION = doi:10.1016/j.nuclphysb.2005.09.010;%%

%\cite{Babusci:1991sk}
\bibitem{Babusci:1991sk}
  D.~Babusci, S.~Bellucci, G.~Giordano, G.~Matone, A.~M.~Sandorfi, and M.~A.~Moinester,
  %``Chiral symmetry and pion polarizabilities,''
  Phys.\ Lett.\ B {\bf 277}, 158 (1992).
%  doi:10.1016/0370-2693(92)90973-8
  %%CITATION = doi:10.1016/0370-2693(92)90973-8;%%

%\cite{Marsiske:1990hx}
\bibitem{Marsiske:1990hx}
  H.~Marsiske {\it et al.} [Crystal Ball Collaboration],
  %``A Measurement of $\pi^0 \pi^0$ Production in Two Photon Collisions,''
  Phys.\ Rev.\ D {\bf 41}, 3324 (1990).
%  doi:10.1103/PhysRevD.41.3324
  %%CITATION = doi:10.1103/PhysRevD.41.3324;%%

%\cite{Uehara:2008ep}
\bibitem{Uehara:2008ep}
  S.~Uehara {\it et al.} (Belle Collaboration9,
  %``High-statistics measurement of neutral pion-pair production in two-photon collisions,''
  Phys.\ Rev.\ D {\bf 78}, 052004 (2008).
%  doi:10.1103/PhysRevD.78.052004
%  [arXiv:0805.3387 [hep-ex]].
  %%CITATION = doi:10.1103/PhysRevD.78.052004;%%

%\cite{Uehara:2009cka}
\bibitem{Uehara:2009cka}
  S.~Uehara {\it et al.} (Belle Collaboration),
  %``High-statistics study of neutral-pion pair production in two-photon collisions,''
  Phys.\ Rev.\ D {\bf 79}, 052009 (2009).
%  doi:10.1103/PhysRevD.79.052009
%  [arXiv:0903.3697 [hep-ex]].
  %%CITATION = doi:10.1103/PhysRevD.79.052009;%%

%\cite{Colangelo:2017fiz}
\bibitem{Colangelo:2017fiz}
  G.~Colangelo, M.~Hoferichter, M.~Procura, and P.~Stoffer,
  %``Dispersion relation for hadronic light-by-light scattering: two-pion contributions,''
  JHEP {\bf 1704}, 161 (2017).
%  doi:10.1007/JHEP04(2017)161
%  [arXiv:1702.07347 [hep-ph]].
  %%CITATION = doi:10.1007/JHEP04(2017)161;%%

%\cite{Bijnens:1999sh}
\bibitem{Bijnens:1999sh}
  J.~Bijnens, G.~Colangelo, and G.~Ecker,
  %``The Mesonic chiral Lagrangian of order p**6,''
  JHEP {\bf 9902}, 020 (1999).
%  doi:10.1088/1126-6708/1999/02/020
%  [hep-ph/9902437].
  %%CITATION = doi:10.1088/1126-6708/1999/02/020;%%

%\cite{Redmer:2018uew}
\bibitem{Redmer:2018uew}
  C.~F.~Redmer (BESIII Collaboration),
  %``Measurement of meson transition form factors at BESIII,''
  arXiv:1810.00654 [hep-ex].
  %%CITATION = ARXIV:1810.00654;%%

%\cite{Guo:2019gjf}
\bibitem{Guo:2019gjf}
  Y.~Guo (BESIII Collaboration),
  %``Two photon physics at BESIII,''
  J.\ Phys.\ Conf.\ Ser.\  {\bf 1137}, no. 1, 012008 (2019).
%  doi:10.1088/1742-6596/1137/1/012008
  %%CITATION = doi:10.1088/1742-6596/1137/1/012008;%%

%\cite{Gasser:1984gg}
\bibitem{Gasser:1984gg}
  J.~Gasser and H.~Leutwyler,
  %``Chiral Perturbation Theory: Expansions in the Mass of the Strange Quark,''
  Nucl.\ Phys.\ B {\bf 250}, 465 (1985).
%  doi:10.1016/0550-3213(85)90492-4
  %%CITATION = doi:10.1016/0550-3213(85)90492-4;%%

%\cite{Donoghue:1989si}
\bibitem{Donoghue:1989si}
  J.~F.~Donoghue and B.~R.~Holstein,
  %``Kaon Transitions and Predictions of Chiral Symmetry,''
  Phys.\ Rev.\ D {\bf 40}, 3700 (1989).
%  doi:10.1103/PhysRevD.40.3700
  %%CITATION = doi:10.1103/PhysRevD.40.3700;%%

%\cite{Guerrero:1997rd}
\bibitem{Guerrero:1997rd}
  F.~Guerrero and J.~Prades,
  %``Kaon polarizabilities in chiral perturbation theory,''
  Phys.\ Lett.\ B {\bf 405}, 341 (1997).
%  doi:10.1016/S0370-2693(97)00656-4
%  [hep-ph/9702303].
  %%CITATION = doi:10.1016/S0370-2693(97)00656-4;%%

%\cite{Backenstoss:1973jx}
\bibitem{Backenstoss:1973jx}
  G.~Backenstoss {\it et al.},
  %``K- mass and k- polarizability from kaonic atoms,''
  Phys.\ Lett.\  {\bf 43B}, 431 (1973).
%  doi:10.1016/0370-2693(73)90391-2
  %%CITATION = doi:10.1016/0370-2693(73)90391-2;%%

%\cite{Moinester:2003rb}
\bibitem{Moinester:2003rb}
  M.~Moinester (COMPASS Collaboration),
  %``Pion and kaon polarizabilities at CERN COMPASS,''
  Czech.\ J.\ Phys.\  {\bf 53}, B169 (2003).
%  [hep-ex/0301024].
  %%CITATION = HEP-EX/0301024;%%

%\cite{Denisov:2018unj}
\bibitem{Denisov:2018unj}
  O.~Y.~Denisov,
  %``Letter of Intent: A New QCD facility at the M2 beam line of the CERN SPS,''
  arXiv:1808.00848 [hep-ex].
  %%CITATION = ARXIV:1808.00848;%%

%\cite{Buenerd:1995dd}
\bibitem{Buenerd:1995dd}
  M.~Bu\'enerd,
  %``Prospects for hadron electromagnetic polarizability measurement by radiative scattering on a nuclear target,''
  Nucl.\ Instrum.\ Meth.\ A {\bf 361}, 111 (1995).
%  doi:10.1016/0168-9002(95)00270-7
  %%CITATION = doi:10.1016/0168-9002(95)00270-7;%%

%\cite{Kowalewski:1984it}
\bibitem{Kowalewski:1984it}
  R.~V.~Kowalewski {\it et al.},
  %``Elastic Pion Compton Scattering,''
  Phys.\ Rev.\ D {\bf 29}, 1000 (1984).
%  doi:10.1103/PhysRevD.29.1000
  %%CITATION = doi:10.1103/PhysRevD.29.1000;%%

%\cite{Abbon:2007pq}
\bibitem{Abbon:2007pq}
  P.~Abbon {\it et al.} (COMPASS Collaboration),
  %``The COMPASS experiment at CERN,''
  Nucl.\ Instrum.\ Meth.\ A {\bf 577}, 455 (2007).
%  doi:10.1016/j.nima.2007.03.026
%  [hep-ex/0703049].
  %%CITATION = doi:10.1016/j.nima.2007.03.026;%%

%\cite{Guskov:2010zna}
\bibitem{Guskov:2010zna}
  A.~Guskov, Analysis of the charged pion polarizability measurement method at COMPASS experiment,
  CERN-THESIS-2010-264.
  %%CITATION = CERN-THESIS-2010-264;%%

%\cite{Nagel:2012cla}
\bibitem{Nagel:2012cla}
  T.~Nagel, Measurement of the Charged Pion Polarizability at COMPASS,
Ph.D.~thesis, TU M\"unchen, 2012, Section 5.2.
%  CERN-THESIS-2012-138.
  %%CITATION = CERN-THESIS-2012-138;%%

%\cite{Friedrich:2012ybb}
\bibitem{Friedrich:2012ybb}
  J.~M.~Friedrich, Chiral Dynamics in Pion-Photon Reactions,
  CERN-THESIS-2012-333.
  %%CITATION = CERN-THESIS-2012-333;%%

%\cite{Akhundov:1994uv}
\bibitem{Akhundov:1994uv}
  A.~A.~Akhundov, S.~Gerzon, S.~Kananov, and M.~A.~Moinester,
  %``Radiative corrections for pion polarizability experiments,''
  Z.\ Phys.\ C {\bf 66}, 279 (1995).
%  doi:10.1007/BF01496602
%  [hep-ph/9410251].
  %%CITATION = doi:10.1007/BF01496602;%%

%\cite{Kaiser:2008hj}
\bibitem{Kaiser:2008hj}
  N.~Kaiser and J.~M.~Friedrich,
  %``Radiative corrections to pion-nucleus bremsstrahlung,''
  Eur.\ Phys.\ J.\ A {\bf 39}, 71 (2009).
%  doi:10.1140/epja/i2008-10688-0
%  [arXiv:0811.1434 [hep-ph]].
  %%CITATION = doi:10.1140/epja/i2008-10688-0;%%

%\cite{Kaiser:2008ss}
\bibitem{Kaiser:2008ss}
  N.~Kaiser and J.~M.~Friedrich,
  %``Cross-sections for low-energy pi- gamma reactions,''
  Eur.\ Phys.\ J.\ A {\bf 36}, 181 (2008).
%  doi:10.1140/epja/i2008-10582-9
%  [arXiv:0803.0995 [nucl-th]].
  %%CITATION = doi:10.1140/epja/i2008-10582-9;%%

%\cite{Friedrich:2016gqb}
\bibitem{Friedrich:2016gqb}
  J.~Friedrich (COMPASS Collaboration),
  %``The pion polarisability and more measurements on chiral dynamics at COMPASS,''
  PoS CD {\bf 15}, 015 (2016).
%  doi:10.22323/1.253.0015
  %%CITATION = doi:10.22323/1.253.0015;%%

%\cite{Antipov:1982kz}
\bibitem{Antipov:1982kz}
  Y.~M.~Antipov {\it et al.},
  %``Measurement of pi- Meson Polarizability in Pion Compton Effect,''
  Phys.\ Lett.\  {\bf 121B}, 445 (1983).
%  doi:10.1016/0370-2693(83)91195-4
  %%CITATION = doi:10.1016/0370-2693(83)91195-4;%%

%\cite{Antipov:1984ez}
\bibitem{Antipov:1984ez}
  Y.~M.~Antipov {\it et al.},
  %``Experimental Evaluation of the Sum of the Electric and Magnetic Polarizabilities of Pions,''
  Z.\ Phys.\ C {\bf 26}, 495 (1985).
%  doi:10.1007/BF01551790
  %%CITATION = doi:10.1007/BF01551790;%%

%\cite{Drechsel:1994kh}
\bibitem{Drechsel:1994kh}
  D.~Drechsel and L.~V.~Fil'kov,
  %``Compton scattering on the pion and radiative pion photoproduction from the proton,''
  Z.\ Phys.\ A {\bf 349}, 177 (1994).
%  doi:10.1007/BF01291177
  %%CITATION = doi:10.1007/BF01291177;%%

%\cite{Kao:2008pf}
\bibitem{Kao:2008pf}
  C.~w.~Kao, B.~E.~Norum, and K.~Wang,
  %``Extraction of the charged pion polarizabilities from radiative charged pion photoproduction in heavy baryon chiral perturbation theory,''
  Phys.\ Rev.\ D {\bf 79}, 054001 (2009).
%  doi:10.1103/PhysRevD.79.054001
%  [hep-ph/0409081].
  %%CITATION = doi:10.1103/PhysRevD.79.054001;%%

%\cite{Alexander:1993rz}
\bibitem{Alexander:1993rz}
  G.~Alexander {\it et al.},
  %``Two photon physics capabilities of KLOE at DAPHNE,''
  Nuovo Cim.\ A {\bf 107}, 837 (1994).
%  doi:10.1007/BF02731100
  %%CITATION = doi:10.1007/BF02731100;%%

%\cite{Ong:1988tr}
\bibitem{Ong:1988tr}
  S.~Ong, P.~Kessler, and A.~Courau,
  %``Azimuthal Correlations in Double Tag Measurements of Photon-Photon Collisions,''
  Mod.\ Phys.\ Lett.\ A {\bf 4}, 909 (1989).
%  doi:10.1142/S0217732389001076
  %%CITATION = doi:10.1142/S0217732389001076;%%

\bibitem{Pasquini:2017}
B.~Pasquini, private communication (2017).

%\cite{Filkov:2008vak}
\bibitem{Filkov:2008vak}
  L.~V.~Fil'kov and V.~L.~Kashevarov,
  %``Comment on `Polarizability of the pion: No conflict between dispersion theory and chiral perturbation theory',''
  Phys.\ Rev.\ C {\bf 81}, 029801 (2010).
%  doi:10.1103/PhysRevC.81.029801
%  [arXiv:0805.4486 [hep-ph]].
  %%CITATION = doi:10.1103/PhysRevC.81.029801;%%

\bibitem{Filkov:2017}
   L.~V.~Fil'kov and V.~L.~Kashevarov,
   %``Dipole Polarizabilities of Charged Pions,''
   Phys.\ Part.\ Nucl.\ {\bf 48}, 117 (2017).

\bibitem{Filkov:2018}
   L.~V.~Fil'kov and V.~L.~Kashevarov,
   %``Dipole Polarizabilities of $\pi^\pm$-Mesons,''
   Int.\ J.\ Mod.\ Phys.: Conf.\ Series {\bf 47}, 1860092 (2018).

%\cite{Pasquini:2009ze}
\bibitem{Pasquini:2009ze}
  B.~Pasquini, D.~Drechsel, and S.~Scherer,
  %``Reply to 'Comment on `The Polarizability of the pion: No conflict between dispersion theory and chiral perturbation theory',''
  Phys.\ Rev.\ C {\bf 81}, 029802 (2010).
%  doi:10.1103/PhysRevC.81.029802
%  [arXiv:0908.4291 [hep-ph]].
  %%CITATION = doi:10.1103/PhysRevC.81.029802;%%

\bibitem{Colangelo:2001sp}
  G.~Colangelo, J.~Gasser, and H.~Leutwyler,
  %``The quark condensate from K(e4) decays,''
  Phys.\ Rev.\ Lett.\  {\bf 86}, 5008 (2001).
  %[arXiv:hep-ph/0103063].
  %%CITATION = PRLTA,86,5008;%%

%\cite{Wess:1971yu}
\bibitem{Wess:1971yu}
  J.~Wess and B.~Zumino,
  %``Consequences of anomalous Ward identities,''
  Phys.\ Lett.\  {\bf 37B}, 95 (1971).
%  doi:10.1016/0370-2693(71)90582-X
  %%CITATION = doi:10.1016/0370-2693(71)90582-X;%%

%\cite{Witten:1983tw}
\bibitem{Witten:1983tw}
  E.~Witten,
  %``Global Aspects of Current Algebra,''
  Nucl.\ Phys.\ B {\bf 223}, 422 (1983).
%  doi:10.1016/0550-3213(83)90063-9
  %%CITATION = doi:10.1016/0550-3213(83)90063-9;%%

%\cite{Fuchs:2000pn}
\bibitem{Fuchs:2000pn}
  T.~Fuchs, B.~Pasquini, C.~Unkmeir, and S.~Scherer,
  %``Virtual Compton scattering off the pseudoscalar meson octet,''
  Czech.\ J.\ Phys.\  {\bf 52}, B135 (2002).
%  doi:10.1007/s10582-001-0051-3
%  [hep-ph/0010218].
  %%CITATION = doi:10.1007/s10582-001-0051-3;%%

%\cite{Ocherashvili:2001aj}
\bibitem{Ocherashvili:2001aj}
  A.~Ocherashvili {\it et al.} (SELEX Collaboration),
  %``First Measurement of $\pi^- e \to \pi^- e\gamma$ Pion Virtual Compton Scattering,''
  Phys.\ Rev.\ C {\bf 66}, 034613 (2002).
%  doi:10.1103/PhysRevC.66.034613
%  [hep-ex/0109003].
  %%CITATION = doi:10.1103/PhysRevC.66.034613;%%

%\cite{Unkmeir:2001gw}
\bibitem{Unkmeir:2001gw}
  C.~Unkmeir, A.~Ocherashvili, T.~Fuchs, M.~A.~Moinester, and S.~Scherer,
  %``Pion generalized dipole polarizabilities by virtual Compton scattering pi e ---> pi e gamma,''
  Phys.\ Rev.\ C {\bf 65}, 015206 (2002).
%  doi:10.1103/PhysRevC.65.015206
%  [hep-ph/0107020].
  %%CITATION = doi:10.1103/PhysRevC.65.015206;%%


\end{thebibliography}
\end{document}